\documentclass{article}
\usepackage[utf8]{inputenc}
\usepackage{settings}
\usepackage{makecell}
\usepackage{tabularx}
\usepackage{changepage}

\newcolumntype{B}{>{\centering\arraybackslash\hsize=.09\hsize}X}
\newcolumntype{K}{>{\centering\arraybackslash\hsize=.12\hsize}X}
\newcolumntype{Y}{>{\centering\arraybackslash}X}

\usepackage{float}
\newfloat{Table}{H}{grf}

\usepackage{microtype}              
\usepackage{ragged2e}
\usepackage{tabularray}
\UseTblrLibrary{booktabs}           
\NewTableCommand\category[1][0pt]{
        \SetRow{abovesep+=#1}
        \SetCell[c=4]{l, font=\small\itshape\bfseries}
\SetTblrStyle{contfoot-text}{font=\footnotesize\itshape}}

\usepackage{xargs}
\newcommandx{\transmat}[2][1={i,j},2={}]{\mathcal{P}_{#1}^{#2}}
\newcommandx{\transmatf}[2][1={i,j},2={}]{\tilde{\mathcal{P}}_{#1}^{#2}}

\newcommandx{\transmatbinary}[2][1={i,j},2={}]{\mathcal{M}_{#1}^{#2}}
\newcommandx{\fitness}[2][1={(n)},2={i}]{F^{#1}_#2}
\newcommandx{\fitnesstemp}[2][1={(n)},2={i}]{\tilde{F}^{#1}_#2}
\newcommandx{\complexity}[2][1={(n)},2={j}]{Q^{#1}_#2}              
\newcommandx{\complexitytemp}[2][1={(n)},2={j}]{\tilde{Q}^{#1}_#2}
\newcommandx{\threshacc}{\Theta^{A}}
\newcommandx{\threshtrans}{\Theta^{T}}

\newcommand{\norm}[1]{\left\lVert#1\right\rVert}

\title{ 
\textbf{The Structure of Occupational Mobility in France}
}

\author{\normalsize Max Sina Knicker$^{1,2}$, Karl Naumann-Woleske$^{1,2,3}$, and Michael Benzaquen$^{1,2,4,}$\footnote{michael.benzaquen@polytechnique.edu}}

\date{
\small
\textit{
$^1$Chair of Econophysics and Complex Systems,  Ecole Polytechnique,  91128 Palaiseau, France\\%
$^2$LadHyX, UMR CNRS 7646, Ecole Polytechnique, Institut Polytechnique de Paris, 91128 Palaiseau, France\\
$^3$Vienna University of Economics and Business (WU), 
Welthandelsplatz 1, 
1020 Vienna, Austria\\%
$^4$Capital Fund Management, 23 rue de l'Université, 75007 Paris, France\\%
}\medskip
\today}

\begin{document}
\maketitle

\captionsetup{margin=10pt,font=footnotesize,labelfont=bf,labelsep=endash,justification=centerlast}

\begin{abstract}
In an era of rapid technological advancements and macroeconomic shifts, worker reallocation is necessary, yet responses to labor market shocks remain sluggish, making it crucial to identify bottlenecks in occupational transitions to understand labor market dynamics and improve mobility. In this study, we analyze French occupational data to uncover patterns of worker mobility and pinpoint specific occupations that act as bottlenecks which impede rapid reallocation. We introduce two metrics, transferability and accessibility, to quantify the diversity of occupational transitions and find that bottlenecks can be explained by a condensation effect of occupations with high accessibility but low transferability. Transferability measures the variety of transitions from an occupation to others, while accessibility assesses the variety of transitions into an occupation. We provide a comprehensive framework for analyzing occupational complexity and mobility patterns, offering insights into potential barriers and pathways for efficient retraining programs. We argue that our approach can inform policymakers and stakeholders aiming to enhance labor market efficiency and support workforce adaptability.

\end{abstract}

\section{Introduction}\label{sec:intro}

Reallocating workers across occupations is crucial for resilient labor markets, allowing economies to adapt to technological innovations and economic fluctuations. However, this process often encounters challenges. Structural barriers, such as skill mismatches and occupational rigidities, can impede the smooth transition of workers \citep{adao2022fast}, and the presence of bottleneck occupations further contributes to sluggish reallocation \citep{bocquet2022network}. As labor markets evolve with automation and digital transformation, understanding the mechanisms that hinder worker transitions becomes increasingly important. For example, innovations like Generative Artificial Intelligence (GenAI) are diminishing the need for repetitive tasks, potentially affecting 44\% of working hours in impacted industries \citep{hbr2023}. At the same time, there is a growing demand for workers in technology-related roles, such as information security specialists \citep{di2023future}, which underscores the need for effective worker reallocation.

In this paper, we investigate the question of how to identify occupations that may impede rapid reallocation of workers across occupations using a fitness and complexity-inspired method. 
Our main contribution is the development of novel, data-driven metrics for accessibility and transferability that quantify occupational mobility and effectively identify bottleneck occupations in the labor market.
Transferability is a measure of the diversity of potential destinations when leaving a given occupation, while accessibility is a measure of the diversity of different occupations from which a given occupation can be accessed. 
Using these metrics, we identify \textit{condenser} occupations, with low transferability and high accessibility, that act as bottlenecks in occupational transitions. 

It is important to note that when we refer to \textit{bottlenecks} or \textit{condenser} occupations, we do so in a descriptive manner. These terms are used to characterize structural features of the observed transition network, highlighting areas of low inter-community mobility or high condensation, without implying any normative judgment about the labor market's optimal functioning.

We apply our framework to the French labor market using employer tax data from 2012 to 2020 to construct year-on-year occupational transition matrices (Section \ref{sec:constructMatrix}). 
The transition matrix exhibits an inherent block structure, with, on average, only one out of six occupations per block exhibiting significant inter-block transitions (Section \ref{sec:clusteringBRIM}). 
The implied steady state of this occupational transition matrix is far from the current distribution of workers across occupations.
Over half of the occupations deviate more than 10\% from the implied steady state distribution (Section \ref{sec:stationaryDist}). 
Furthermore, transitions into and out of the employed labor force account for $\sim40\%$ of all inter-community transitions, suggesting the existence of important barriers hindering direct movement between occupational communities (Section \ref{sec:entryExit}).

To identify the occupations acting as bottlenecks to reallocation we introduce two metrics:  transferability and accessibility.
Transferability quantifies the diversity of outside options for a given occupation, while accessibility quantifies the diversity of pathways to a given occupation (Section \ref{sec:fitnessandcomplex}). 
This enables us to classify occupations into four categories: (1) Hub, (2) Diffuser, (3) Channel, and (4) Condenser occupations.
Overall, 93\% of occupations are condensers, which emerge as bottleneck occupations due to their low transferability but high accessibility: many workers can transition into these occupations but they offer limited follow-on opportunities. 
This aligns with human capital theory, which posits that as workers gain experience and accumulate occupation-specific skills, their incentives to transition to different occupations decrease, resulting in lower mobility over time \citep{becker2009human, marginson2019limitations}. Our framework complements this theory by introducing metrics that quantify the structural properties of actual transitions, offering a novel perspective on the constraints in labor reallocation while remaining consistent with the underlying principles of human capital theory.

Furthermore, we find that transferability is negatively correlated with the number of employees in a given occupation. 
In Section~\ref{ssec:bottlenecks}, we show that our accessibility metric successfully identifies occupations with high betweenness centrality, a metric commonly used in the literature to determine bottlenecks.
In particular, occupations with high accessibility show high skill similarity with occupations inside \textit{and} outside of the community, making them the entry-points for inter-community transitions. 

Our main contribution to the literature is the additional information gained from the transferability metric.
Specifically, occupations with high transferability have a high skill similarity to occupations outside of the community, thus serving as the origins of inter-community occupational transitions. 
Together with the negative relationship between transferability and the number of employees in an occupation, it is these occupations that may be at the heart of slow labor reallocation. 
To demonstrate how these metrics can effectively inform policy interventions, Section~\ref{sec:policy} illustrates that leveraging both transferability and accessibility metrics can help identify occupations for targeted policies aimed at reducing frictions in the labor market.

\section{Literature Review}

At the macro-level, structural shifts may be attributed to technological advancements \citep{atalay2018new} or macroeconomic events such as the  ``China Shock" \citep{autor2021persistence} or the COVID-19 pandemic \citep{cortes2023heterogeneous}.
Empirical evidence suggests that worker reallocation following structural shocks is a slow process \citep{dix2023globalization, autor2021persistence}. 
Understanding the reasons for slow worker reallocation is crucial, as it may influence policymakers' decisions on labor market interventions, particularly in the context of pandemic-induced job market shifts, the green transition, and the impacts of automation. 

Several factors have been identified as contributing to slow worker reallocation. 
First, skills are often imperfectly transferable between occupations, leading to mismatches that impede transitions \citep{adao2022fast, bocquet2022network, buera2022skill}. 
\cite{adao2022fast} finds that, in occupations with strong skill specificity, the pace of labor market adjustments slows down because the transition relies less on reallocating older workers and more on gradually integrating younger workers into these roles.
\cite{bocquet2022network} was the first to highlight and quantify the role of bottleneck or ``bridge occupations" on the speed of reallocation. He finds that reallocation is significantly
slowed down due to the structure of the occupation network and emphasizes that the impact of skill
frictions can be effectively modeled within a search and matching framework, making the underlying
effects more tractable.

Second, local labor markets are often imperfectly connected on an occupational and sectoral level \citep{schmutz2020search, dix2023globalization}. 
\cite{kline2013place} emphasize geographical barriers, suggesting that labor markets in highly competitive and productive cities are not well-connected, leading to a scarcity of occupational transitions. 
\cite{schmutz2020search} develop a general equilibrium search and matching model to incorporate these geographical frictions. 

Third, recent approaches emphasize analyzing labor flows and their network patterns to understand occupational mobility, which tracks workers' transitions through a complex network based on various attributes \citep{schmutte2014free, del2021occupational, bocquet2022network, joyez2022occupation}. 

Incorporating skill data into analyses of labor flows has revealed that skill-relatedness not only facilitates labor transitions between occupations but also promotes regional diversification and local industry growth \citep{neffke2013skill, neffke2017inter, jara2018role, o2022modular}. 
A recent study by \cite{aufiero2024mapping} investigates the relationship between job fitness, skill coherence, and wages in the US labor market by employing network-based tools from the Economic Complexity framework.
Going beyond these studies, we here relate network properties to slow worker reallocation by developing a set of metrics to measure and understand the patterns that lead to slow reallocation.

\section{Occupational Transition Network of France}\label{sec:occupationNetwork}

Using micro-data covering all employees in France, we construct an occupational transition network for the period 2012-2020 (Section \ref{sec:constructMatrix}) to identify bottlenecks in labor flows by examining the  labor market's implied steady state, community structure, and entry and exit dynamics. 
We observe a scarce transition matrix with a strong community structure, where only 9\% of all transitions occur across communities (Section \ref{sec:clusteringBRIM}), and with over half of all occupations deviating by over 10\% from their implied steady state (Section \ref{sec:stationaryDist}).
Furthermore, we show that, on average, 40\% of the total inflow to each occupational community originates from outside the labor market. 
These results indicate that there are significant constraints in the reallocation of workers in the French labor market, which appears to be out of its equilibrium state. 

\begin{figure}[t!]
    \centering
    \includegraphics[width=0.8\textwidth]{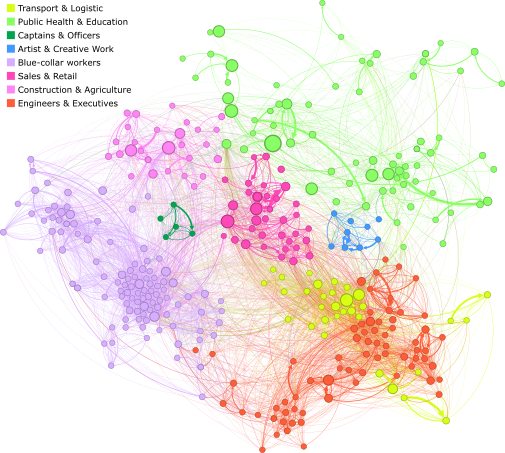}
    \caption{\textbf{Occupational Transition Network of France}: Graph visualization of the weighted and directed labor market network derived from the transition probability matrix $\transmat$ (see eq.~\ref{eq:pij}), computed from data spanning the years 2012 to 2020. Each node symbolizes an occupation, with links that illustrate transitions between them. Node sizes correspond to the occupation's workforce size, while node color indicates its BRIM community~\citep{Blondel} (see Section~\ref{sec:clusteringBRIM}). Line widths are proportional to the transition probability $\transmat[i,j]$. The layout is based on the OpenOrd algorithm \citep{martin2011openord}, and the graph was generated using Gephi \citep{bastian2009gephi}.}
    \label{fig:occ_net_2020}
\end{figure}

\subsection{Constructing the Occupational Transition Network}\label{sec:constructMatrix}

The object of analysis is the right-stochastic occupational transition matrix $\transmat[i,j][T]$, where each entry 
\begin{equation}\label{eq:pij}
    \transmat[i,j][T] = \mathbb{P}(X_i^{T} = i | X_j^{T-1} = j)
\end{equation}
corresponds to the realized transition rate from occupation $j$ to occupation $i$ for the period (T-1, T). 
The matrix is normalized such that $\sum_i \transmat[i,j][T] = 1$, and can be interpreted an observation of the underlying probability of a transition from $j$ to $i$.
The matrix $\transmat$ defines a weighted directed network of occupational transitions that is visually represented in Figure \ref{fig:occ_net_2020}.

To estimate $\transmat$, we use the BTS-POSTES dataset provided by the National Institute of Statistics and Economic Studies of France (INSEE).\footnote{Access to this confidential data was gained through the Centre d’accès sécurisé aux données (Ref. 10.34724/CASD). The persistence identifier for the dataset of the year 2020 is \url{https://doi.org/10.34724/CASD.21.5012.V1}}  
The dataset covers all employees in France for a sequence of two-year snapshots, such that one can track an individual through the years $T-1$ and $T$. 
To construct $\transmat$, we consider an individual's change in \textit{primary} occupation within a $(T-1,T)$ year period. 
The primary occupation is defined as occupation with the highest net salary provided that at least 30 days have been worked in the job and the salary is at least the minimum wage.\footnote{The indicator is defined by INSEE and is called PPS. In case of a tie, the job with the highest number of hours worked is taken. The minimum salary threshold refers to the minimum wage for full-time, full-month employment in France, ensuring that primary occupations considered in the analysis meet a baseline for economic activity.}
If an individual does not have a primary occupation we classify them as out of the occupational network.
Given each individual's primary jobs in $T-1$ and $T$, we then consider their change in occupation.\footnote{An occupation is defined by the nomenclatures of Professions and Socio-professional Categories of Salaried Jobs of Private and Public Employers (PCS-ESE). Here we use the 4-digit PCE-ESE nomenclature}
We thus focus on the network of year-on-year occupational transitions.

\begin{figure}[htb!]
    \centering
    \includegraphics[width=1\textwidth]{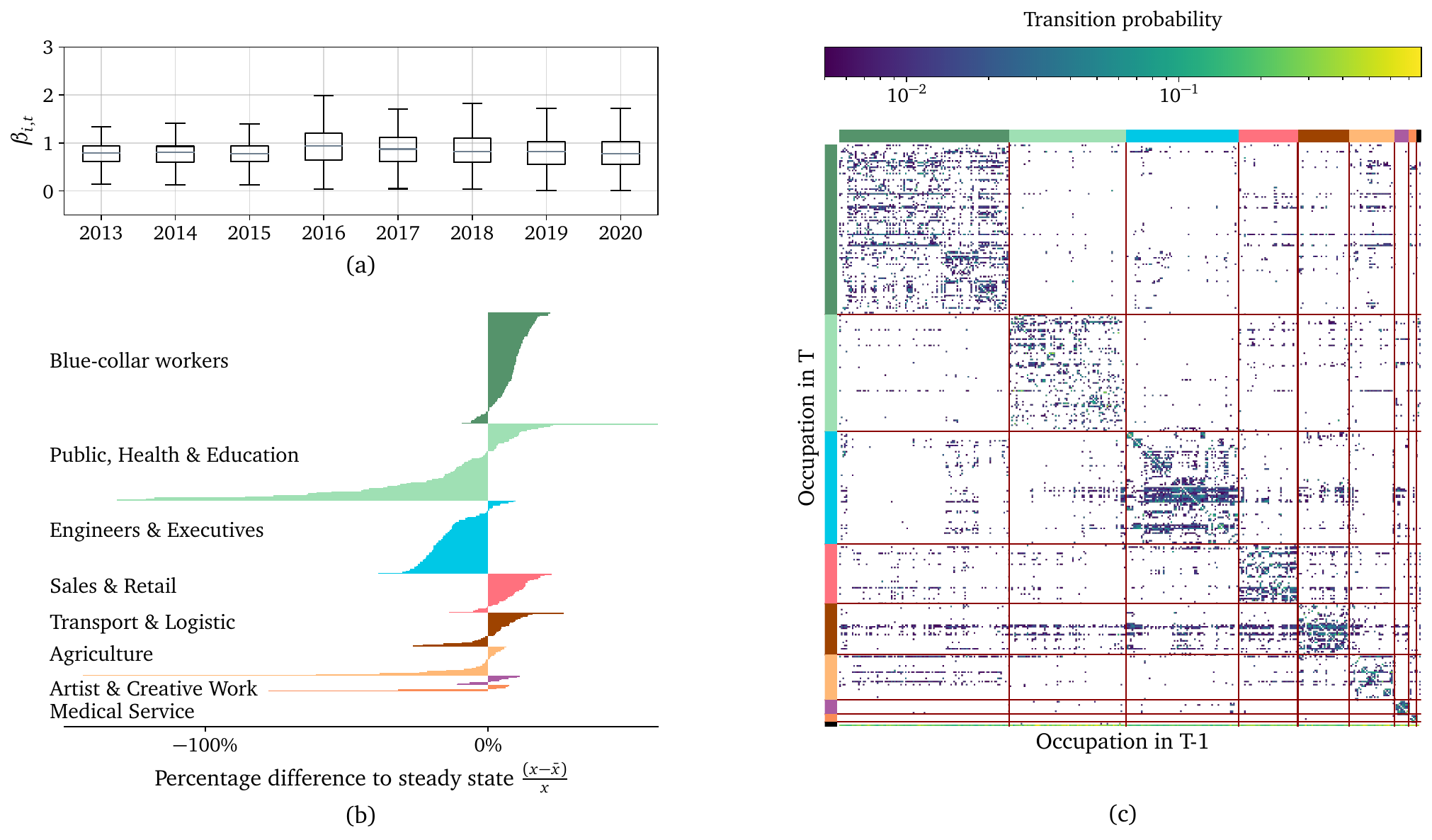}
    \caption{\textbf{Occupational Transition Matrix and Steady State Distribution} (a) Temporal stability analysis from 2012-2020 where each box corresponds to the distribution of standardized coefficient $\beta_{i,t}$ (see Appendix~\ref{sec:tmpstability}) estimated by the regression in eq.~\ref{eq:tempstabocc}. The grey line in each box corresponds to the median of the distribution. (b) Bar plot of percentage difference between the actual share of workers in each occupation $x$ and the share of workers implied by the steady state $\bar{x}$ given by the eigenvector corresponding to the eigenvalue $\lambda = 1$ (see Section~\ref{sec:stationaryDist}). (c)~Transition probability matrix $\transmat$ constructed from labor flow data (BTS-Postes dataset), where each entry represents the fraction of individuals in~$(T-1)$ (columns) who transitioned to occupation in~$(T)$ (rows). Self-loops are excluded. Red lines and colored bars indicate the BRIM communities.}
    \label{fig:transition_prop}
\end{figure}

We aggregate all year-on-year observations for the years 2012-2020. 
In total, we account for 253.3 million person-year observations, among which 41.2 million (16.2\%) undergo primary-to-primary occupational transitions and 184.9 million (73.0 \%) remain in the same occupation, 10.9 million (4.3\%) leave the network and 16.3 million (6.5\%) enter the occupational network. 
The box-plot in Figure \ref{fig:transition_prop}(a) shows the regression coefficients of the year-specific transition matrices in comparison to 2012, implying that it is reasonable to assume temporal stability for the purposes of this paper (see Appendix \ref{sec:tmpstability}).\footnote{We perform the regression $\transmat[{i,j}][t] = \alpha_{i,t} + \beta_{i,t} \transmat[{i,j}][t^\prime=2012] + \epsilon_{i,t}\quad \forall t\in\{2013,\dots,2020\}$ and assume temporal stability if $\alpha_{i,t} \rightarrow 0$ and $\beta_{i,t} \rightarrow 1$ (for more details see Appendix \ref{sec:tmpstability} Figure \ref{fig:temporalstabilityall})}  
The combination of the matrices across the years yields 372 distinct occupation codes (PCS-ESE), as we retain the occupations that are present in all years. 
Consequently, we removed observations of occupation codes not present in all years from the dataset, which account for 10.0 million person-year datums (3.7\%).
The construction of $\transmat$ naturally excludes job changes where an individual has an extended unemployment spell, whether this be due to actively searching for a job, training, leaving or entering the workforce, or entering extended periods of leave such as maternity leave. 
Furthermore, we  do not account for changes in companies or location if the individuals remain within the same occupational code.

To mitigate noise effects, links between two occupations are removed if the flow is below a fraction $\Theta=1\%$ from the originating occupation.\footnote{The results provided in Section \ref{sec:taxonomysec} remain robust with small changes in the order of a few percent in $\Theta$.}
For $\Theta=0\%$ each occupation connects to $278$ other occupations on average, with many connections due to a few individuals transitioning. 
With the filtering process we remove $5.3\%$ of observations from the data which results in a total of $6900$ existing links, with $18$ links per occupation on average. 
This magnitude aligns with estimates provided by France Travail's ROME job classification system,\footnote{See the official website \url{https://www.data.gouv.fr/en/datasets/repertoire-operationnel-des-metiers-et-des-emplois-rome/}.} where an average of $13$ transitions per occupation is deemed possible, considering similar skill and work environments.

Note that in studying occupational transitions, two types of networks are commonly considered: the skill-based network (see e.g. \cite{bocquet2022network, grigsby2022skill}) and the observed transition network, which is the focus of this study. The skill-based network uses measures of skill similarity to map potential pathways between occupations based on shared competencies. A key advantage of this approach is its focus on identifying potential mobility opportunities based on skill relatedness, providing insights into the set of occupations that workers might feasibly transition into if they possess the relevant skills \citep{neffke2013skill}. However, skill-based networks can only explain part of the mobility structure and have an ex-ante nature, meaning they do not account for realized choices made by workers and firms in the labor market. Therefore our study focuses on the observed transition network, which reflects actual labor market flows based on realized worker movements and goes beyond a skill-based network, revealing structural patterns that are not accounted for by skill similarity alone. This choice provides a more empirical, agnostic perspective, capturing the dynamic interplay between worker preferences, employer demands, wage offers and economic conditions, which shapes the real mobility patterns observed in the labor market.

\subsection{Occupational Communities}\label{sec:clusteringBRIM}
Both the occupational transition network (Figure \ref{fig:occ_net_2020}) and the occupational transition matrix (Figure \ref{fig:transition_prop}) reveal a community structure indicating barriers to the free reallocation of workers among jobs. As expected, workers are most often \textit{trapped} within an occupational community: on average, 91\% of transitions happen within a given community and only 1 out of 6 occupations per community (on average 17\%) have over 10\% of their outgoing transitions to different communities.\footnote{We define an occupation with an inter-community transition as one where over 10\% of transitions occur outside its community. If we set the threshold at 5\%, this would result in 2 out of 5 occupations per community exhibiting inter-community transitions.} 

To identify the occupational communities, we utilize the BRIM algorithm (See Appendix~\ref{appx:brimcluster}), which is designed to maximize the modularity of \citet{barber2007modularity}, due to its rapid convergence and its ability to incorporate interdependence between communities across the network \citep[see][and Appendix \ref{appx:brim}]{platig2016bipartite}. To quantify the intra-community density compared to inter-community density of flows we use the same network modularity as in \citet{barber2007modularity}.\footnote{Modularity serves as a metric for assessing the clustering of networks or graphs, quantifying the effectiveness of partitioning a network into distinct communities. Networks exhibiting high modularity contain tightly connected nodes within communities while showing limited connections between nodes belonging to different communities. The equation to calculate modularity (see eq.~\ref{eq:modularity}) is a bipartite expansion of the modularity proposed by \citet{newmanGirvan2004modularity}.}

Employing the BRIM algorithm results in 8 communities and a modularity of 0.45. The communities identified through the maximization of modularity broadly correspond to (1) Blue-collar workers, (2) Public, Health \& Education, (3) Engineers \& Executives, (4) Sales \& Retail (5) Transport \& Logistic (6) Agriculture, (7) Artist \& Creative Work (8) Medical Service (see Appendix \ref{appx:brimcluster} for tables with exact community membership). 
Although communities could also be given by the INSEE's 1-digit PCS codes, we choose the BRIM communities over the INSEE partition since they have a higher modularity of 0.45, compared to 0.36 when using the 1-digit PCS.\footnote{The INSEE's PCS 372 occupational types are grouped into six broad subgroups, represented by the red separators in Figure \ref{fig:transition_prop} (a): (1) Agriculture (Agriculteurs), (2) Craftsmen, traders, and entrepreneurs (Artisans, commerçants et chefs d'entreprises), (3) Managers and higher intellectual professions (Cadres et professions intellectuelles supérieures), (4) Intermediate Occupations (Professions intermédiaires), (5) Employees (Employés), and (6) Workers (Ouvriers). Note that within the Agriculture category there is only one single occupation.} 
The community membership structure and community size distribution determined by BRIM is robust across different clustering algorithms.\footnote{We employed the BiLouvain \citep{blondel2008fast} and infomap \citep{rosvall2008maps} community detection methods.}

The low inter-community mobility represents \textit{bottlenecks} in occupational flows in the French labor market \citep[see also][]{bocquet2022network}. Similarly, \citet{neffke2013skill} also find an inherent community structure within the occupational transition network for Sweden, suggesting frictional reallocation.
They attribute this phenomenon primarily to the skill-relatedness of different jobs: individuals are more likely to transition to jobs that require similar skill-sets, whereas different skill-sets represent a higher barrier to making a given transition thus potentially requiring additional training for some transitions.

\subsection{Stationary Distribution of Occupations}\label{sec:stationaryDist}

Assuming that the right-stochastic transition matrix $\transmat$ remains constant (consistent with our findings), it can be thought of as the transition matrix of a Markov chain, which implies a steady state distribution of occupational employment shares to which the system converges. 
The implied steady state distribution $\bar{x}$ over the different occupations is given by the eigenvector of $\transmat$ associated with the eigenvalue $\lambda=1$, while the rate of convergence for individual occupations is given by the eigenvector associated with the second largest eigenvalue. 

While the steady-state distribution does not necessarily represent an optimal or desirable outcome from an economic perspective, it serves as a useful benchmark for understanding the structural dynamics of the labor market. The steady-state distribution provides a baseline that reflects the equilibrium implied by the observed transition matrix. It allows us to quantify deviations from this baseline and identify occupation communities where flows appear constrained or misaligned with the tendency of the network. By benchmarking against the steady-state distribution, we can identify patterns of excess or underemployment, informing where interventions might be necessary to facilitate smoother labor market transitions.

We analyze the implied steady-state distribution for each BRIM community and find that the current state of the labor market deviates from the implied steady state. This deviation can be explained by the block structure of the occupational transition matrix due to limited inter-community flows (see Figure \ref{fig:transition_prop}). 

Figure \ref{fig:transition_prop}(b) shows the percentage difference between the data and the implied steady state for the occupation share in each occupation. Deviation of the current distribution from the implied steady state suggests either a slower convergence rate compared to structural changes or insufficient time for the system to reach equilibrium. 
These observations indicate labor market constraints that impede the system from reaching a steady state. Moreover, the majority of transitions occurs within communities, which explains the over and under-representation of workers in different communities in comparison to the occupation's implied steady state.

Over half of all occupations (223 out of 372 occupations) exhibit deviations of more than 10\% from the steady state distribution. Specifically, 88.1\% of occupations classified as Blue-collar workers, 84.2\% of Sales \& Retail, and 69.7\% of Transport \& Logistics occupations experience excess employment. In contrast, the communities of Engineers \& Executives and Public, Health \& Education are underemployed, with 86.1\% and 64.0\% of their occupations, respectively, having fewer workers than the steady-state distribution suggests. In other communities, the number of occupations with excess employment and underemployment are balanced (see Figure \ref{fig:transition_prop}(b)).
The half-time until convergence of the system to its implied equilibrium is given by the spectral gap of $\transmat$, which yields $\frac{1}{1-\lambda_2} \approx 31$ years. Putting this timescale into perspective is challenging, but if worker reallocation were fast, we would expect an order of magnitude of few years rather than few decades. For example, \cite{bocquet2022network} finds that the speed of labor market reallocation after an asymmetric shock is around 1 year. This is because they assumes a new transition matrix due to the shock and measures transition time using a normalized cumulative impulse response. In contrast, we assume a stable transition matrix and measure the timescale using the spectral gap between the first and second eigenvalues of the transition matrix. The timescale of convergence of 31 years suggests that it is slower than small structural shifts in the labor market, implying that worker reallocation to the implied steady state in the French labor market tends to be sluggish.

\subsection{Entry and Exit from the Occupation Network}\label{sec:entryExit}

Since there is low inter-community mobility within our network, the only other option for entry and exit from a given community is by entry and exit from the occupational network.
Over the period of 2012-2020, 16.3 million (2.0 million p.a.) individuals entered the occupational network, while 10.9 million (1.4 million p.a.) exited, resulting in a net inflow of 600,000 workers per year on average.

\begin{figure}[t!]
    \centering
    \includegraphics[trim={0cm 0cm 0cm 0cm},clip,width=1.0\textwidth]{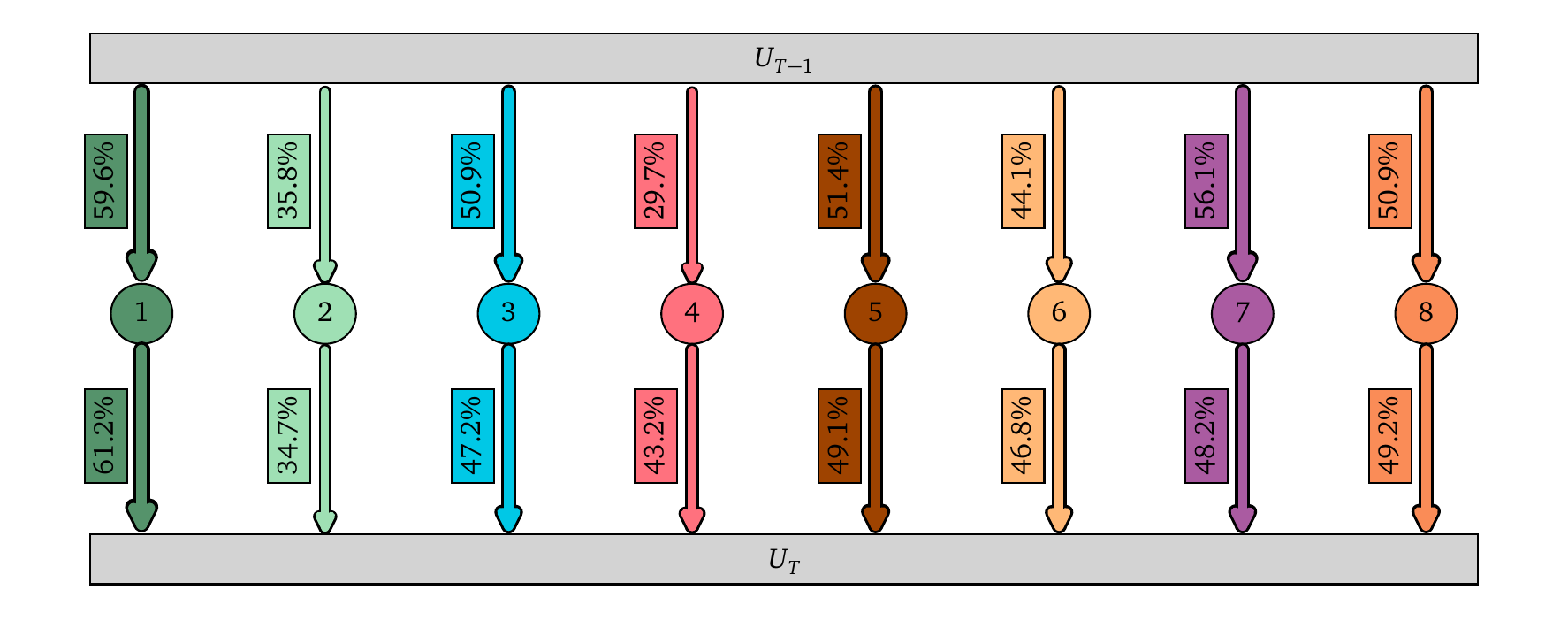}
    \caption{\textbf{Inflow and Outflow Dynamics of the Labor Market}: Each node in the network represents an occupational BRIM community: (1) Blue-collar workers, (2) Public, Health \& Education, (3) Engineers \& Executives, (4) Sales \& Retail (5) Transport \& Logistic (6) Agriculture, (7) Artist and Creative Work (8) Medical Service and (U) Out of the occupation network. Flows indicate transitions to and from outside the labor market normalized by the total number of transitions from and to the community.}
    \label{fig:entryexitnetwork}
\end{figure}

Figure~\ref{fig:entryexitnetwork} shows the inflow and outflow dynamics for each occupational community from and to outside of the labor market.
The width of link from node $i$ to $j$ in Fig.~\ref{fig:entryexitnetwork} corresponds to the proportion of people transitioning from $i$ to $j$ out of the total number of individuals leaving $i$ and the total number of individuals transitioning into $j$.

On average 47.3\% of transitioning workers enter an occupational community from outside the occupational network, while 47.5\% of those leaving an occupational community enter a state of unemployment (see Figure \ref{fig:entryexitnetwork}). 
This underscores the substantial influence of  unemployment on inter-community transitions. The prevalence of an interim unemployment phase preceding inter-community transitions suggests considerable bottlenecks for direct transitions between communities. We leave the study of the unemployment dynamics to future work, as our sample only includes observations in the period starting $T-1$ to the end of $T$, implying that we cannot ascertain where those unemployed at the end of $T$ potentially re-enter the occupational network in $T+1$.

When separating the occupational network across different age groups, we observe that for individuals under the age of 30, the number of workers entering the occupational network is 25\% greater than the number of workers exiting, while for workers over 60, the total number of workers exiting exceeds the number of people entering by 100\% (see Appendix \ref{appx:entryExit} Figure \ref{fig:entryexit}).
This implies a strong role for job-entry and retirement exit dynamics across the unemployment transitions. To adjust for these effects, we note that excluding individuals under the age of 18 entering the occupational network, and those above the age of 64 leaving the occupational network, the  ratio between transitions into and out of the labor force with the total number of direct occupation-to-occupation transitions shrinks from $14.2\%$ to $6.3\%$, such that we can safely neglect these dynamics  in the present analysis.

\section{Identifying Bottlenecks: A Data-driven Taxonomy}\label{sec:taxonomysec}

The occupational transition network developed in Section \ref{sec:occupationNetwork} shows evidence of a strongly clustered structure, with a majority of transitions occurring within a community. 
In this Section we apply methods from the fitness and complexity literature to develop a taxonomy of occupations, classifying them either as (i) Hubs, (ii) Condensers, (iii) Diffusers, or (iv) Channels, based on their accessibility and transferability (Section \ref{sec:fitnessandcomplex}). 
Accessibility (fitness) describes the diversity of occupations from which a given occupation can be reached, while transferability (inverse complexity) describes the diversity of occupations that can be reached from a given occupation.
We show that the majority of occupations in the French labor market exhibit a higher accessibility than transferability, resulting in workers becoming \textit{condensed} into these occupations, with limited opportunities for transitioning elsewhere, thus Condenser and Channel occupations emerge as bottleneck occupations. We find that the accessibility metric correctly identifies bottlenecks as it shows high correlation with betweenness centrality (\ref{ssec:bottlenecks}). Additionally transferability correlated with closeness centrality showing that entering high transferability occupation helps workers to diffuse into different communities of the network. Furthermore, we show that our metrics provide information not otherwise contained in factors such as skill similarities across occupations (Section \ref{sec:skilleff}). See Appendices \ref{sec:gendereff} and \ref{sec:ageeff} for gender and age effects. Future research could be directed towards understanding how the pressure originating from high accessibility but low transferability might, to some extent,  explain the occupational wage-distribution.

\subsection{Accessibility and Transferability of Occupations}\label{sec:fitnessandcomplex}

We quantify the accessibility and transferability of an occupation by measuring the diversity of transitions to and from that occupation using economic fitness and complexity metrics that describe a hierarchy as nested patterns in a system \citep{tacchella2012new}. 
Economic fitness and complexity has previously been used to quantify the productivity and technology composition of economies \citep{tacchella2012new, zaccaria2014taxonomy, hidalgo2009building}, with a particular focus on economic and technological growth and development using patent and export data \citep{morrison2017economic,sbardella2018green,straccamore2023urban}.
Fitness and complexity have only recently been related to labor context \citep{aufiero2024mapping}, yet in a different manner. 

The fitness, $\fitness$, of an occupation $i$ is an extensive measure defined as the sum over the complexities of the occupations $j$ that lead to the given occupation $i$. 
The fitness can be interpreted as an occupation's \textit{accessibility}, quantifying the ease and extent of workers transitioning from other occupations to occupation $i$.\footnote{Note that an occupation is more accessible when it is reached by workers from occupations that have limited options for transition. Thus, an occupation accessible to a wide range of other occupations contributes less to the transferability of that occupation. Instead, higher accessibility is associated with occupations that are predominantly transitioned into by workers from other occupations without dispersing elsewhere.}
The complexity, $\complexity$, of occupation $j$ indicates the difficulty of transferring from occupation $j$ to other occupations. 
The inverse complexity ${1}/{\complexity}$ of an occupation thus serves as an indicator of the occupational opportunities offered by $j$, and we refer to it as the \textit{transferability} of an occupation.

Following \citet{tacchella2012new}, these two metrics are determined as the fixed point solution to 
\begin{align}
    \fitnesstemp &= \sum_{j} \transmatf \complexity[(n-1)][j] & \complexitytemp[(n)][j] &= \left( \sum_i \transmatf \frac{1}{\fitness[(n-1)][i]} \right)^{-1}\label{eq:fc}\\
    \fitness &= \frac{\fitnesstemp}{\sum_i\fitnesstemp} & \complexity &= \frac{\complexitytemp}{\sum_j\complexitytemp}\label{eq:fcnorm}
\end{align}

with $i \in \mathbb{V}^{T-1}$, $j \in \mathbb{V}^{T}$ where $\mathbb{V}^T$ is the space of occupations in year $T$, and $n$ is the iteration.\footnote{The initialization is done via $\tilde{F}_i^{(0)} = 1 \;\forall i$ and $\tilde{Q}^{(0)}_j = 1 \;\forall j$. We use 200 iterations and confirm that  that the algorithm converges (the fixed point solution is reached).} 
$\transmatf$ denotes the transition probability matrix conditional on a worker making a transition, that is we exclude self-loops i.e. $\mathrm{Tr}(\transmatf) = \sum_{i}\tilde{P}_{i,i} = 0$. 
Note that the fitness of an occupation after one iteration $\fitness[{(1)}]$, is the weighted number of occupations from which individuals transitioned into occupation $i$. 
Meanwhile, the complexity of an occupation after one iteration $\complexity[{(1)}]$, is determined by the inverse of the weighted count of occupations to which individuals transition starting from occupation $j$. 
It is important to note that these metrics only have an interpretation in relative terms, but not in absolute terms as there is no unit attached.
For a schematic representation see Figure \ref{fig:schematic_example} in Appendix \ref{appx:fc}.

Figure \ref{fig:taxonomypanel}(a) shows the distribution of accessibility (y-axis) and transferability (x-axis) for the 371 occupations in the French labor market (defined in Section \ref{sec:occupationNetwork}). 
For the interpretation of accessibility and transferability we have to focus on relative terms, i.e. the comparison between two occupations. 
From Figure \ref{fig:taxonomypanel}(a) we note that the accessibility of occupations spans seven orders of magnitude, while the transferability spans six. 
The majority of occupations show a high degree of accessibility relative to their degree of transferability, suggesting that they \textit{condense} the flow of labor: moving between these occupations slowly reduces the outside options of the workers. 
In contrast to this, there is only a handful of occupations where relative transferability remains high while relative accessibility is also high.
Let us stress again that our analysis focuses solely on an average year-on-year transition matrix due to data limitations. 
Specifically, we do not directly observe individuals' residence times in a given occupation, which may affect the diversity of outside options (transferability) of the occupation, thus potentially affecting the dispersion of transferability and accessibility of the occupational space.  

\begin{figure}[t!]
\centering
\includegraphics[width=0.9\textwidth]{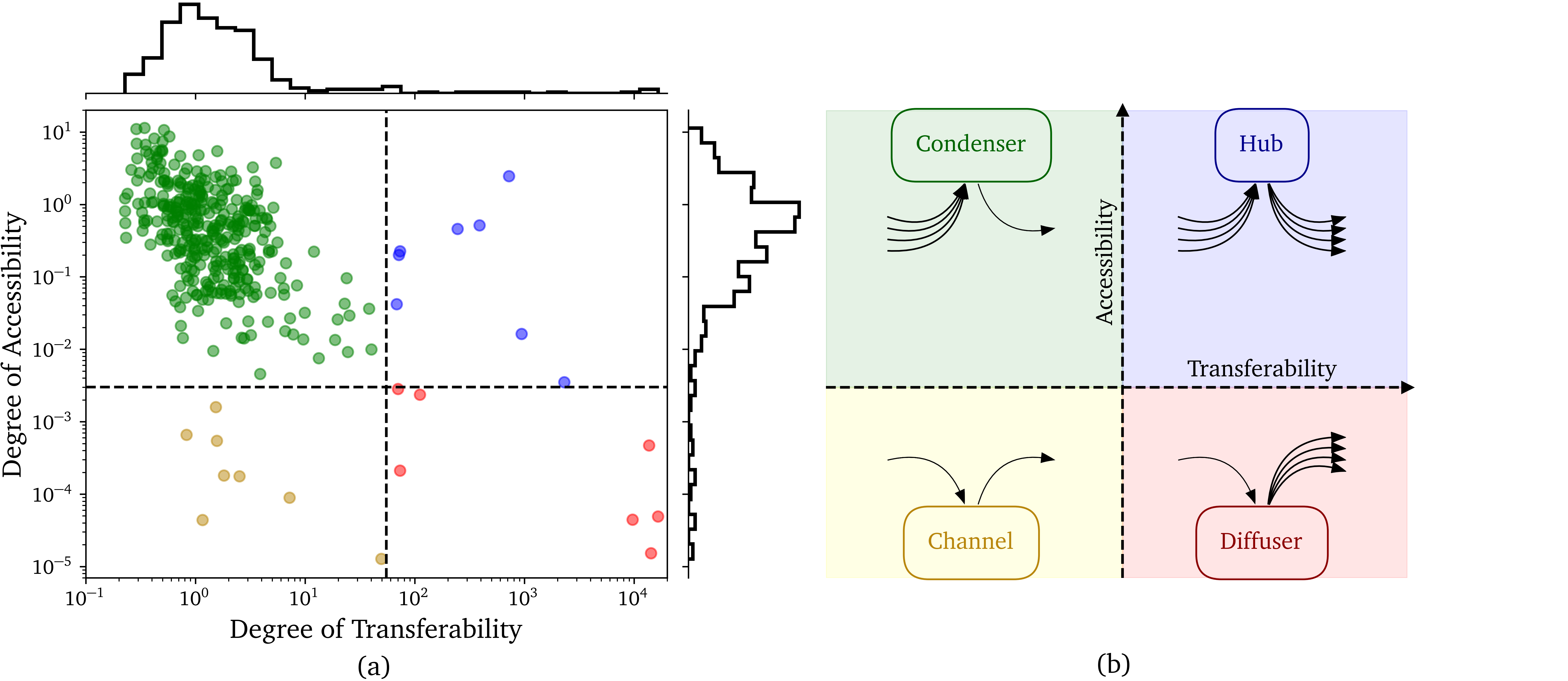}
\caption{\textbf{Accessibility and Transferability Space}. (a) Each point represents a single occupation and is located in the space by its corresponding accessibility (y-axis) and transferability (x-axis) metric. The dashed lines denote the mean-shift clustering boundaries at $\Theta^A= 0.0031$ and $\Theta^T=49.9$. The histograms show the distribution along the accessibility (horizontal) and transferability (vertical) dimensions. (b) Proposal of a new taxonomy based on the degree of accessibility and transferability of an occupation into four categories. The arrows schematically represent the potential number of incoming and outgoing transitions for the different categories.}
\label{fig:taxonomypanel}
\end{figure}

To study this phenomena of condensation, and identify associated bottlenecks in labor reallocation, we propose a taxonomy that classifies occupations into four groups (see Figure \ref{fig:taxonomypanel}(b)):
\begin{itemize}
    \item \textbf{Hubs} are characterized by a high transferability and accessibility. These occupations are broadly accessible to individuals from many occupations while also providing access to a variety of other occupations and often do not require specialized training or education. For example, sellers of cultural goods (554G).
    \item \textbf{Condensers} are characterized by a high accessibility but low transferability. While accessible from a broad range of occupations, these occupations allow for transitions only to a few other occupations, thus funnelling individuals, e.g. care givers (526A).
    \item \textbf{Diffusers} are characterized by a low accessibility but high transferability. As the inverse of the Condenser, these occupations are hard to reach, often due to specialized training, but exhibit a diversity of opportunities, such as officers and technical flight managers of the merchant navy (389C).
    \item \textbf{Channels} are characterized by both a low accessibility and a low transferability, representing occupations with only a few occupations to transition from and into. For example, skilled operators on welding machines (623D).
\end{itemize}

The classification of occupations into this taxonomy depends on two thresholds, $\threshacc$ and $\threshtrans$, that separate the low and high regimes for the accessibility and transferability of occupations.\footnote{Thresholds are determined by  a mean-shift clustering algorithm \citep{comaniciu2002mean}. Specifically, we use the MeanShift implementation from the scikit-learn package in Python with a bandwidth of 3 \citep{pedregosa2011scikit}. To ensure robustness, we employed both spectral clustering and k-means clustering (for each dimension separately), which resulted in similar threshold values \citep{pedregosa2011scikit}.} 
Figure \ref{fig:taxonomypanel}(a) shows that the defined taxonomy clusters are well partitioned in the data of the French labor market.\footnote{We perform a significance testing with a random network, preserving both in- and out-degree distributions as in the labor market, to confirm that the cluster sizes of the French labor market are significant. This suggests that the findings are intricacies specific to the labor market's flows, and not merely due to the distribution of in- and out-degrees of occupations.} 
Appendix \ref{appx:taxonomy} provides a comprehensive table of occupations within each cluster.

The cluster of Condenser occupations contains the largest number of occupations (348 out of 371), describing 94\% of the total person-year datums in the dataset and 94\% of estimated total wages in 2020, thus indicating the economic significance of this phenomenon.\footnote{We compute estimated total wages as the sum over occupations' median wage multiplied by the number of employees in the occupation}
The categories of Channel (8 occupations), Hub (8 occupations), and Diffusers (7 occupations) make up the remaining set of occupations.
The large share of condensers could be interpreted as a general trend of specialization: over the course of a career one specializes in terms of experience and skill, therewith inadvertently reducing the transferability to other occupations.
In fact, for 3 out of 5 occupations (59.6\%) the expected transferability of the next occupation is lower than a person's current occupation. This aligns with human capital theory, which posits that as workers accumulate specific human capital, the returns to staying in their current role increase, making transitions less attractive.

However, it is important to note that while human capital theory provides one explanation, our analysis focuses solely on occupational transitions between the current and subsequent occupations across all workers, without tracing individual career paths through the labor market.

\begin{figure}[t]
\centering
\includegraphics[width=0.9\textwidth]{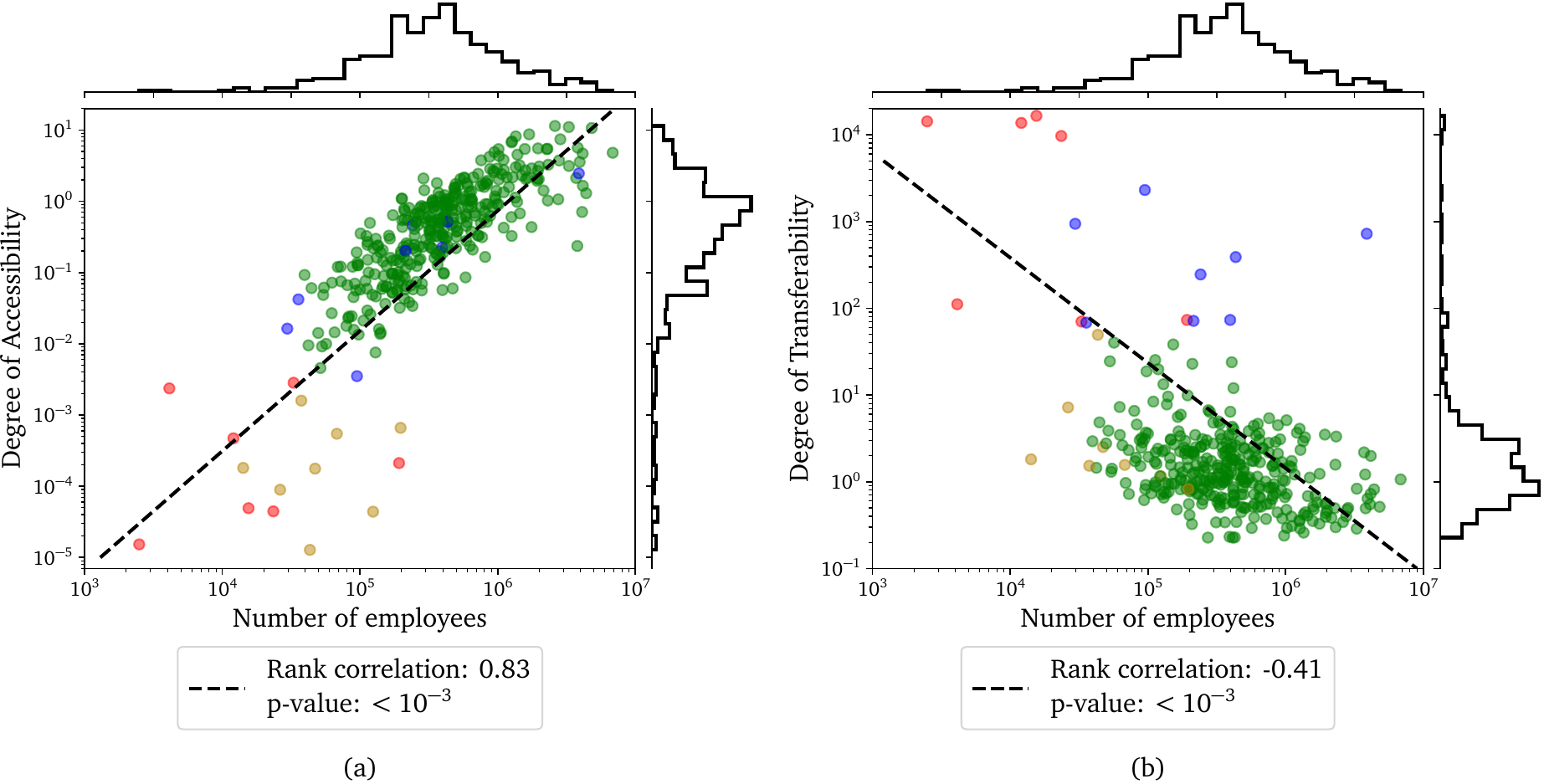}
\caption{\textbf{Occupation Size Effect on Accessibility and Transferability.} Scatter plot depicts the relationship between the accessibility (a), and transferability (b) and the number of employees in a given occupation. A dashed line serves as a visual guide and represents the linear fit of the points. There is a significant and positive rank correlation of 0.83 between accessibility and number of employees and a significant and negative rank correlation of -0.41 observed between number of employees and transferability.}
\label{fig:sizevsAT}
\end{figure}

Figure \ref{fig:sizevsAT} shows the relation between the number of employees in a given occupation to the occupation's accessibility and transferability metrics. 
As expected, highly accessible occupations have a higher number of employees given the diversity of occupations, both from within and across communities, that transfer into these occupations. 
By contrast, more transferable occupations, i.e. those with a more diverse set of outside options, are generally smaller. 
This highlights the condensation effect: within a community of occupations, there are inroads via highly-accessible occupations, but only a smaller fraction of individuals transition into highly transferable occupations that allow for a higher degree of inter-community mobility.

\subsection{Identifying Bottlenecks in the French Labor Market}\label{ssec:bottlenecks}
\begin{figure}[t]
\centering
\includegraphics[width=0.9\textwidth]{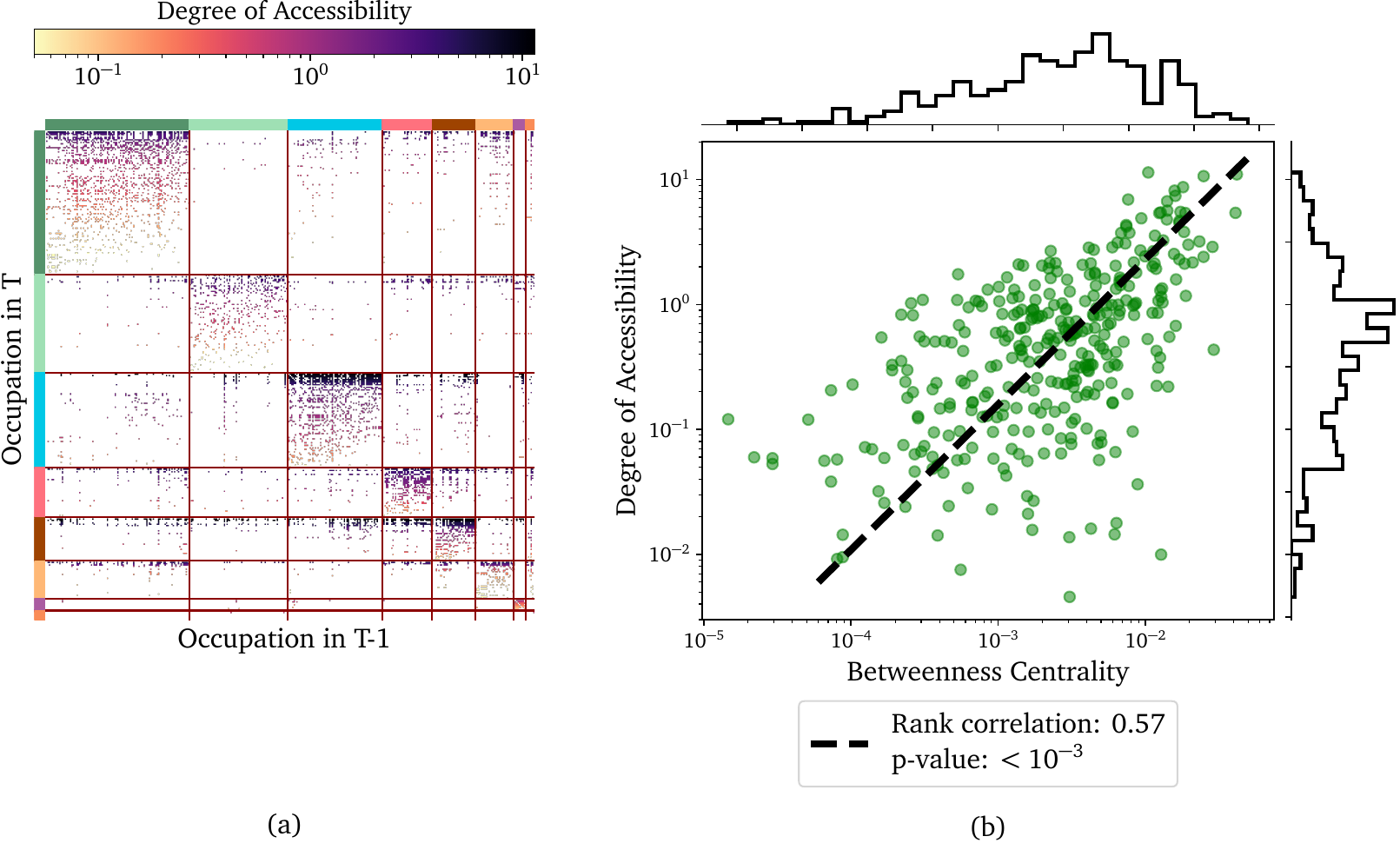}
\caption{\textbf{Correlation of Accessibility and Betweenness Centrality}. (a) Heatmap of the occupational transition matrix, with rows colored according to the accessibility of each respective occupation. (b) Scatter plot depicting the relationship between the accessibility of an occupation and its betweenness centrality, a measure of bottlenecks, showing a significant positive rank correlation of 0.57. Only condenser occupations are shown. A dashed line serves as a visual guide and represents the linear fit of the points.}
\label{fig:bottlenacksaccessibility}
\end{figure}

The implication of the nested and clustered structure of the occupational transition matrix shown in Section 2 is the existence of bottlenecks, that is, there being only a few occupations that have a high share of inter-community transitions as compared to intra-community transitions. 
The accessibility of an occupation captures this particular phenomenon. 
Figure \ref{fig:bottlenacksaccessibility}(a) shows the occupational transition matrix coloured by the degree of accessibility (dark having high accessibility) and split into the 8 BRIM communities (Section \ref{sec:clusteringBRIM}).
One can clearly see that, intuitively, those occupations who receive a higher share of inflow from inter-community transition are also those with a higher accessibility, thus these occupations represent the handful of entry-points into different occupational communities.
By contrast, those occupations with a low accessibility receive workers almost exclusively from within their given community (See also Figure \ref{fig:intervsintra} in Appendix \ref{appx:intravsinter}).
Prior research has used the betweenness centrality \citep{freeman1977set}, i.e. the fraction of shortest paths between any two nodes that pass through a given node, to identify bottlenecks in network structures \citep[see][for an application to labor markets]{bocquet2022network}.
Focusing on the condenser occupations, comprising 94\% of the employed workforce, Figure \ref{fig:bottlenacksaccessibility}(b) shows that our accessibility metric captures the betweenness centrality quite well, with a significant Rank Correlation of 0.57. 
Accessibility and betweenness centrality capture bottlenecks in an ex-post fashion: they consider the diversity of the incoming transitions, thus identifying the occupations in a community that are the \textit{target} of workers from other communities (i.e. they consider incoming and outgoing links). 
However, this does not isolate the ex-ante perspective of the worker: what can they reach from a given occupation.
 Here  we capture this additional information through the transferability metric, i.e. we can not only identify bottlenecks from the perspective of where workers arrive in $T$ (accessibility), but also from the perspective of where workers start in $T-1$ (transferability). 

\begin{figure}[t]
\centering
\includegraphics[width=0.9\textwidth]{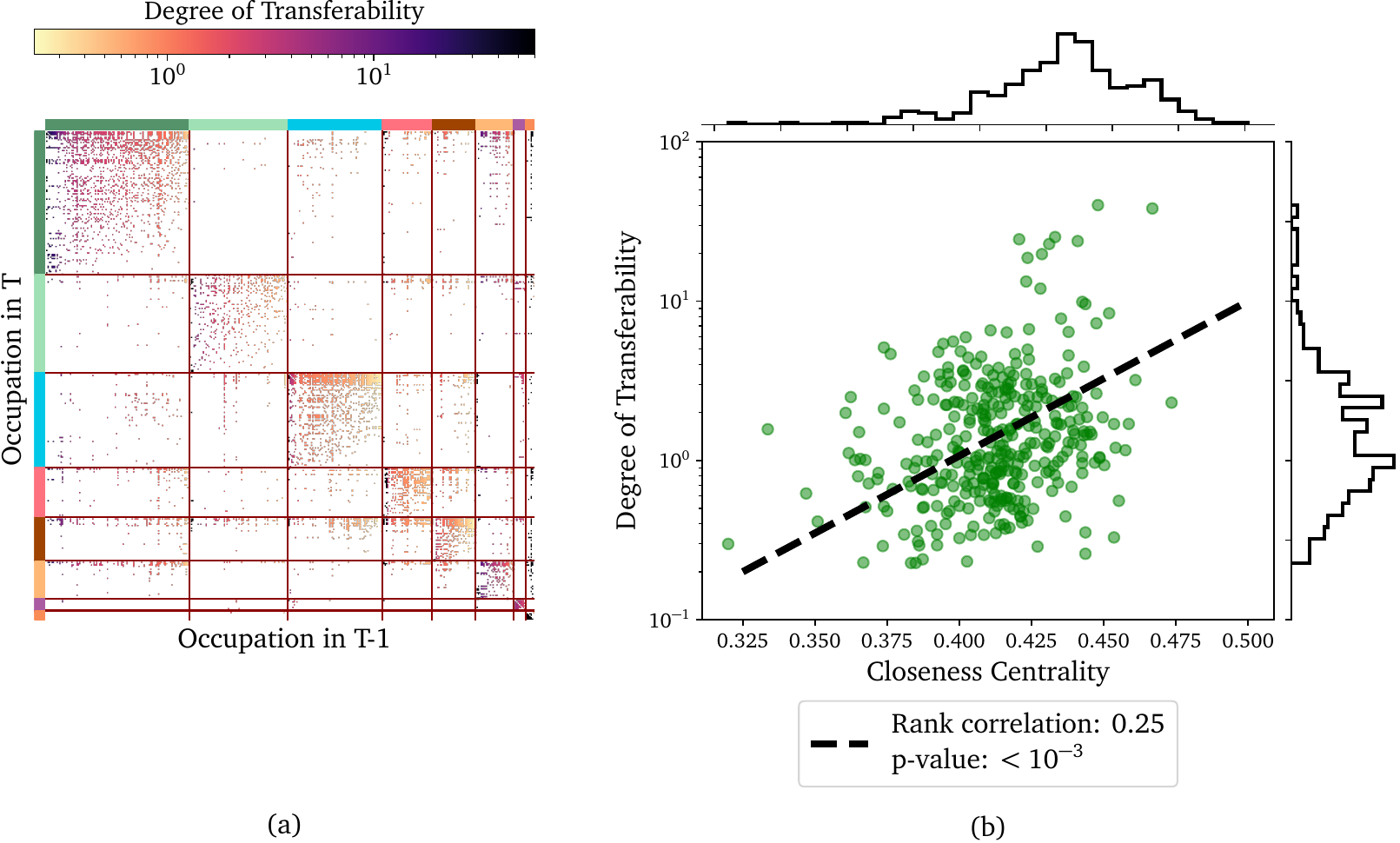}
\caption{\textbf{Correlation of Transferability and Closeness Centrality}. (a)  Heatmap of the occupational transition matrix, with columns colored according to the transferability of each respective occupation. (b)~Scatter plot depicting the relationship between the transferability of an occupation and its closeness centrality, a measure of a nodes ability to spread information in a graph, revealing a significant positive rank correlation of 0.27. Only condenser occupations are shown. A dashed line serves as a visual guide and represents the linear fit of the points.}
\label{fig:bottlenackstransferability}
\end{figure}

Figure \ref{fig:bottlenackstransferability}(a) shows the occupational transition matrix colour-coded by the transferability of a given occupation. 
The figure shows that it is those occupations with a high transferability that have a higher dispersion in possible destinations, both within and across communities. 
As with high-accessibility occupations, there are only a few occupations per community that show a high degree of transferability, and, as highlighted in Section \ref{sec:fitnessandcomplex}, high-transferability occupations tend to have a smaller number of employees. 
Furthermore, higher transferability is positively correlated (although in weaker fashion than for accessibility) with the ratio of inter- to intra-community transitions (see Figure \ref{fig:intervsintra} in Appendix \ref{appx:intravsinter}). Thus, occupations with high transferability are more likely to facilitate transitions between communities, serving as exit points to other communities.
The consequence of this being that these occupations reflect the here-presented definition of bottlenecks: they limit the degree to which there is free mobility between occupations.
This is not intended to suggest any occupation should be able to transition to any other occupation (e.g. in the sense that some occupations require specialized training), but rather that this is, empirically, the identified structure of the labor market.

\subsection{Skill Effects on Accessibility and Transferability}\label{sec:skilleff}

To get a better understanding of the drivers behind accessibility and transferability we consider the skills required in each occupation. 
Skill-similarity between different occupations has been shown to be a significant influence on occupational transitions \citep{gathmann2010general, neffke2013skill} and has been used to reconstruct labor flow networks \citep{del2021occupational}.
We compute the skill-similarity between two occupations as the cosine similarity between their respective binary skill-vectors (derived from the ROME classification system) with 1 (resp. 0) for every skill present (resp. absent) in a given occupation.
Occupations with low average skill similarity tend to be highly specialized, such as Tax, Treasury, Customs controllers (451C), while occupations with high average skill similarity include occupations such as Childcare workers (431C).
Based on the skill similarity, we compute the average similarity of an occupation i to occupations in its own community $d_{\mathrm{Intra}, i}$ and all occupations in all other communities $d_{\mathrm{Inter}, i}$ (See Appendix \ref{appx:skills} eq.~\ref{eq:interintraskill}). 
From this we can construct two Inter- \& Intra- community skill similarity scores: the product $d^p = d_{\mathrm{Intra}} \cdot d_{\mathrm{Inter}}$, and ratio $d^r = d_{\mathrm{Intra}} / d_{\mathrm{Inter}}$.
Thus occupations with a high inter and intra-community similarity score highly, while those with only unique skills have a low score. 

\begin{figure}[t]
    \centering
    \includegraphics[width=1\textwidth]{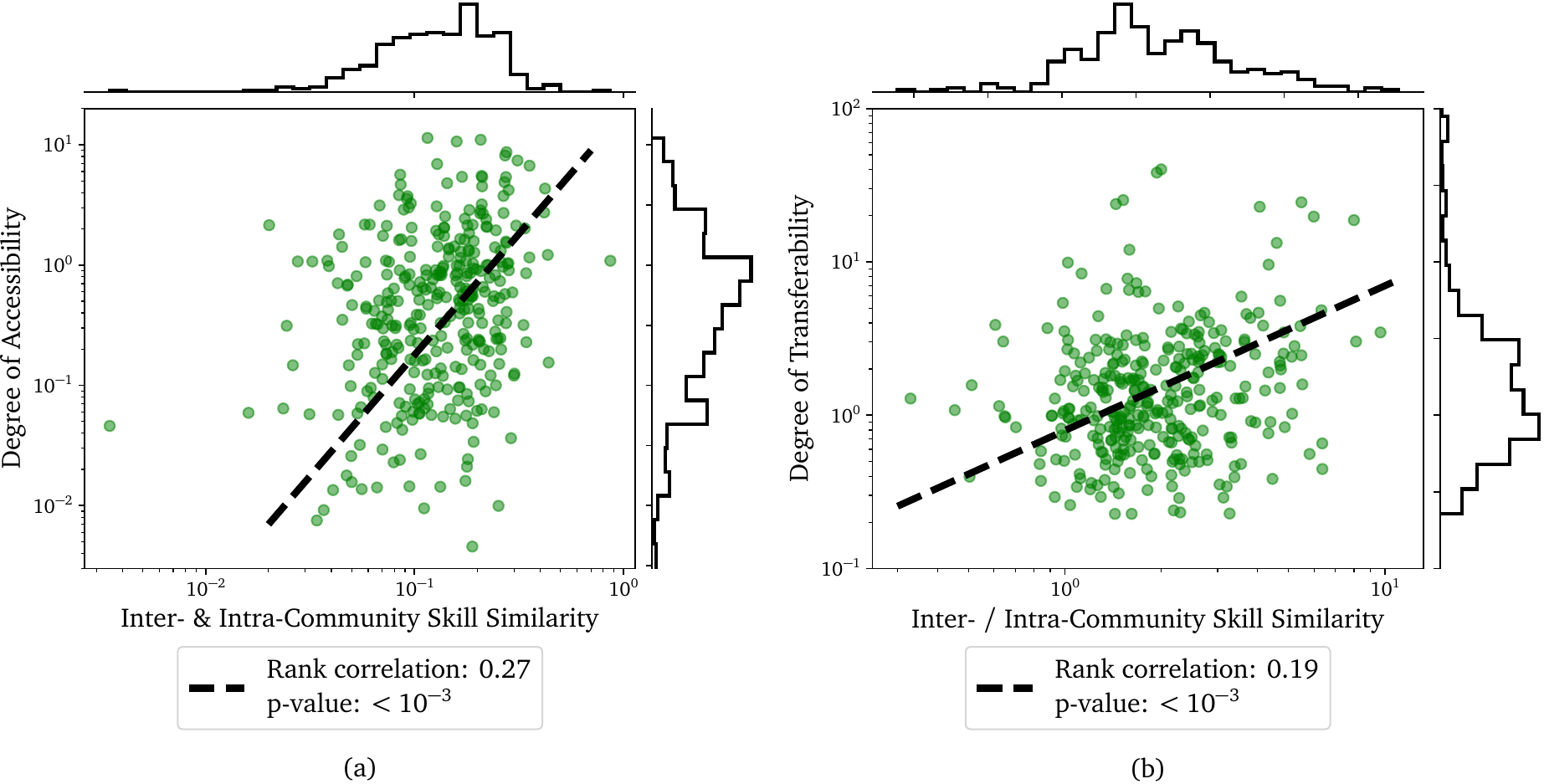}
    \caption{\textbf{Skill Effect on Accessibility and Transferability.} Scatter plot depicts the relationship between (a) $d^p$ the product of inter- and intra-community skill similarity  and the degree of accessibility and (b) $d^r$ the ratio of inter- and intra-community skill similarity and the degree of transferability for each occupation. A dashed line serves as a visual guide and represents the linear fit of the points. There is a significant and positive  rank correlation of 0.27 observed between the product of inter- and intra-community of skill similarity and accessibility and a significant and positive rank correlation of 0.19 between the  ratio of inter- and intra-community skill similarity and the degree of transferability for each occupation.}
    \label{fig:skill}
\end{figure}

Finally, Figure~\ref{fig:skill} shows the relationship between the accessibility and the product of inter-intra skill similarity, $d^p$ (a) and the relationship between the transferability and the ratio $d^r$ (b). 
In line with the discussion of Section \ref{ssec:bottlenecks}, occupations with a higher accessibility score show evidence of a higher overall skill-similarity, as signified by the significantly positive rank correlation.
Thus, occupations from outside a given community have sufficient overlap in skills for workers to transition into these particular occupations, making them entry-points into new occupational communities.
By contrast, considering the relationship between the ratio of intra-community similarity to inter-community similarity, we see that the degree of transferability is positively related to the similarity of skills to occupations outside of the current community, suggesting that the transferability maps those occupations acting as springboards to inter-community transitions.
It is the identification of the occupation of departure that sets apart our accessibility-transferability approach from simply considering betweenness centrality as identifiers of bottlenecks.

\section{Informing Policy using Accessibility and Transferability} \label{sec:policy}

In the previous sections, we have shown that there are several occupations that can act as bottlenecks between occupational communities, and we can identify these using accessibility and transferability metrics.
These occupations may be part of the reason we observe a significant deviation between the realized distribution of occupations and the steady state implied by the occupational transition matrix.\footnote{We here explicitly do not make a statement about whether the implied steady state is desirable.} While the steady-state distribution does not necessarily represent an optimal labor market outcome, it allows us to quantify deviations from this baseline applying labor market interventions.
In this section, we show how to improve the speed of convergence, i.e. to reduce frictions, through a small rebalancing of the occupational flows. 
In particular, for a given occupation rebalancing outgoing flows away from high accessibility and towards high transferability targets leads to faster convergence towards the implied steady state. 

For the occupational transition matrix, the rate of convergence to the steady state is determined by the second eigenvalue $\lambda_2$\footnote{Note that the rate of convergence is determined by the spectral gap between the first and second eigenvalues. Since the transition probability matrix is stochastic, the first eigenvalue is $\lambda_1=1$.}, with a lower eigenvalue indicating a faster rate of convergence. 
Therefore, our objective is to minimize $\Delta \lambda_2 = \lambda_2^\star - \lambda_2$, where $\lambda_2^\star$ is the second eigenvalue of the transition matrix following a small perturbation $\transmat[][\star] = \transmat[][] + \epsilon V$. 
Here $V$ is a $N\times N$ matrix indicating the location of the interventions.
Specifically, we consider the simple case where for a chosen occupation $k$, we increase the flow to occupation $l^+$ in equal magnitude to a decrease of the flow to occupation $l^-$, thus preserving the normalization of $\transmat$.
Mathematically, one can write
\begin{align}
  (V)_{i,j} &= 
  \begin{cases}
      -1, \quad i=l^-, j=k\\
      1, \quad i=l^+, j=k\\
      0, \quad \mathrm{else}
  \end{cases}
\end{align}
The constant $\epsilon\ll1$ gives the size of the intervention, which is assumed to be small so as to keep fixed the implied steady state distribution and the distributions of the accessibility and transferability metrics. 
We consider here the case of $\epsilon=0.01$, whose natural interpretation is that we reallocate 1\% of the outgoing labor flows for occupation $k$.\footnote{To assess the effects of the implemented strategies on these metrics, we compute the rank correlation between the metrics of the original and perturbed networks. The transferability metrics show a significant average rank correlation of 0.961, while the accessibility metrics exhibit a significant average rank correlation of 0.975 across all policies. These findings indicate that even with minor perturbations, the relative values of the accessibility and transferability metrics remain stable. Furthermore, the mean absolute percentage difference between the steady states of the perturbed matrices and the unperturbed matrix is below $10^{-4}$.  }

\begin{figure}[htp]
    \centering
    \includegraphics[width=1\textwidth]{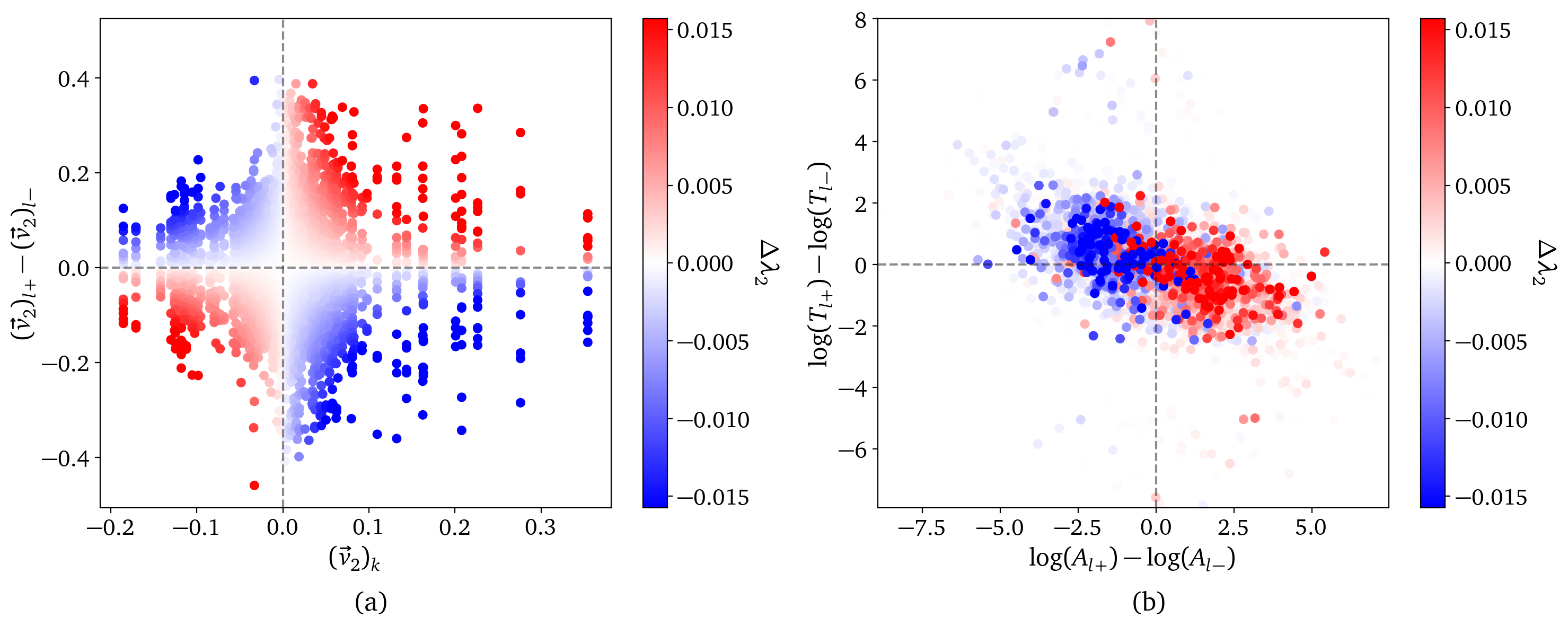}
    \caption{\textbf{Reducing Constraints through Effective Policy Making} (a) Scatter plot showing the eigenvector value difference between the destination and source occupation $(\vec v_2)_{l^+} - (\vec v_2)_{l^-}$ against the eigenvector value of the target occupation $(\vec v_2)_{k}$ (see eq. \ref{eq:changelambda}). The color coding indicates the change in the second eigenvalue, $\Delta \lambda_2$. (b) Scatter plot showing the relation between log-difference of source and target occupation in transferability $\log(T_{l^+})-\log(T_{l^-})$ and accessibility $\log(A_{l^+})-\log(A_{l^-})$. The color coding indicates the change in the second eigenvalue, $\Delta \lambda_2$.}
    \label{fig:perturbation_main}
\end{figure}

Using matrix perturbation theory, the first-order change in the second eigenvalue can be approximated by:\footnote{Note that, to apply first-order matrix perturbation, the matrix must be symmetric. The occupational transition matrix has a symmetry value of $\frac{||S||_F - ||A||_F }{ ||S||_F + ||A||_F} = 0.9$, where $\transmat[][] = \frac{\transmat[][]+\transmat[][T]}{2} + \frac{\transmat[][]-\transmat[][T]}{2} = S + A$. This indicates that the occupational transition matrix is almost symmetric. Furthermore, our numerical results demonstrate that the predictions from first-order perturbation theory align well with the numerical findings (see Figure \ref{fig:perturbation_main}(a) and Appendix \ref{appx:policyperturbation}, Figure \ref{fig:confirmation_theory}).}
\begin{align}
    \Delta \lambda_2 &= \vec v_2^TV\vec v_2 = (\vec v_2^T)_k \cdot ((\vec v_2)_{l^+} - (\vec v_2)_{l^-}), \label{eq:changelambda}
\end{align}
where we have used the second eigenvector, $\vec{v}_2$, of the unperturbed transition matrix (see Appendix \ref{appx:policyperturbation}).
The second eigenvector $\vec v_2$ describes the slowest mode by which the system approaches its stationary distribution. 
States with large magnitudes, $|(\vec v_2)_i|\rightarrow 1$, dominate the system's slow dynamics, as they correspond to states with the highest contribution to this relaxation mode.
Sign differences in the entries indicate partitioning of states, as they represent opposite directions in the evolution of the system's slow mode, with states in different partitions exhibiting divergent behavior as the system converges to equilibrium.

If the value of the eigenvector of the target occupation, $(\vec v_2)_k$, has a different sign than the difference in eigenvector values of the source and destination occupations, $((\vec v_2)_{l^+} - (\vec v_2)_{l^-})$, the change in second eigenvalue, $\Delta \lambda_2$, is negative:
\begin{align}
    \Delta \lambda_2 < 0 \; \mathrm{if} \;
    \begin{cases}
      (\vec v_2)_{l^+}>(\vec v_2)_{l^-} \wedge (\vec v_2)_{k} < 0\\
      (\vec v_2)_{l^+}<(\vec v_2)_{l^-} \wedge (\vec v_2)_{k} > 0\label{eq:caseslambda}
    \end{cases}
\end{align}

Figure \ref{fig:perturbation_main}(a) shows the distribution of $(\vec v_2)_{k}$ versus $(\vec v_2)_{l^+}<(\vec v_2)_{l^-}$ and the consequent distribution of eigenvalue changes $\Delta\lambda_2$. Positive and negative changes in the adjustment speed $\Delta \lambda_2$ align with calculated regimes (see eq. \ref{eq:caseslambda}).
While the bulk of the distribution is concentrated around 0, there are several significant outliers in the $(\vec v_2)_{l^+}<(\vec v_2)_{l^-} \wedge (\vec v_2)_{k} > 0$ quadrant, hinting at room for policy interventions to improve labor reallocation. 
Improvements in adjustment speed (measured by $\Delta \lambda_2$) can be achieved by considering accessibility and transferability metrics. Figure \ref{fig:perturbation_main}(b) illustrates the relationship between the log difference in accessibility and transferability metrics of the flow-increase occupation $l^+$ and the flow-decrease occupation $l^-$ on, along with the difference in the second eigenvalue, $\Delta \lambda_2$. 
Each point represents a policy that constitutes a perturbation $V$ for all possible combinations of target $k$ and destination $l^+$ occupation.\footnote{For each link in the occupational transition matrix from occupation $k$ to $l^+$ where $\transmat>0.01$ a perturbation $V$ is applied. With this, the target and destination occupations are given. The source occupation is then chosen at random from existing links of the target occupation $k$ with a flow greater than epsilon, ensuring that $\transmat \geq 0 \;\; \forall i,j$.}
The eigenvalue difference is more likely to be negative when the accessibility of the flow-increase occupation is lower than that of the flow-decrease occupation, with a significant correlation coefficient of 0.26 (Appendix \ref{appx:policyperturbation}).
Simultaneously, we find that $\Delta \lambda_2 < 0$ when the transferability of the destination occupation is higher than the transferability of the source occupation, with a significant correlation coefficient of -0.13 (Appendix \ref{appx:policyperturbation}).
In terms of our taxonomy, this implies an improvement in the speed of convergence to the implied steady state when we shift the target of outgoing individuals to diffuser occupations in place of condenser occupations. 
Practically, this implies fewer individuals transitioning into already large occupations (see Section \ref{sec:fitnessandcomplex}) with a low transferability and limited further mobility. 
For example, increasing the flow from ``Executives in charge of economic, financial and commercial studies (372A)" to ``Banking Operations Executives (376B)", while reducing the flow to ``IT project managers, IT managers (388C)", has a strongly positive impact on the convergence speed.

Historically, the French state's labor market policies have focused on on-the-job training together with technical and vocational education \citep{gazier2019opportunities}, which links to the latter objective: increasing the transferability of currently low-transferability occupations.
While we do not focus here on the \textit{how} of implementing labor market policy, the above results present an intuition about which occupations to target with policy, presenting a set of channels that may optimally reduce frictions under the assumption that the implied steady state is indeed a desirable target.
At this point, we should however also note that while we categorize different occupations by their accessibility and transferability, our dataset does not distinguish between the supply (unemployed and their characteristics) and the demand (vacancies) drivers of the transition rates between occupations.
Nonetheless, integrating information with accessibility and transferability measures can enhance the effectiveness of a policy aimed at alleviating bottlenecks in the labor market. 
    
\section{Conclusion}\label{sec:conclusion}

In this study, we analyzed the labor flow network of all economically active workers in the French labor market using micro-data from French employers (BTS-POSTES). Here the observed flows represent realized transitions within the labor market, as opposed to potential transitions based on skill similarity between occupations. Realized transitions provide a more accurate reflection of the complex dynamics emerging from the interaction of workers' preferences, employers' hiring decisions, and the prevailing economic conditions. While skill-based networks can help map potential pathways based on related skills, our findings reveal that these predicted transitions only partially overlap with observed flows. The observed transition network, therefore, provides a richer empirical perspective, capturing structural constraints and revealing patterns of mobility that extend beyond what can be inferred from skill similarity alone.
We demonstrated the presence of significant constraints in the labor market evidenced by strong communities and a lack of inter-community transitions, hindering reallocation of workers (Section \ref{sec:occupationNetwork}). 
We then employed a quantitative method to assess the accessibility and transferability of occupations, using an economic fitness and complexity framework, to identify precise bottlenecks in the labor market (Section \ref{sec:taxonomysec}). 
We showed that the emergence of bottlenecks can be explained by a condensation effect due to occupations with high accessibility but low transferability, that is in line with the principles of human capital theory, where workers' accumulation of occupation-specific skills reduces their incentives to transition to other roles. 
Finally, in Section \ref{sec:policy}, we introduced a quantitative approach with the aim of evaluating policy consequences using the rate of convergence to the stationary distribution, that might aid policymakers in assessing potential policy initiatives.

While our approach provides valuable insights into the structure of the labor market, it also has certain limitations. The analysis relies on year-to-year snapshots of transitions without longitudinal panel data to capture the full trajectory of workers over time.
This could introduce potential biases, as residence times within different occupations vary, which could slightly influence the degree of transferability of an occupation. 
Future research with panel data could extend this work by analyzing the time-dependent condensation effect of individual workers transitioning to more specialized occupations, and link this to labor market outcomes such as lifetime earnings and wage progressions. 
Additionally, it would be valuable to explore whether our metrics could be used to explain wage differences in occupations as a mismatch between transferability and accessibility affecting workers' and firms' bargaining power.

\section{Declarations}
\subsection*{Availability of data and materials}
The data supporting the findings of this study are available from the National Institute of Statistics and Economic Studies (INSEE), authorized by the Comité du Secret Statistique France, and accessed via the Secure Data Access Center (CASD). Due to licensing restrictions, the data used in this study are not publicly available. However, they can be obtained from the authors upon reasonable request and with permission from the relevant institutions.

\subsection*{Competing interests}
The authors declare that they have no competing interests.

\subsection*{Acknowledgments}
We extend our sincere gratitude to Jean-Philippe Bouchaud for his insightful ideas. We  also thank  Léonard Bocquet, Morgan Frank, Xavier Gabaix, Dany Lang, Antoine Mandel, José Moran, Frank Neffke and  Maria del Rio-Chanona  for fruitful discussions.
This research was conducted within the Econophysics \& Complex Systems Research Chair, under the aegis of the Fondation du Risque, the Fondation de l’École polytechnique, the École polytechnique and Capital Fund Management. This work is supported by a public grant overseen by the French National Research Agency (ANR) as part of the ``Investissements d’Avenir" programme (reference: ANR-10-EQPX-17 Centre d’accès
 sécurisé aux données – CASD).

\bibliography{Refs.bib}

\newpage

\appendix

\section{Methods \& Data}\label{appx:methodsanddata}

\subsection{Skill Data and the ROME Classification}\label{appx:skills}
In addition to the annual declaration of social data, we utilize occupational skill data, specifically the Operational Directory of Trades and Jobs (Répertoire Opérationnel des Métiers et des Emplois, ROME). This directory, created as a reference system by the French Employment Center (Pôle emploi), categorizes occupations based on their skills and professional fields. The dataset encompasses skills ranging from know-how and hard skills to soft skills for over a thousand occupation labels. We convert these skills into vectors of approximately 800 dimensions, where each entry represents the extent to which a given skill is associated with that occupation. However, the ROME occupation classification comprises over 5000 occupations and differs from the 4-digit occupation classification PCS. Consequently, the mapping from ROME classification to PCS classification is not a one-to-one correspondence. In cases where a single PCS code is corresponds to multiple ROME occupations, the skill vector for that PCS occupation is calculated as the element-wise average of the skill vectors associated with the corresponding ROME occupations. Finally, as a distance metric between two PCS codes, we use a simple cosine similarity of the skill vectors for each occupation.
\begin{equation}\label{eq:distanceskill}
    d_{i,j} = \frac{\mathbf{S}_i\cdot \mathbf{S}_j}{\norm{\mathbf{S}_i}\norm{\mathbf{S}_j}}
\end{equation}
where $\textbf{S}_{k}$ are the skill vectors of the PCS codes. 
In Section \ref{sec:skilleff}, we use the ratio and product of inter- and intra-community skill similarities, $d_{\mathrm{Intra}}$ and $d_{\mathrm{Inter}}$. For a given occupation i, this metric calculates the average skill similarity to all occupations j within (intra) and outside (inter) occupation i's community.
\begin{align}
    d_{\mathrm{Intra},i} &= \langle \tilde{d}_{i,j} \rangle_{c^i=c^j} \\
    d_{\mathrm{Inter},i} &= \langle \tilde{d}_{i,j} \rangle_{c^i\neq c^j}
    \label{eq:interintraskill}
\end{align}
where $c^i$ is the BRIM community of occupation i.

\subsection{Brim Algorithm}\label{appx:brim}
To identify the communities within our bipartite network, we employ the modularity maximization algorithm BRIM, which seeks to maximize the Barber modularity defined as follows \citep{barber2007modularity, platig2016bipartite}:
\begin{equation} \label{eq:modularity}
    Q = \frac{1}{E} \sum^{N_R}_{i=1} \sum^{N_C}_{\alpha=1} (M_{i, \alpha} - P_{i,\alpha}) \delta(a_i, a_\alpha)
\end{equation}
with the total number of links $E$, $M_{i, \alpha}$ the biadjacency matrix, $P_{i,\alpha}$ the probability that a link exists by chance and the membership variable that defines the block that node i belongs to $a_i$. This specific algorithm proved effective for the paper's objectives, particularly due to its consideration of the interdependence between communities in the different parts of the network, resulting in a well-defined partition for the labor flow network. Nonetheless, we also tested a partitioning by the BiLouvain algorithm to validate the robustness of our findings.

\subsection{Nestedness}\label{appx:nestedness}

In this study, we employ the NODF (Nestedness based on Overlap and Decreasing Fill) measure, a widely utilized metric for nestedness analysis developed by \cite{almeida2008consistent}. This metric is chosen due to its relevance to a structural property in complex networks, wherein the neighborhood of a given node is a subset of the neighborhoods of better-connected nodes. In the context of the fitness and complexity metric, our focus lies in ranking occupations based on their transferability and accessibility. It has been demonstrated that various social systems exhibit this hierarchical behavior in their rankings \citep{ren2020bridging, laudati2023different}, and the nestedness metric serves as a quantitative measure for this phenomenon. The values span a range from 0 to 100, where 100 represents a fully nested network, and 0 reflects a non-nested pattern resembling a checkerboard like structure. This metric is based on two key properties: decreasing fill and paired overlap. Paired overlap is essentially the percentage of 1's in a given column (or row) that align with identical positions in another column. Then, NODF is the average of all paired values across columns (n) and rows (m):
\begin{equation} 
    \mathrm{NODF} = \frac{\sum N_{\mathrm{paired}}}{\left( \frac{n(n-1)}{2}\right) + \left(\frac{m(m-1)}{2} \right)}
\end{equation}
In the realm of fitness and complexity research, where scholars examine import/export matrices of countries and their products, diverse values of the NODF measure are observed. For instance, the world trade network for Bovine exhibits a NODF value of $0.12$, the typewriter network has a NODF value of $0.35$, and on a company-product level, the NODF value is also $0.12$ \citep{ren2020bridging, laudati2023different}.

\section{Temporal Stability}\label{sec:tmpstability}

The structure of the occupation transitions in France has remained stable over the sample period of 2012-2020, despite events such as COVID-19.
To test temporal stability we focus on link-stability over time, both in existence and magnitude of the transition rates.
This can be captured in a simple regression comparing the current transition rate to that of 2012 as a baseline year: 
\begin{equation}
\transmat[{i,j}][t] = \alpha_{i,t} + \beta_{i,t} \transmat[{i,j}][2012] + \epsilon_{i,t}\quad \forall t\in\{2013,\dots,2020\},
\label{eq:tempstabocc}
\end{equation}

such that the hypothesis of temporal stability can be captured by the joint test of $H_0:\alpha_{i,T}=0,\beta_{i,T}=1$. 
That is, we expect the magnitude of transitions to be the same as 2012 without any general trend ($\alpha_{i,T}$).

\begin{figure}[htb!]
    \centering
    \includegraphics[width=1\textwidth]{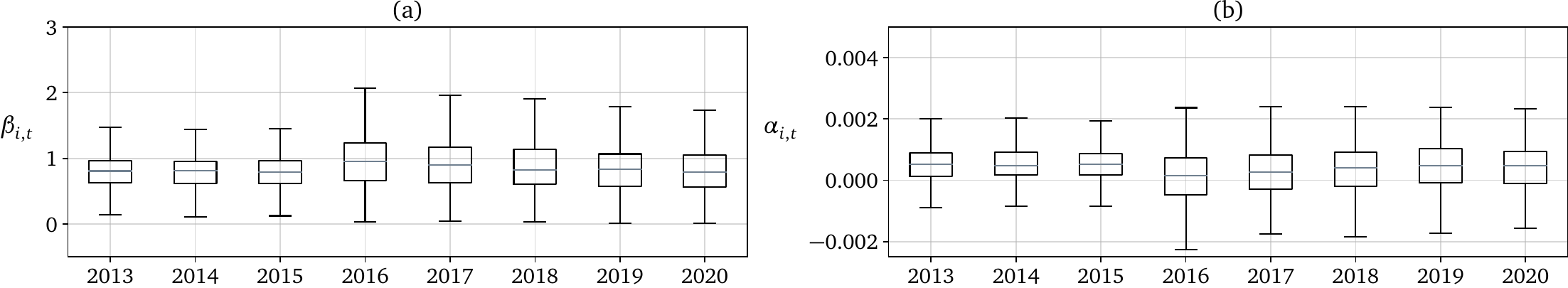}
    \caption{\textbf{Regression Parameters of Linear Regression in eq. \ref{eq:tempstabocc}.} (a) Box plot of the coefficient $\beta_{i,t}$ for occupation i and year t. (b) Box plot of the intersection coefficient $\alpha_{i,t}$ for occupation i and year t. }
    \label{fig:temporalstabilityall}
\end{figure}

Our findings indicate that the coefficient $\alpha_{i,t}\to 0$, and the coefficient $\beta_{i,t} \to 1$ remains stable over time (Figure \ref{fig:temporalstabilityall}). This stability underscores the persistence of transition probabilities across years and allows as to combine all observations over the period of 2012-2020.

\section{Entry and Exit from the Occupation Network}\label{appx:entryExit}

\begin{figure}[htb!]
    \centering
    \includegraphics[width=1.0\textwidth]{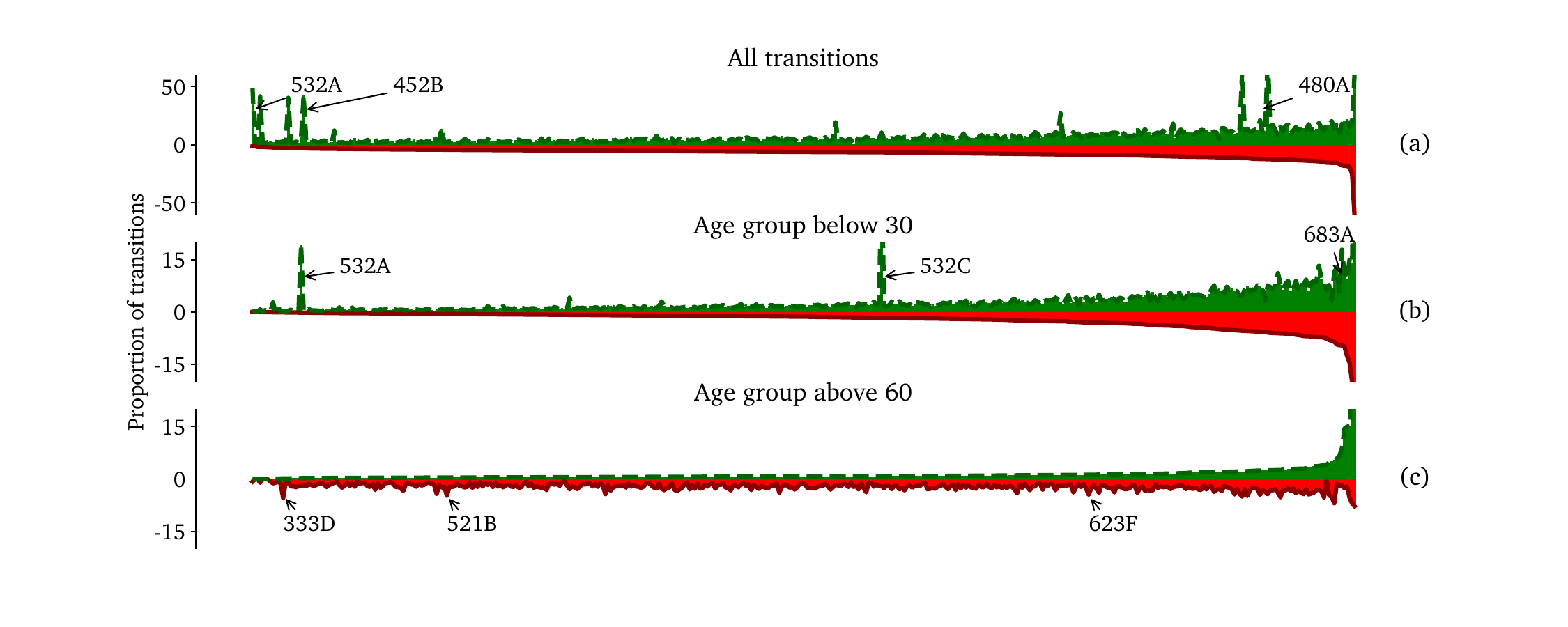}
    \caption{\textbf{Entry and Exit Dynamics in the Labor Market}: Distribution of inflows (green) and outflows (red) for each occupation (on the x-axis) over the sample period for (a) all transitions, (b) workers below 30 years old, and (c) workers above 60 years old. The y-axis represents the proportion of transitions relative to the total number of workers in each occupation. Throughout the sample period, there is a slightly higher influx of workers into the labor market, amounting to 2\% more inflows. Among the age group below 30, the influx is 25\% higher, while for the age group below 60, outflows exceed inflows by a factor of two.}
    \label{fig:entryexit}
\end{figure}

Figure \ref{fig:entryexit} illustrates the proportion of workers transitioning into (green) and out of (red) specific occupations across different age groups. There is variability in the inflow and outflow across occupations. In panel (a), which depicts the entire network distribution, occupations like Gendarmes (532A) or Chief Warrant Officers (452B) exhibit significantly higher inflows than outflows, whereas Foremen and supervisors in agriculture, forestry (480A) experience higher outflows than inflows.

In Figure \ref{fig:entryexit} (b), displaying distributions only for workers under 30, occupations such as Apprentice bakers and butchers (683A) show more outflows than inflows, while Non-commissioned military officers have higher inflows than outflows. Overall, in the under-30 age group the inflow of workers is 25\% higher than the outflow.

Conversely, for workers above 60, the outflow exceeds the inflow by a factor of two. The occupations with the highest outflow are for example Administrative executives of France Telecom (333D), France Telecom employees (521B) Qualified metal machining operators (623F).

\section{BRIM Communities}\label{appx:brimcluster}

\begin{longtblr}[
caption = {Occupations in the BRIM Community ``Blue Collar Workers"},
  label = {table:sources}
                    ]{colsep  = 4pt,
                      colspec = {@{} X[1, j, cmd=\RaggedRight] X[8,j] 
                                     @{}},
                      rows    = {font=\small},
                      row{1}  = {font=\small\bfseries},
                      rowsep  = 0.3pt,
                      rowhead = 1,
                      }
    \toprule
                PCS-ESE & Occupation Name in French          \\
    \midrule
477D      & Techniciens de l'environnement et du traitement des pollutions                                                                                                         \\
 623F      & Opérateurs qualifiés d'usinage des métaux travaillant à l'unité ou en petite série, moulistes qualifiés                                                                \\
 623E      & Soudeurs manuels                                                                                                                                                       \\
 477A      & Techniciens de la logistique, du planning et de l'ordonnancement                                                                                                       \\
 621A      & Chefs d'équipe du gros oeuvre et des travaux publics                                                                                                                   \\
 621B      & Ouvriers qualifiés du travail du béton                                                                                                                                 \\
 623D      & Opérateurs qualifiés sur machine de soudage                                                                                                                            \\
 465C      & Photographes                                                                                                                                                           \\
 476A      & Assistants techniques, techniciens de l'imprimerie et de l'édition                                                                                                     \\
 475A      & Techniciens de recherche-développement et des méthodes de production des industries de transformation                                                                  \\
 475B      & Techniciens de production et de contrôle-qualité des industries de transformation                                                                                      \\
 476B      & Techniciens de l'industrie des matériaux souples, de l'ameublement et du bois                                                                                          \\
 624B      & Monteurs, metteurs au point très qualifiés d'ensembles mécaniques travaillant à l'unité ou en petite série                                                             \\
 534A      & Agents civils de sécurité et de surveillance                                                                                                                           \\
 624C      & Monteurs qualifiés d'ensembles mécaniques travaillant en moyenne ou en grande série                                                                                    \\
 624D      & Monteurs qualifiés en structures métalliques                                                                                                                           \\
 623G      & Opérateurs qualifiés d'usinage des métaux sur autres machines (sauf moulistes)                                                                                         \\
 487B      & Responsables du tri, de l'emballage, de l'expédition et autres responsables de la manutention                                                                          \\
 622E      & Autres monteurs câbleurs en électronique                                                                                                                               \\
 623A      & Chaudronniers-tôliers industriels, opérateurs qualifiés du travail en forge, conducteurs qualifiés d'équipement de formage, traceurs qualifiés                         \\
 477C      & Techniciens d'installation et de maintenance des équipements non industriels (hors informatique et télécommunications)                                                 \\
 621G      & Mineurs de fond qualifiés et autres ouvriers qualifiés des industries d'extraction (carrières, pétrole, gaz...)                                                        \\
 477B      & Techniciens d'installation et de maintenance des équipements industriels (électriques, électromécaniques, mécaniques, hors informatique)                               \\
 621E      & Autres ouvriers qualifiés des travaux publics                                                                                                                          \\
 621D      & Ouvriers des travaux publics en installations électriques et de télécommunications                                                                                     \\
 621C      & Conducteurs qualifiés d'engins de chantiers du bâtiment et des travaux publics                                                                                         \\
 481A      & Conducteurs de travaux (non cadres)                                                                                                                                    \\
 481B      & Chefs de chantier (non cadres)                                                                                                                                         \\
 484A      & Agents de maîtrise en fabrication : agroalimentaire, chimie, plasturgie, pharmacie.                                                                                    \\
 484B      & Agents de maîtrise en fabrication : métallurgie, matériaux lourds et autres industries de transformation                                                               \\
 485A      & Agents de maîtrise et techniciens en production et distribution d'énergie, eau, chauffage                                                                              \\
 485B      & Agents de maîtrise en fabrication des autres industries (imprimerie, matériaux souples, ameublement et bois)                                                           \\
 486B      & Agents de maîtrise en maintenance, installation en électricité et électronique                                                                                         \\
 474B      & Techniciens de recherche-développement et des méthodes de fabrication en construction mécanique et travail des métaux                                                  \\
 474A      & Dessinateurs en construction mécanique et travail des métaux                                                                                                           \\
 473C      & Techniciens de fabrication et de contrôle-qualité en électricité, électromécanique et électronique                                                                     \\
473B      & Techniciens de recherche-développement et des méthodes de fabrication en électricité, électromécanique et électronique                                                 \\
 486D      & Agents de maîtrise en maintenance, installation en mécanique                                                                                                           \\
 622A      & Opérateurs qualifiés sur machines automatiques en production électrique ou électronique                                                                                \\
 472C      & Métreurs et techniciens divers du bâtiment et des travaux publics                                                                                                      \\
 486E      & Agents de maîtrise en entretien général, installation, travaux neufs (hors mécanique, électromécanique, électronique)                                                  \\
 487A      & Responsables d'entrepôt, de magasinage                                                                                                                                 \\
 622C      & Monteurs câbleurs qualifiés en électricité                                                                                                                             \\
 623B      & Tuyauteurs industriels qualifiés                                                                                                                                       \\
 474C      & Techniciens de fabrication et de contrôle-qualité en construction mécanique et travail des métaux                                                                      \\
 479B      & Experts salariés de niveau technicien, techniciens divers                                                                                                              \\
 624F      & Ouvriers qualifiés des traitements thermiques et de surface sur métaux                                                                                                 \\
 673B      & Ouvriers de production non qualifiés travaillant par formage de métal                                                                                                  \\
 673A      & Ouvriers de production non qualifiés travaillant par enlèvement de métal                                                                                               \\
 672A      & Ouvriers non qualifiés de l'électricité et de l'électronique                                                                                                           \\
 671C      & Ouvriers non qualifiés des travaux publics et du travail du béton                                                                                                      \\
 655A      & Autres agents et ouvriers qualifiés (sédentaires) des services d'exploitation des transports                                                                           \\
 653A      & Magasiniers qualifiés                                                                                                                                                  \\
 673C      & Ouvriers non qualifiés de montage, contrôle en mécanique et travail des métaux                                                                                         \\
 652A      & Ouvriers qualifiés de la manutention, conducteurs de chariots élévateurs, caristes                                                                                     \\
 651A      & Conducteurs d'engin lourd de levage                                                                                                                                    \\
 644A      & Conducteurs de véhicule de ramassage des ordures ménagères                                                                                                             \\
 624E      & Ouvriers qualifiés de contrôle et d'essais en mécanique                                                                                                                \\
 641A      & Conducteurs routiers et grands routiers                                                                                                                                \\
 637D      & Ouvriers qualifiés divers de type artisanal                                                                                                                            \\
 636C      & Boulangers, pâtissiers (sauf activité industrielle)                                                                                                                    \\
 651B      & Conducteurs d'engin lourd de manoeuvre                                                                                                                                 \\
 674A      & Ouvriers de production non qualifiés : chimie, pharmacie, plasturgie                                                                                                   \\
 674B      & Ouvriers de production non qualifiés de la transformation des viandes                                                                                                  \\
 674C      & Autres ouvriers de production non qualifiés : industrie agroalimentaire                                                                                                \\
 685A      & Ouvriers non qualifiés divers de type artisanal                                                                                                                        \\
 684B      & Ouvriers non qualifiés de l'assainissement et du traitement des déchets                                                                                                \\
 684A      & Nettoyeurs                                                                                                                                                             \\
 682A      & Métalliers, serruriers, réparateurs en mécanique non qualifiés                                                                                                         \\
 681B      & Ouvriers non qualifiés du second oeuvre du bâtiment                                                                                                                    \\
 681A      & Ouvriers non qualifiés du gros oeuvre du bâtiment                                                                                                                      \\
 676E      & Ouvriers non qualifiés divers de type industriel                                                                                                                       \\
 676D      & Agents non qualifiés des services d'exploitation des transports                                                                                                        \\
 676C      & Ouvriers du tri, de l'emballage, de l'expédition, non qualifiés                                                                                                        \\
 676A      & Manutentionnaires non qualifiés                                                                                                                                        \\
 675C      & Ouvriers de production non qualifiés de l'imprimerie, presse, édition                                                                                                  \\
 675B      & Ouvriers de production non qualifiés du travail du bois et de l'ameublement                                                                                            \\
 675A      & Ouvriers de production non qualifiés du textile et de la confection, de la tannerie-mégisserie et du travail du cuir                                                   \\
 674E      & Ouvriers de production non qualifiés : industrie lourde du bois, fabrication des papiers et cartons                                                                    \\
674D      & Ouvriers de production non qualifiés : métallurgie, production verrière, céramique, matériaux de construction                                                          \\
 635A      & Tailleurs et couturières qualifiés, ouvriers qualifiés du travail des étoffes (sauf fabrication de vêtements), ouvriers qualifiés de type artisanal du travail du cuir \\
 634D      & Mécaniciens qualifiés de maintenance, entretien : équipements non industriels                                                                                          \\
 628G      & Ouvriers qualifiés divers de type industriel                                                                                                                           \\
 634B      & Métalliers, serruriers qualifiés                                                                                                                                       \\
 634C      & Mécaniciens qualifiés en maintenance, entretien, réparation : automobile                                                                                               \\
 628A      & Mécaniciens qualifiés de maintenance, entretien : équipements industriels                                                                                              \\
 627F      & Ouvriers de la composition et de l'impression, ouvriers qualifiés de la brochure, de la reliure et du façonnage du papier-carton                                       \\
 627D      & Ouvriers qualifiés de scierie, de la menuiserie industrielle et de l'ameublement                                                                                       \\
 627C      & Ouvriers qualifiés du travail industriel du cuir                                                                                                                       \\
 627B      & Ouvriers qualifiés de la coupe des vêtements et de l'habillement, autres opérateurs de confection qualifiés                                                            \\
 627A      & Opérateurs qualifiés du textile et de la mégisserie                                                                                                                    \\
 626C      & Opérateurs et ouvriers qualifiés des industries lourdes du bois et de la fabrication du papier-carton                                                                  \\
 628C      & Régleurs qualifiés d'équipements de fabrication (travail des métaux, mécanique)                                                                                        \\
 626B      & Autres opérateurs et ouvriers qualifiés : métallurgie, production verrière, matériaux de construction                                                                  \\
 625H      & Ouvriers qualifiés des autres industries (eau, gaz, énergie, chauffage)                                                                                                \\
 625G      & Autres ouvriers de production qualifiés ne travaillant pas sur machine : industrie agroalimentaire (hors transformation des viandes)                                   \\
 625F      & Autres opérateurs travaillant sur installations ou machines : industrie agroalimentaire (hors transformation des viandes)                                              \\
 625D      & Opérateurs de la transformation des viandes                                                                                                                            \\
 625C      & Autres opérateurs et ouvriers qualifiés de la chimie (y.c. pharmacie) et de la plasturgie                                                                              \\
 625B      & Ouvriers qualifiés et agents qualifiés de laboratoire : agroalimentaire, chimie, biologie, pharmacie                                                                   \\
 625A      & Pilotes d'installation lourde des industries de transformation : agroalimentaire, chimie, plasturgie, énergie                                                          \\
 624G      & Autres mécaniciens ou ajusteurs qualifiés (ou spécialité non reconnue)                                                                                                 \\
 626A      & Pilotes d'installation lourde des industries de transformation : métallurgie, production verrière, matériaux de construction                                           \\
 628D      & Régleurs qualifiés d'équipements de fabrication (hors travail des métaux et mécanique)                                                                                 \\
 628B      & Electromécaniciens, électriciens qualifiés d'entretien : équipements industriels                                                                                       \\
 632A      & Maçons qualifiés                                                                                                                                                       \\
 632C      & Charpentiers en bois qualifiés                                                                                                                                         \\
 634A      & Carrossiers d'automobiles qualifiés                                                                                                                                    \\
 633D      & Electriciens, électroniciens qualifiés en maintenance, entretien : équipements non industriels                                                                         \\
 632D      & Menuisiers qualifiés du bâtiment                                                                                                                                       \\
 633B      & Dépanneurs qualifiés en radiotélévision, électroménager, matériel électronique (salariés)                                                                              \\
 632E      & Couvreurs qualifiés                                                                                                                                                    \\
 632F      & Plombiers et chauffagistes qualifiés                                                                                                                                   \\
 633A      & Electriciens qualifiés de type artisanal (y.c. bâtiment)                                                                                                               \\
 632K      & Ouvriers qualifiés d'entretien général des bâtiments                                                                                                                   \\
 628E      & Ouvriers qualifiés de l'assainissement et du traitement des déchets                                                                                                    \\
 632G      & Peintres et ouvriers qualifiés de pose de revêtements sur supports verticaux                                                                                           \\
 632H      & Soliers moquetteurs et ouvriers qualifiés de pose de revêtements souples sur supports horizontaux                                                                      \\
 632J      & Monteurs qualifiés en agencement, isolation
    
\end{longtblr}

\begin{longtblr}[
caption = {Occupations in the BRIM Community ``Engineers \& Executives"},
  label = {table:sources}
                    ]{colsep  = 4pt,
                      colspec = {@{} X[1, j, cmd=\RaggedRight] X[8,j] 
                                     @{}},
                      rows    = {font=\small},
                      row{1}  = {font=\small\bfseries},
                      rowsep  = 0.3pt,
                      rowhead = 1,
                      }
    \toprule
                PCS-ESE & Occupation Name in French          \\
    \midrule
    542A      & Secrétaires                                                                                                                                                 \\
 387B      & Ingénieurs et cadres de la logistique, du planning et de l'ordonnancement                                                                                   \\
 387A      & Ingénieurs et cadres des achats et approvisionnements industriels                                                                                           \\
 386E      & Ingénieurs et cadres de fabrication des autres industries (imprimerie, matériaux souples, ameublement et bois)                                              \\
 386D      & Ingénieurs et cadres de la production et de la distribution d'énergie, eau                                                                                  \\
 386C      & Ingénieurs et cadres d'étude, recherche et développement des autres industries (imprimerie, matériaux souples, ameublement et bois)                         \\
 386B      & Ingénieurs et cadres d'étude, recherche et développement de la distribution d'énergie, eau                                                                  \\
 385C      & Ingénieurs et cadres technico-commerciaux des industries de transformations (biens intermédiaires)                                                          \\
 385B      & Ingénieurs et cadres de fabrication des industries de transformation (agroalimentaire, chimie, métallurgie, matériaux lourds)                               \\
 385A      & Ingénieurs et cadres d'étude, recherche et développement des industries de transformation (agroalimentaire, chimie, métallurgie, matériaux lourds)          \\
 384C      & Ingénieurs et cadres technico-commerciaux en matériel mécanique professionnel                                                                               \\
 384B      & Ingénieurs et cadres de fabrication en mécanique et travail des métaux                                                                                      \\
 387C      & Ingénieurs et cadres des méthodes de production                                                                                                             \\
 384A      & Ingénieurs et cadres d'étude, recherche et développement en mécanique et travail des métaux                                                                 \\
 383B      & Ingénieurs et cadres de fabrication en matériel électrique, électronique                                                                                    \\
 383A      & Ingénieurs et cadres d'étude, recherche et développement en électricité, électronique                                                                       \\
 382D      & Ingénieurs et cadres technico-commerciaux en bâtiment, travaux publics                                                                                      \\
 382C      & Ingénieurs, cadres de chantier et conducteurs de travaux (cadres) du bâtiment et des travaux publics                                                        \\
 382B      & Architectes salariés                                                                                                                                        \\
 461C      & Secrétaires de niveau supérieur (non cadres, hors secrétaires de direction)                                                                                 \\
 376G      & Cadres de l'immobilier                                                                                                                                      \\
 376F      & Cadres des services techniques des organismes de sécurité sociale et assimilés                                                                              \\
 376E      & Cadres des services techniques des assurances                                                                                                               \\
 376D      & Chefs d'établissements et responsables de l'exploitation bancaire                                                                                           \\
 376C      & Cadres commerciaux de la banque                                                                                                                             \\
 383C      & Ingénieurs et cadres technico-commerciaux en matériel électrique ou électronique professionnel                                                              \\
 376B      & Cadres des opérations bancaires                                                                                                                             \\
 387D      & Ingénieurs et cadres du contrôle-qualité                                                                                                                    \\
 387F      & Ingénieurs et cadres techniques de l'environnement                                                                                                          \\
 461D      & Maîtrise et techniciens des services financiers ou comptables                                                                                               \\
 461E      & Maîtrise et techniciens administratifs des services juridiques ou du personnel                                                                              \\
 461F      & Maîtrise et techniciens administratifs des autres services administratifs                                                                                   \\
 462C      & Acheteurs non classés cadres, aides-acheteurs                                                                                                               \\
 462E      & Autres professions intermédiaires commerciales (sauf techniciens des forces de vente)                                                                       \\
 463B      & Techniciens commerciaux et technico-commerciaux, représentants en biens d'équipement, en biens intermédiaires, commerce interindustriel (hors informatique) \\
 463C      & Techniciens commerciaux et technico-commerciaux, représentants en biens de consommation auprès d'entreprises                                                \\
 463D      & Techniciens commerciaux et technico-commerciaux, représentants en services auprès d'entreprises ou de professionnels (hors banque, assurance, informatique) \\
 463E      & Techniciens commerciaux et technico-commerciaux, représentants auprès de particuliers (hors banque, assurance, informatique)                                \\
 464A      & Assistants de la publicité, des relations publiques                                                                                                         \\
    465A      & Concepteurs et assistants techniques des arts graphiques, de la mode et de la décoration salariés                                                           \\
 387E      & Ingénieurs et cadres de la maintenance, de l'entretien et des travaux neufs                                                                                 \\
 467A      & Chargés de clientèle bancaire                                                                                                                               \\
 467C      & Professions intermédiaires techniques et commerciales des assurances                                                                                        \\
 543B      & Employés qualifiés des services comptables ou financiers                                                                                                    \\
 467D      & Professions intermédiaires techniques des organismes de sécurité sociale                                                                                    \\
 472A      & Dessinateurs en bâtiment, travaux publics                                                                                                                   \\
 472B      & Géomètres, topographes                                                                                                                                      \\
 473A      & Dessinateurs en électricité, électromécanique et électronique                                                                                               \\
 388E      & Ingénieurs et cadres spécialistes des télécommunications                                                                                                    \\
 388D      & Ingénieurs et cadres technico-commerciaux en informatique et télécommunications                                                                             \\
 388C      & Chefs de projets informatiques, responsables informatiques                                                                                                  \\
 388B      & Ingénieurs et cadres d'administration, maintenance, support et services aux utilisateurs en informatique                                                    \\
 388A      & Ingénieurs et cadres d'étude, recherche et développement en informatique                                                                                    \\
 467B      & Techniciens des opérations bancaires                                                                                                                        \\
 376A      & Cadres des marchés financiers                                                                                                                               \\
 382A      & Ingénieurs et cadres d'étude du bâtiment et des travaux publics                                                                                             \\
 375A      & Cadres de la publicité                                                                                                                                      \\
 354A      & Artistes plasticiens                                                                                                                                        \\
 478D      & Techniciens des télécommunications et de l'informatique des réseaux                                                                                         \\
 353A      & Directeurs de journaux, administrateurs de presse, directeurs d'éditions (littéraire, musicale, audiovisuelle et multimédia)                                \\
 352A      & Journalistes (y c. rédacteurs en chef)                                                                                                                      \\
 375B      & Cadres des relations publiques et de la communication                                                                                                       \\
 555A      & Vendeurs par correspondance, télévendeurs                                                                                                                   \\
 545D      & Employés des services techniques des organismes de sécurité sociale et assimilés                                                                            \\
 545C      & Employés des services techniques des assurances                                                                                                             \\
 545B      & Employés des services commerciaux de la banque                                                                                                              \\
 545A      & Employés administratifs des services techniques de la banque                                                                                                \\
 544A      & Employés et opérateurs d'exploitation en informatique                                                                                                       \\
 543H      & Employés administratifs non qualifiés                                                                                                                       \\
 543G      & Employés administratifs qualifiés des autres services des entreprises                                                                                       \\
 543F      & Employés qualifiés des services commerciaux des entreprises (hors vente)                                                                                    \\
 543E      & Employés qualifiés des services du personnel et des services juridiques                                                                                     \\
 231A      & Chefs de grande entreprise de 500 salariés et plus                                                                                                          \\
 543C      & Employés non qualifiés des services comptables ou financiers                                                                                                \\
 478C      & Techniciens d'installation, de maintenance, support et services aux utilisateurs en informatique                                                            \\
 478B      & Techniciens de production, d'exploitation en informatique                                                                                                   \\
 461B      & Secrétaires de direction, assistants de direction (non cadres)                                                                                              \\
 374D      & Cadres commerciaux des petites et moyennes entreprises (hors commerce de détail)                                                                            \\
 372A      & Cadres chargés d'études économiques, financières, commerciales                                                                                              \\
 372B      & Cadres de l'organisation ou du contrôle des services administratifs et financiers                                                                           \\
 374C      & Cadres commerciaux des grandes entreprises (hors commerce de détail)  \\
  372C      & Cadres spécialistes des ressources humaines et du recrutement                                                                                               \\
 374B      & Chefs de produits, acheteurs du commerce et autres cadres de la mercatique                                                                                  \\
 372E      & Juristes                                                                                                                                                    \\
 373B      & Cadres des autres services administratifs des grandes entreprises                                                                                           \\
 373A      & Cadres des services financiers ou comptables des grandes entreprises                                                                                        \\
 373D      & Cadres des autres services administratifs des petites et moyennes entreprises                                                                               \\
 373C      & Cadres des services financiers ou comptables des petites et moyennes entreprises                                                                            \\
 478A      & Techniciens d'étude et de développement en informatique

\end{longtblr}

\begin{longtblr}[
caption = {Occupations in the BRIM Community ``Public, Health \& Education"},
  label = {table:sources}
                    ]{colsep  = 4pt,
                      colspec = {@{} X[1, j, cmd=\RaggedRight] X[8,j] 
                                     @{}},
                      rows    = {font=\small},
                      row{1}  = {font=\small\bfseries},
                      rowsep  = 0.3pt,
                      rowhead = 1,
                      }
    \toprule
                PCS-ESE & Occupation Name in French          \\
    \midrule
 372F      & Cadres de la documentation, de l'archivage (hors fonction publique)                                                    \\
 525A      & Agents de service des établissements primaires                                                                         \\
 524C      & Agents administratifs des collectivités locales                                                                        \\
 524B      & Agents administratifs de l'Etat et assimilés (sauf Poste, France Télécom)                                              \\
 523C      & Adjoints administratifs des collectivités locales                                                                      \\
 523D      & Adjoints administratifs des hôpitaux publics                                                                           \\
 523B      & Adjoints administratifs de l'Etat et assimilés (sauf Poste, France Télécom)                                            \\
 422B      & Professeurs de lycée professionnel                                                                                     \\
 422A      & Professeurs d'enseignement général des collèges                                                                        \\
 421B      & Professeurs des écoles                                                                                                 \\
 372D      & Cadres spécialistes de la formation                                                                                    \\
 422C      & Maîtres auxiliaires et professeurs contractuels de l'enseignement secondaire                                           \\
 351A      & Bibliothécaires, archivistes, conservateurs et autres cadres du patrimoine                                             \\
 342B      & Professeurs et maîtres de conférences                                                                                  \\
 342G      & Ingénieurs d'étude et de recherche de la recherche publique                                                            \\
 526D      & Aides médico-psychologiques                                                                                            \\
 526A      & Aides-soignants                                                                                                        \\
 525D      & Agents de service hospitaliers                                                                                         \\
 525C      & Agents de service de la fonction publique (sauf écoles, hôpitaux)                                                      \\
 311D      & Psychologues, psychanalystes, psychothérapeutes (non médecins)                                                         \\
 331A      & Personnels de direction de la fonction publique (Etat, collectivités locales, hôpitaux)                                \\
 332A      & Ingénieurs de l'Etat (y.c. ingénieurs militaires) et assimilés                                                         \\
 332B      & Ingénieurs des collectivités locales et des hôpitaux                                                                   \\
 333A      & Magistrats*                                                                                                            \\
 525B      & Agents de service des autres établissements d'enseignement                                                             \\
 333E      & Autres personnels administratifs de catégorie A de l'Etat (hors Enseignement, Patrimoine, Impôts, Trésor, Douanes)     \\
 333F      & Personnels administratifs de catégorie A des collectivités locales et hôpitaux publics (hors Enseignement, Patrimoine) \\
 341A      & Professeurs agrégés et certifiés de l'enseignement secondaire                                                          \\
 422E      & Surveillants et aides-éducateurs des établissements d'enseignement                                                     \\
 342C      & Professeurs agrégés et certifiés en fonction dans l'enseignement supérieur                                             \\
 342D      & Personnel enseignant temporaire de l'enseignement supérieur                                                            \\
 342F      & Directeurs et chargés de recherche de la recherche publique                                                            \\
 343A      & Psychologues spécialistes de l'orientation scolaire et professionnelle                                                 \\
 423B      & Formateurs et animateurs de formation continue                                                                         \\
 563A      & Assistantes maternelles, gardiennes d'enfants, familles d'accueil                                                      \\
 425A      & Sous-bibliothécaires, cadres intermédiaires du patrimoine                                                              \\
 424A      & Moniteurs et éducateurs sportifs, sportifs professionnels                                                              \\
 434G      & Educateurs de jeunes enfants                                                                                           \\
 435A      & Directeurs de centres socioculturels et de loisirs                                                                     \\
 435B      & Animateurs socioculturels et de loisirs                                                                                \\
 441A      & Clergé séculier                                                                                                        \\
 441B      & Clergé régulier                                                                                                        \\
  451C      & Contrôleurs des Impôts, du Trésor, des Douanes et assimilés*                                                           \\
 434F      & Educateurs techniques spécialisés, moniteurs d'atelier                                                                 \\
 451E      & Autres personnels administratifs de catégorie B de l'Etat (hors Enseignement, Patrimoine, Impôts, Trésor, Douanes)     \\
 451H      & Professions intermédiaires administratives des hôpitaux                                                                \\
 472D      & Techniciens des travaux publics de l'Etat et des collectivités locales                                                 \\
 479A      & Techniciens des laboratoires de recherche publique ou de l'enseignement                                                \\
 621F      & Ouvriers qualifiés des travaux publics (salariés de l'Etat et des collectivités locales)                               \\
 564B      & Employés des services divers                                                                                           \\
 563C      & Employés de maison et personnels de ménage chez des particuliers                                                       \\
 563B      & Aides à domicile, aides ménagères, travailleuses familiales                                                            \\
 451G      & Professions intermédiaires administratives des collectivités locales                                                   \\
 434E      & Moniteurs éducateurs                                                                                                   \\
 422D      & Conseillers principaux d'éducation                                                                                     \\
 431F      & Infirmiers en soins généraux, salariés                                                                                 \\
 431A      & Cadres infirmiers et assimilés                                                                                         \\
 431B      & Infirmiers psychiatriques                                                                                              \\
 431C      & Puéricultrices                                                                                                         \\
 431D      & Infirmiers spécialisés (autres qu'infirmiers psychiatriques et puéricultrices)                                         \\
 628F      & Agents qualifiés de laboratoire (sauf chimie, santé)                                                                   \\
 434C      & Conseillers en économie sociale familiale                                                                              \\
 434D      & Educateurs spécialisés                                                                                                 \\
 432B      & Masseurs-kinésithérapeutes rééducateurs, salariés                                                                      \\
 432D      & Autres spécialistes de la rééducation, salariés                                                                        \\
 434B      & Assistants de service social                                                                                           \\
 434A      & Cadres de l'intervention socio-éducative                                                                               \\
 433A      & Techniciens médicaux                                                                                                   
    
\end{longtblr}

\begin{longtblr}[
caption = {Occupations in the BRIM Community ``Sales \& Retail"},
  label = {table:sources}
                    ]{colsep  = 4pt,
                      colspec = {@{} X[1, j, cmd=\RaggedRight] X[8,j] 
                                     @{}},
                      rows    = {font=\small},
                      row{1}  = {font=\small\bfseries},
                      rowsep  = 0.3pt,
                      rowhead = 1,
                      }
    \toprule
                PCS-ESE & Occupation Name in French          \\
    \midrule
 433D      & Préparateurs en pharmacie                                                             \\
 468A      & Maîtrise de restauration : salle et service                                           \\
 462D      & Animateurs commerciaux des magasins de vente, marchandiseurs (non cadres)             \\
 462B      & Maîtrise de l'exploitation des magasins de vente                                      \\
 462A      & Chefs de petites surfaces de vente                                                    \\
 468B      & Maîtrise de l'hébergement : hall et étages                                            \\
 541D      & Standardistes, téléphonistes                                                          \\
 541C      & Agents d'accueil non qualifiés                                                        \\
 488A      & Maîtrise de restauration  : cuisine/production                                        \\
 636A      & Bouchers (sauf industrie de la viande)                                                \\
 636B      & Charcutiers (sauf industrie de la viande)                                             \\
 377A      & Cadres de l'hôtellerie et de la restauration                                          \\
 636D      & Cuisiniers et commis de cuisine                                                       \\
 374A      & Cadres de l'exploitation des magasins de vente du commerce de détail                  \\
 233D      & Chefs d'entreprise de services, de 10 à 49 salariés                                   \\
 233C      & Chefs d'entreprise commerciale, de 10 à 49 salariés                                   \\
 233B      & Chefs d'entreprise de l'industrie ou des transports, de 10 à 49 salariés              \\
 233A      & Chefs d'entreprise du bâtiment et des travaux publics, de 10 à 49 salariés            \\
 232A      & Chefs de moyenne entreprise, de 50 à 499 salariés                                     \\
 220X      & Commerçants et assimilés, salariés de leur entreprise                                 \\
 433B      & Opticiens lunetiers et audioprothésistes salariés                                     \\
 488B      & Maîtrise de restauration  : gestion d'établissement                                   \\
 627E      & Ouvriers de la photogravure et des laboratoires photographiques et cinématographiques \\
 210X      & Artisans salariés de leur entreprise                                                  \\
 554C      & Vendeurs en droguerie, bazar, quincaillerie, bricolage                                \\
 554B      & Vendeurs en ameublement, décor, équipement du foyer                                   \\
 554A      & Vendeurs en alimentation                                                              \\
 554E      & Vendeurs en habillement et articles de sport                                          \\
 554F      & Vendeurs en produits de beauté, de luxe (hors biens culturels) et optique             \\
 554G      & Vendeurs de biens culturels (livres, disques, multimédia, objets d'art)               \\
 553C      & Autres vendeurs non spécialisés                                                       \\
 554H      & Vendeurs de tabac, presse et articles divers                                          \\
 556A      & Vendeurs en gros de biens d'équipement, biens intermédiaires                          \\
 561C      & Serveurs, commis de restaurant, garçons non qualifiés                                 \\
 561D      & Aides de cuisine, apprentis de cuisine et employés polyvalents de la restauration     \\
 561B      & Serveurs, commis de restaurant, garçons qualifiés                                     \\
 562A      & Manucures, esthéticiens                                                               \\
 551A      & Employés de libre service du commerce et magasiniers                                  \\
 552A      & Caissiers de magasin                                                                  \\
 541B      & Agents d'accueil qualifiés, hôtesses d'accueil et d'information                       \\
 561E      & Employés de l'hôtellerie : réception et hall                                          \\
 561F      & Employés d'étage et employés polyvalents de l'hôtellerie                              \\
 562B      & Coiffeurs salariés                                                                    \\
    
\end{longtblr}

\begin{longtblr}[
caption = {Occupations in the BRIM Community ``Transport \& Logistic"},
  label = {table:sources}
                    ]{colsep  = 4pt,
                      colspec = {@{} X[1, j, cmd=\RaggedRight] X[8,j] 
                                     @{}},
                      rows    = {font=\small},
                      row{1}  = {font=\small\bfseries},
                      rowsep  = 0.3pt,
                      rowhead = 1,
                      }
    \toprule
                PCS-ESE & Occupation Name in French          \\
    \midrule
			642A & Conducteurs de taxi \\
			642B & Conducteurs de voiture particulière \\
			546A & Contrôleurs des transports (personnels roulants) \\
			546B & Agents des services commerciaux des transports de voyageurs et du tourisme \\
			546C & Employés administratifs d'exploitation des transports de marchandises \\
			546D & Hôtesses de l'air et stewards \\
			546E & Autres agents et hôtesses d'accompagnement (transports, tourisme) \\
			654B & Conducteurs qualifiés d'engins de transport guidés (sauf remontées mécaniques) \\
			656B & Matelots de la marine marchande \\
			656C & Capitaines et matelots timoniers de la navigation fluviale \\
			643A & Conducteurs livreurs, coursiers \\
			641B & Conducteurs de véhicule routier de transport en commun \\
			692A & Marins-pêcheurs et ouvriers de l'aquaculture \\
			389C & Officiers et cadres navigants techniques de la marine marchande \\
			389B & Officiers et cadres navigants techniques et commerciaux de l'aviation civile \\
			389A & Ingénieurs et cadres techniques de l'exploitation des transports \\
			451A & Professions intermédiaires de la Poste \\
			466A & Responsables commerciaux et administratifs des transports de voyageurs et du tourisme (non cadres) \\
			521A & Employés de la Poste \\
			333C & Cadres de la Poste* \\
			466C & Responsables d'exploitation des transports de voyageurs et de marchandises (non cadres) \\
			526E & Ambulanciers salariés \\
			480B & Maîtres d'équipage de la marine marchande et de la pêche \\
			466B & Responsables commerciaux et administratifs des transports de marchandises (non cadres) 
    
\end{longtblr}

\begin{longtblr}[
caption = {Occupations in the BRIM Community ``Agriculture"},
  label = {table:sources}
                    ]{colsep  = 4pt,
                      colspec = {@{} X[1, j, cmd=\RaggedRight] X[8,j] 
                                     @{}},
                      rows    = {font=\small},
                      row{1}  = {font=\small\bfseries},
                      rowsep  = 0.3pt,
                      rowhead = 1,
                      }
    \toprule
                PCS-ESE & Occupation Name in French          \\
    \midrule
 480A      & Contremaîtres et agents d'encadrement (non cadres) en agriculture, sylviculture                     \\
 691E      & Ouvriers agricoles sans spécialisation particulière                                                 \\
 691A      & Conducteurs d'engin agricole ou forestier                                                           \\
 631A      & Jardiniers                                                                                          \\
 381B      & Ingénieurs et cadres d'étude et développement de l'agriculture, la pêche, les eaux et forêts        \\
 691B      & Ouvriers de l'élevage                                                                               \\
 691C      & Ouvriers du maraîchage ou de l'horticulture                                                         \\
 691D      & Ouvriers de la viticulture ou de l'arboriculture fruitière                                          \\
 381C      & Ingénieurs et cadres de production et d'exploitation de l'agriculture, la pêche, les eaux et forêts \\
 471A      & Techniciens d'étude et de conseil en agriculture, eaux et forêt                                     \\
 471B      & Techniciens d'exploitation et de contrôle de la production en agriculture, eaux et forêt            
\end{longtblr}

\begin{longtblr}[
caption = {Occupations in the BRIM Community ``Artist and Creative Work"},
  label = {table:sources}
                    ]{colsep  = 4pt,
                      colspec = {@{} X[1, j, cmd=\RaggedRight] X[8,j] 
                                     @{}},
                      rows    = {font=\small},
                      row{1}  = {font=\small\bfseries},
                      rowsep  = 0.3pt,
                      rowhead = 1,
                      }
    \toprule
                PCS-ESE & Occupation Name in French          \\
    \midrule
 465B      & Assistants techniques de la réalisation des spectacles vivants et audiovisuels salariés         \\
 354G      & Professeurs d'art (hors établissements scolaires)                                               \\
 354E      & Artistes de la danse                                                                            \\
 354C      & Artistes dramatiques                                                                            \\
 637C      & Ouvriers et techniciens des spectacles vivants et audiovisuels                                  \\
 353B      & Directeurs, responsables de programmation et de production de l'audiovisuel et des spectacles   \\
 354F      & Artistes du cirque et des spectacles divers                                                     \\
 353C      & Cadres artistiques et technico-artistiques de la réalisation de l'audiovisuel et des spectacles \\
 354B      & Artistes de la musique et du chant                                                              
\end{longtblr}

\begin{longtblr}[
caption = {Occupations in the BRIM Community ``Medical Service"},
  label = {table:sources}
                    ]{colsep  = 4pt,
                      colspec = {@{} X[1, j, cmd=\RaggedRight] X[8,j] 
                                     @{}},
                      rows    = {font=\small},
                      row{1}  = {font=\small\bfseries},
                      rowsep  = 0.3pt,
                      rowhead = 1,
                      }
    \toprule
                PCS-ESE & Occupation Name in French          \\
    \midrule
			311C & Chirurgiens dentistes \\
			433C & Autres spécialistes de l'appareillage médical salariés \\
			344D & Pharmaciens salariés \\
			344C & Internes en médecine, odontologie et pharmacie \\
			344A & Médecins hospitaliers sans activité libérale \\
			526B & Assistants dentaires, médicaux et vétérinaires, aides de techniciens médicaux \\
			344B & Médecins salariés non hospitaliers \\
    
\end{longtblr}

\section{Economic Fitness and Complexity}\label{appx:fc}

\begin{figure}[t]
    \centering
    \includegraphics[width=0.9\textwidth]{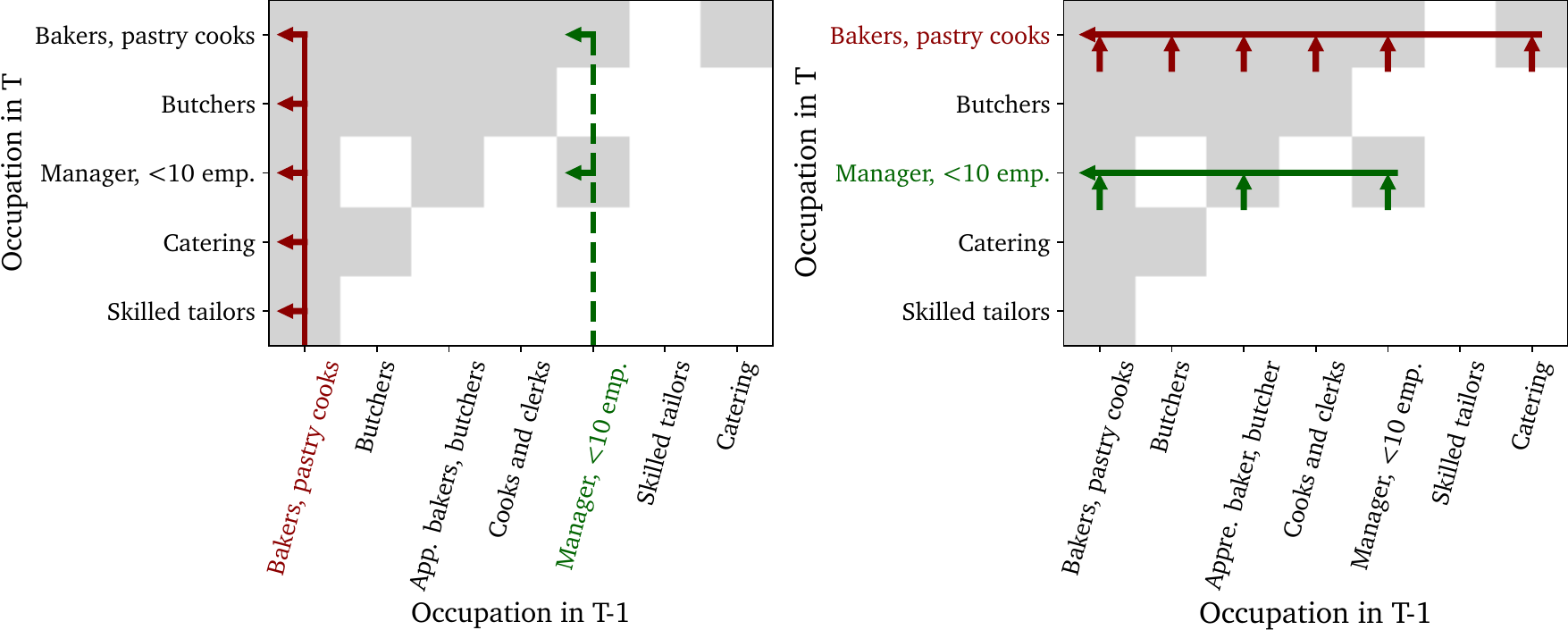}
    \caption{\textbf{Schematic Visualization of the Degree of Transferability and Accessibility.} An illustrative representation of a transitional occupation matrix $\transmatf$, where a black square signifies a connection between two occupations ($\transmatf>10^{-2}$), while a white square denotes no connection. (left) Transferability of an occupation is linked to the number of occupations a worker can transition to from a given occupation in time T-1. (right) Accessibility of an occupation is associated with the number of occupations from which a worker could have transitioned to a given occupation in time T. In this example, '. }
    \label{fig:schematic_example}
\end{figure}

Figure \ref{fig:schematic_example} shows a schematic example of transferability (left) and accessibility (right) based on  $\transmatf$.
A black square indicates a link between two occupations, i.e. $\transmatf>10^{-2}$.
In Figure \ref{fig:schematic_example} bakers and pastry cooks exhibit higher transferability than manager (left panel) as they have more opportunities for transitions.
Similarly, the occupation of bakers and pastry cooks is more accessible than the occupation of a manager, i.e. more different occupations lead to bakers and pastry cooks than to a manager.

\begin{figure}[t]
    \centering
    \includegraphics[width=0.9\textwidth]{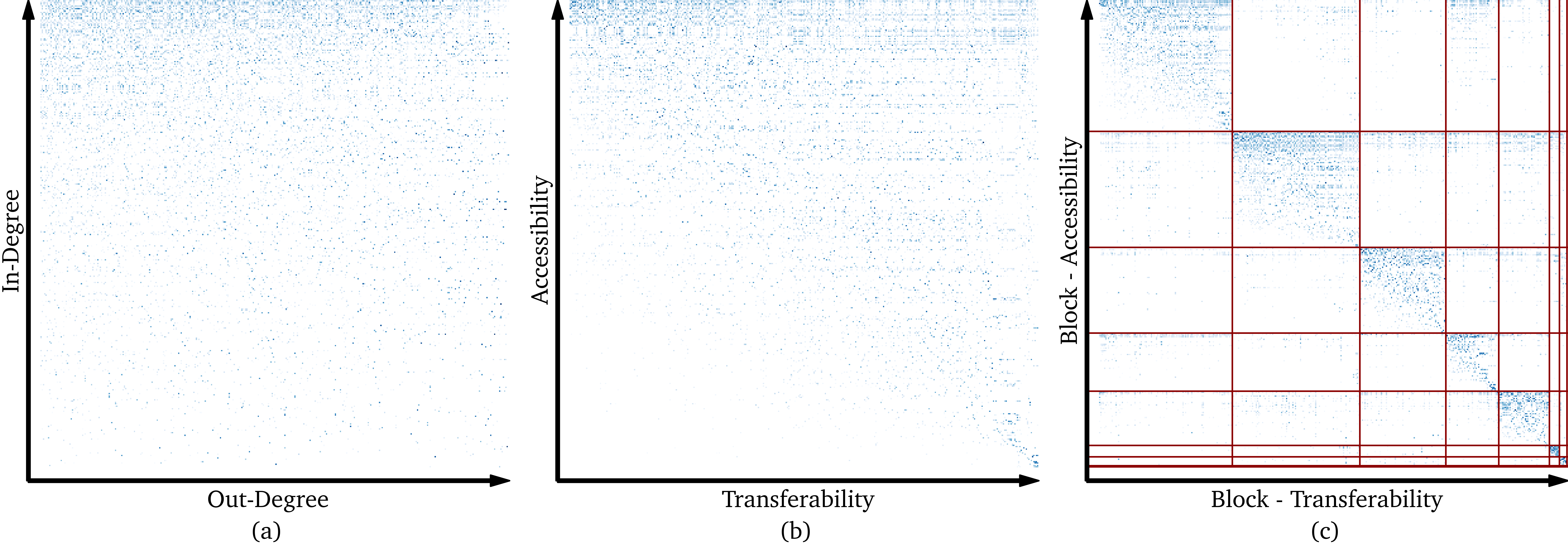}
    \caption{\textbf{Sorted, Transition Probability Matrix.} X-axis corresponds to occupations in (T-1) and y-axis to occupations in (T). (a) Diversification and Ubiquity ordering (b) Accessibility and Transferability ordering (c) In-Block Accessibility and Transferability ordering. Each Block, detected by the BRIM algorithm (red lines), has a higher NODF than the raw matrix. The intensity of the coloring corresponds to higher probability flow from (T-1) to (T).}
    \label{fig:fitness_complexity}
\end{figure}

We need to make sure that the fitness and complexity algorithm converges. For the iterative process to find a fixed point for Accessibility and Transferability, the underlying transition matrix needs to exhibit a certain nestedness \citep{tacchella2012new}. 
Ordering the axes of the binary transition matrix based only on the in- and out-degree of the respective occupation does not reveal a nested structure (Figure \ref{fig:fitness_complexity} (a)). 
However, when sorting the x-axis by transferability and the y-axis by accessibility, an inherent nested structure within the labor market becomes apparent (see Figure \ref{fig:fitness_complexity} (b)). 
In this arrangement, the bottom-right corner of the matrix contains on the x-axis occupations characterized by the highest transferability and on the y-axis occupations with lowest accessibility (see Figure \ref{fig:fitness_complexity} (b)). 
This nestedness reveals a distinct hierarchy in labor flows, suggesting that occupations positioned higher on the y-axis are broadly accessible to a diverse array of occupations. Nestedness can be quantified through the NODF (Nestedness based on Overlap and Decreasing Fill), which determines the degree to which the neighborhood of a given node is a subset of better-connected nodes' neighborhoods. The values range from 0 to 100, where 100 indicates full nestedness, and 0 signifies a non-nested pattern resembling a checkerboard structure. For the French labor market, the $\text{NODF} = 10.01$, falling within a comparable order of magnitude to other economic systems \citep{ren2020bridging}. For instance, in the company-product space, NODF values are around 12 \citep{laudati2023different}.

Despite comparable high nestedness, the Accessibility-Transferability metrics do not consider the labor market's inherent community structure. Subsequently, we evaluate the in-block accessibility and transferability of the communities given by the BRIM algorithm (see Figure \ref{fig:fitness_complexity} (c) and Section \ref{sec:clusteringBRIM}). As expected, the nestedness within each block exceeds the overall nestedness of the unclustered transition matrix ($\langle\text{NODF}\rangle = 12.56$). As already found in Section \ref{sec:clusteringBRIM} most transitions occur within these communities, but some extend beyond them (see Figure \ref{fig:fitness_complexity} (c)). Transitions outside the community are mainly associated with occupations having high transferability, suggesting a bottleneck effect for individuals in low-transferability occupations (Condenser and Channel occupation) aspiring to transition beyond their community. These findings already suggests a potential avenue for increasing reallocation speed since one can identify precisely occupations that cause frictions (see Section \ref{sec:policy}). 

It is reassuring to note that each community, whether defined by the 1-digit PCS code or by the BRIM algorithm, incorporates Hub occupations. These occupations serve as potential gateways within each community, facilitating access to desired occupations outside the community. Interestingly, this method is flexible enough to study the change of transferability and accessibility metrics and the corresponding clusters over time. When comparing accessibility and transferability metrics across different transition matrices for various years, it's important to note that direct value comparison is not possible since these metrics are relative within each year. Instead, one need to rely on a rank comparisons between different years.

\section{Taxonomy Clusters}\label{appx:taxonomy}

\begin{longtblr}[
caption = {Occupations in the cluster ``Hub"},
  label = {table:sources}
                    ]{colsep  = 4pt,
                      colspec = {@{} X[1, j, cmd=\RaggedRight] X[8,j] 
                                     X[1, j, cmd=\RaggedRight] X[1,j] @{}},
                      rows    = {font=\small},
                      row{1}  = {font=\small\bfseries},
                      rowsep  = 0.3pt,
                      rowhead = 1,
                      }
    \toprule
PCS-ESE & Occupation Name in French &  $T_i$ (log) &  $A_i$ (log)         \\
  \midrule
                 389B      & Officiers et cadres navigants techniques et commerciaux de l'aviation civile                        &           3.36 &          -2.45 \\
                 627E      & Ouvriers de la photogravure et des laboratoires photographiques et cinématographiques               &           2.97 &          -1.79 \\
                 431F      & Infirmiers en soins généraux, salariés                                                              &           2.86 &           0.39 \\
                 623E      & Soudeurs manuels                                                                                    &           2.59 &          -0.29 \\
                 465A      & Concepteurs et assistants techniques des arts graphiques, de la mode et de la décoration salariés   &           2.39 &          -0.34 \\
                 636A      & Bouchers (sauf industrie de la viande)                                                              &           1.86 &          -0.65 \\
                 554G      & Vendeurs de biens culturels (livres, disques, multimédia, objets d'art)                             &           1.85 &          -0.7  \\
                 381C      & Ingénieurs et cadres de production et d'exploitation de l'agriculture, la pêche, les eaux et forêts &           1.83 &          -1.38 
    
\end{longtblr}
\begin{longtblr}[
caption = {Occupations in the cluster ``Diffuser"},
  label = {table:sources}
                    ]{colsep  = 4pt,
                      colspec = {@{} X[1, j, cmd=\RaggedRight] X[8,j] 
                                     X[1, j, cmd=\RaggedRight] X[1,j] @{}},
                      rows    = {font=\small},
                      row{1}  = {font=\small\bfseries},
                      rowsep  = 0.3pt,
                      rowhead = 1,
                      }
    \toprule
                PCS-ESE & Occupation Name in French &  $T_i$ (log) &  $A_i$ (log)         \\
    \midrule
                 656B      & Matelots de la marine marchande                                 &           4.22 &          -4.31 \\
                 480B      & Maîtres d'équipage de la marine marchande et de la pêche        &           4.15 &          -4.82 \\
                 389C      & Officiers et cadres navigants techniques de la marine marchande &           4.14 &          -3.33 \\
                 656C      & Capitaines et matelots timoniers de la navigation fluviale      &           3.99 &          -4.35 \\
                 441B      & Clergé régulier                                                 &           2.05 &          -2.63 \\
                 546D      & Hôtesses de l'air et stewards                                   &           1.86 &          -3.67 \\
                 441A      & Clergé séculier                                                 &           1.85 &          -2.55 
    
\end{longtblr}
\begin{longtblr}[
caption = {Occupations in the cluster ``Channel"},
  label = {table:sources}
                    ]{colsep  = 4pt,
                      colspec = {@{} X[1, j, cmd=\RaggedRight] X[8,j] 
                                     X[1, j, cmd=\RaggedRight] X[1,j] @{}},
                      rows    = {font=\small},
                      row{1}  = {font=\small\bfseries},
                      rowsep  = 0.3pt,
                      rowhead = 1,
                      }
    \toprule
                PCS-ESE & Occupation Name in French &  $T_i$ (log) &  $A_i$ (log)         \\
    \midrule
                 692A      & Marins-pêcheurs et ouvriers de l'aquaculture                                    &           1.69 &          -4.89 \\
                 465C      & Photographes                                                                    &           0.86 &          -4.05 \\
                 623D      & Opérateurs qualifiés sur machine de soudage                                     &           0.4  &          -3.75 \\
                 354A      & Artistes plasticiens                                                            &           0.26 &          -3.74 \\
                 636B      & Charcutiers (sauf industrie de la viande)                                       &           0.19 &          -3.26 \\
                 480A      & Contremaîtres et agents d'encadrement (non cadres) en agriculture, sylviculture &           0.18 &          -2.8  \\
                 431B      & Infirmiers psychiatriques                                                       &           0.06 &          -4.36 \\
                 554H      & Vendeurs de tabac, presse et articles divers                                    &          -0.08 &          -3.18 
    
\end{longtblr}
\begin{longtblr}[
caption = {Occupations in the cluster ``Condenser"},
  label = {table:sources}
                    ]{colsep  = 4pt,
                      colspec = {@{} X[1, j, cmd=\RaggedRight] X[8,j] 
                                     X[1, j, cmd=\RaggedRight] X[1,j] @{}},
                      rows    = {font=\small},
                      row{1}  = {font=\small\bfseries},
                      rowsep  = 0.3pt,
                      rowhead = 1,
                      }
    \toprule
PCS-ESE & Occupation Name in French &  $T_i$ (log) &  $A_i$ (log)         \\
  \midrule
 471B      & Techniciens d'exploitation et de contrôle de la production en agriculture, eaux et forêt                                                                               &           1.6  &          -2    \\
 471A      & Techniciens d'étude et de conseil en agriculture, eaux et forêt                                                                                                        &           1.58 &          -1.44 \\
 635A      & Tailleurs et couturières qualifiés, ouvriers qualifiés du travail des étoffes (sauf fabrication de vêtements), ouvriers qualifiés de type artisanal du travail du cuir &           1.4  &          -1.53 \\
 674E      & Ouvriers de production non qualifiés : industrie lourde du bois, fabrication des papiers et cartons                                                                    &           1.39 &          -2.04 \\
 627F      & Ouvriers de la composition et de l'impression, ouvriers qualifiés de la brochure, de la reliure et du façonnage du papier-carton                                       &           1.38 &          -1.02 \\
 675A      & Ouvriers de production non qualifiés du textile et de la confection, de la tannerie-mégisserie et du travail du cuir                                                   &           1.36 &          -1.37 \\
 627B      & Ouvriers qualifiés de la coupe des vêtements et de l'habillement, autres opérateurs de confection qualifiés                                                            &           1.3  &          -1.59 \\
 627C      & Ouvriers qualifiés du travail industriel du cuir                                                                                                                       &           1.27 &          -1.87 \\
 626C      & Opérateurs et ouvriers qualifiés des industries lourdes du bois et de la fabrication du papier-carton                                                                  &           1.12 &          -2.12 \\
 691E      & Ouvriers agricoles sans spécialisation particulière                                                                                                                    &           1.08 &          -0.65 \\
 691B      & Ouvriers de l'élevage                                                                                                                                                  &           1    &          -1.5  \\
 627A      & Opérateurs qualifiés du textile et de la mégisserie                                                                                                                    &           0.98 &          -1.86 \\
 485B      & Agents de maîtrise en fabrication des autres industries (imprimerie, matériaux souples, ameublement et bois)                                                           &           0.92 &          -1.12 \\
 675C      & Ouvriers de production non qualifiés de l'imprimerie, presse, édition                                                                                                  &           0.89 &          -1.79 \\
 476A      & Assistants techniques, techniciens de l'imprimerie et de l'édition                                                                                                     &           0.86 &          -1.57 \\
 628F      & Agents qualifiés de laboratoire (sauf chimie, santé)                                                                                                                   &           0.82 &          -0.81 \\
 691A      & Conducteurs d'engin agricole ou forestier                                                                                                                              &           0.82 &          -1.75 \\
 691D      & Ouvriers de la viticulture ou de l'arboriculture fruitière                                                                                                             &           0.81 &          -1.25 \\
 691C      & Ouvriers du maraîchage ou de l'horticulture                                                                                                                            &           0.8  &          -1.16 \\
 642A      & Conducteurs de taxi                                                                                                                                                    &           0.77 &          -1.01 \\
 354C      & Artistes dramatiques                                                                                                                                                   &           0.75 &          -0.52 \\
 333E      & Autres personnels administratifs de catégorie A de l'Etat (hors Enseignement, Patrimoine, Impôts, Trésor, Douanes)                                                     &           0.73 &           0.57 \\
 344A      & Médecins hospitaliers sans activité libérale                                                                                                                           &           0.71 &          -0.04 \\
 627D      & Ouvriers qualifiés de scierie, de la menuiserie industrielle et de l'ameublement                                                                                       &           0.7  &          -0.65 \\
 451A      & Professions intermédiaires de la Poste                                                                                                                                 &           0.68 &          -0.3  \\
 354E      & Artistes de la danse                                                                                                                                                   &           0.68 &          -1.22 \\
 432D      & Autres spécialistes de la rééducation, salariés                                                                                                                        &           0.67 &          -0.66 \\
 344C      & Internes en médecine, odontologie et pharmacie                                                                                                                         &           0.67 &          -0.63 \\
 622E      & Autres monteurs câbleurs en électronique                                                                                                                               &           0.66 &          -1.62 \\
 621C      & Conducteurs qualifiés d'engins de chantiers du bâtiment et des travaux publics                                                                                         &           0.65 &          -0.19 \\
 632G      & Peintres et ouvriers qualifiés de pose de revêtements sur supports verticaux                                                                                           &           0.65 &          -0.21 \\
 623A      & Chaudronniers-tôliers industriels, opérateurs qualifiés du travail en forge, conducteurs qualifiés d'équipement de formage, traceurs qualifiés                         &           0.63 &          -0.38 \\
 641B      & Conducteurs de véhicule routier de transport en commun                                                                                                                 &           0.61 &          -0.49 \\
 434E      & Moniteurs éducateurs                                                                                                                                                   &           0.61 &          -0.18 \\
 381B      & Ingénieurs et cadres d'étude et développement de l'agriculture, la pêche, les eaux et forêts                                                                           &           0.59 &          -2.34 \\
 354B      & Artistes de la musique et du chant &      0.58 &		  -0.8	 \\
 431A      & Cadres infirmiers et assimilés                                                                                                                                         &           0.58 &          -0.08 \\
 434D      & Educateurs spécialisés                                                                                                                                                 &           0.57 &           0.2  \\
 651B      & Conducteurs d'engin lourd de manoeuvre                                                                                                                                 &           0.57 &          -0.9  \\
 554F      & Vendeurs en produits de beauté, de luxe (hors biens culturels) et optique                                                                                              &           0.56 &           0.14 \\
 472A      & Dessinateurs en bâtiment, travaux publics                                                                                                                              &           0.55 &          -0.19 \\
 623B      & Tuyauteurs industriels qualifiés                                                                                                                                       &           0.55 &          -1.24 \\
 632A      & Maçons qualifiés                                                                                                                                                       &           0.54 &           0.32 \\
 624C      & Monteurs qualifiés d'ensembles mécaniques travaillant en moyenne ou en grande série                                                                                    &           0.54 &          -0.29 \\
 628C      & Régleurs qualifiés d'équipements de fabrication (travail des métaux, mécanique)                                                                                        &           0.54 &          -0.83 \\
 354F      & Artistes du cirque et des spectacles divers                                                                                                                            &           0.54 &          -1.14 \\
 622C      & Monteurs câbleurs qualifiés en électricité                                                                                                                             &           0.54 &          -0.29 \\
 432B      & Masseurs-kinésithérapeutes rééducateurs, salariés                                                                                                                      &           0.53 &          -1.28 \\
 626B      & Autres opérateurs et ouvriers qualifiés : métallurgie, production verrière, matériaux de construction                                                                  &           0.53 &          -0.08 \\
 643A      & Conducteurs livreurs, coursiers                                                                                                                                        &           0.52 &           0.51 \\
 634C      & Mécaniciens qualifiés en maintenance, entretien, réparation : automobile                                                                                               &           0.51 &          -0.01 \\
 633B      & Dépanneurs qualifiés en radiotélévision, électroménager, matériel électronique (salariés)                                                                              &           0.5  &          -1.8  \\
 465B      & Assistants techniques de la réalisation des spectacles vivants et audiovisuels salariés                                                                                &           0.49 &          -0.08 \\
 425A      & Sous-bibliothécaires, cadres intermédiaires du patrimoine                                                                                                              &           0.49 &          -0.69 \\
 386E      & Ingénieurs et cadres de fabrication des autres industries (imprimerie, matériaux souples, ameublement et bois)                                                         &           0.49 &          -0.72 \\
 684B      & Ouvriers non qualifiés de l'assainissement et du traitement des déchets                                                                                                &           0.49 &          -0.61 \\
 451E      & Autres personnels administratifs de catégorie B de l'Etat (hors Enseignement, Patrimoine, Impôts, Trésor, Douanes)                                                     &           0.49 &          -0.08 \\
 546B      & Agents des services commerciaux des transports de voyageurs et du tourisme                                                                                             &           0.48 &          -0.05 \\
 628D      & Régleurs qualifiés d'équipements de fabrication (hors travail des métaux et mécanique)                                                                                 &           0.48 &          -1.19 \\
 621G      & Mineurs de fond qualifiés et autres ouvriers qualifiés des industries d'extraction (carrières, pétrole, gaz...)                                                        &           0.48 &          -1.61 \\
 673C      & Ouvriers non qualifiés de montage, contrôle en mécanique et travail des métaux                                                                                         &           0.48 &          -0.08 \\
 672A      & Ouvriers non qualifiés de l'électricité et de l'électronique                                                                                                           &           0.48 &          -0.63 \\
 642B      & Conducteurs de voiture particulière                                                                                                                                    &           0.47 &          -1.03 \\
 353A      & Directeurs de journaux, administrateurs de presse, directeurs d'éditions (littéraire, musicale, audiovisuelle et multimédia)                                           &           0.47 &          -1.03 \\
 523B      & Adjoints administratifs de l'Etat et assimilés (sauf Poste, France Télécom)                                                                                            &           0.47 &           0.05 \\
 633A      & Electriciens qualifiés de type artisanal (y.c. bâtiment)                                                                                                               &           0.46 &          -0.04 \\
 628A      & Mécaniciens qualifiés de maintenance, entretien : équipements industriels                                                                                              &           0.45 &          -0.08 \\
 633D      & Electriciens, électroniciens qualifiés en maintenance, entretien : équipements non industriels                                                                         &           0.45 &          -1.16 \\
 623G      & Opérateurs qualifiés d'usinage des métaux sur autres machines (sauf moulistes)                                                                                         &           0.45 &          -0.37 \\
 311C      & Chirurgiens dentistes                                                                                                                                                  &           0.44 &          -1.85 \\
 353C      & Cadres artistiques et technico-artistiques de la réalisation de l'audiovisuel et des spectacles                                                                        &           0.44 &          -0.24 \\
 625D      & Opérateurs de la transformation des viandes                                                                                                                            &           0.43 &          -0.23 \\
 628E      & Ouvriers qualifiés de l'assainissement et du traitement des déchets                                                                                                    &           0.43 &          -0.9  \\
 632C      & Charpentiers en bois qualifiés                                                                                                                                         &           0.43 &          -0.91 \\
 476B      & Techniciens de l'industrie des matériaux souples, de l'ameublement et du bois                                                                                          &           0.42 &          -1.84 \\
 682A      & Métalliers, serruriers, réparateurs en mécanique non qualifiés                                                                                                         &           0.42 &          -0.63 \\
 673A      & Ouvriers de production non qualifiés travaillant par enlèvement de métal                                                                                               &           0.4  &          -0.66 \\
 654B      & Conducteurs qualifiés d'engins de transport guidés (sauf remontées mécaniques)                                                                                         &           0.4  &          -1.24 \\
 526E      & Ambulanciers salariés                                                                                                                                                  &           0.4  &          -1.07 \\
 634D      & Mécaniciens qualifiés de maintenance, entretien : équipements non industriels                                                                                          &           0.4  &          -0.8  \\
 422A      & Professeurs d'enseignement général des collèges                                                                                                                        &           0.4  &          -0.26 \\
 634A      & Carrossiers d'automobiles qualifiés                                                                                                                                    &           0.4  &          -1.01 \\
 625C      & Autres opérateurs et ouvriers qualifiés de la chimie (y.c. pharmacie) et de la plasturgie                                                                              &           0.39 &          -0.16 \\
 624D      & Monteurs qualifiés en structures métalliques                                                                                                                           &           0.39 &          -0.83 \\
 632D      & Menuisiers qualifiés du bâtiment                                                                                                                                       &           0.39 &          -0.35 \\
 624G      & Autres mécaniciens ou ajusteurs qualifiés (ou spécialité non reconnue)                                                                                                 &           0.39 &          -1.35 \\
 466C      & Responsables d'exploitation des transports de voyageurs et de marchandises (non cadres)                                                                                &           0.38 &           0.06 \\
 342B      & Professeurs et maîtres de conférences                                                                                                                                  &           0.38 &          -0.2  \\
 342D      & Personnel enseignant temporaire de l'enseignement supérieur                                                                                                            &           0.38 &          -0.3  \\
 623F      & Opérateurs qualifiés d'usinage des métaux travaillant à l'unité ou en petite série, moulistes qualifiés                                                                &           0.38 &          -0.43 \\
 422E      & Surveillants et aides-éducateurs des établissements d'enseignement                                                                                                     &           0.37 &           0.34 \\
 477D      & Techniciens de l'environnement et du traitement des pollutions                                                                                                         &           0.36 &          -0.79 \\
 353B      & Directeurs, responsables de programmation et de production de l'audiovisuel et des spectacles                                                                          &           0.36 &          -0.92 \\
 434B      & Assistants de service social                                                                                                                                           &           0.36 &          -0.48 \\
 624E      & Ouvriers qualifiés de contrôle et d'essais en mécanique                                                                                                                &           0.36 &          -1.15 \\
 644A      & Conducteurs de véhicule de ramassage des ordures ménagères                                                                                                             &           0.35 &          -1.02 \\
 634B      & Métalliers, serruriers qualifiés                                                                                                                                       &           0.35 &          -0.74 \\
 625A      & Pilotes d'installation lourde des industries de transformation : agroalimentaire, chimie, plasturgie, énergie                                                          &           0.35 &          -0.75 \\
 625F      & Autres opérateurs travaillant sur installations ou machines : industrie agroalimentaire (hors transformation des viandes)                                              &           0.35 &          -0.29 \\
 354G      & Professeurs d'art (hors établissements scolaires)                                                                                                                      &           0.35 &          -1.02 \\
 624B      & Monteurs, metteurs au point très qualifiés d'ensembles mécaniques travaillant à l'unité ou en petite série                                                             &           0.34 &          -0.95 \\
 351A      & Bibliothécaires, archivistes, conservateurs et autres cadres du patrimoine                                                                                             &           0.34 &          -0.82 \\
 433A      & Techniciens médicaux                                                                                                                                                   &           0.34 &          -0.31 \\
 474A      & Dessinateurs en construction mécanique et travail des métaux                                                                                                           &           0.34 &          -0.69 \\
 641A      & Conducteurs routiers et grands routiers                                                                                                                                &           0.34 &           0.33 \\
 485A      & Agents de maîtrise et techniciens en production et distribution d'énergie, eau, chauffage                                                                              &           0.33 &          -0.06 \\
 674C      & Autres ouvriers de production non qualifiés : industrie agroalimentaire                                                                                                &           0.32 &           0.46 \\
 624F      & Ouvriers qualifiés des traitements thermiques et de surface sur métaux                                                                                                 &           0.32 &          -1.23 \\
 632J      & Monteurs qualifiés en agencement, isolation                                                                                                                            &           0.32 &          -0.24 \\
 431C      & Puéricultrices                                                                                                                                                         &           0.32 &          -1.26 \\
 675B      & Ouvriers de production non qualifiés du travail du bois et de l'ameublement                                                                                            &           0.31 &          -1.19 \\
 341A      & Professeurs agrégés et certifiés de l'enseignement secondaire                                                                                                          &           0.3  &           0.22 \\
 621E      & Autres ouvriers qualifiés des travaux publics                                                                                                                          &           0.3  &          -0.27 \\
 632E      & Couvreurs qualifiés                                                                                                                                                    &           0.29 &          -0.89 \\
 343A      & Psychologues spécialistes de l'orientation scolaire et professionnelle                                                                                                 &           0.29 &          -0.45 \\
 474C      & Techniciens de fabrication et de contrôle-qualité en construction mécanique et travail des métaux                                                                      &           0.28 &           0.21 \\
 673B      & Ouvriers de production non qualifiés travaillant par formage de métal                                                                                                  &           0.28 &          -1.64 \\
 546A      & Contrôleurs des transports (personnels roulants)                                                                                                                       &           0.28 &          -1.23 \\
 631A      & Jardiniers                                                                                                                                                             &           0.27 &          -0.72 \\
 486D      & Agents de maîtrise en maintenance, installation en mécanique                                                                                                           &           0.27 &          -0.83 \\
 344D      & Pharmaciens salariés                                                                                                                                                   &           0.26 &          -0.53 \\
 342G      & Ingénieurs d'étude et de recherche de la recherche publique                                                                                                            &           0.26 &          -0.13 \\
 484B      & Agents de maîtrise en fabrication : métallurgie, matériaux lourds et autres industries de transformation                                                               &           0.26 &          -1.11 \\
 628B      & Electromécaniciens, électriciens qualifiés d'entretien : équipements industriels                                                                                       &           0.25 &          -0.5  \\
 473A      & Dessinateurs en électricité, électromécanique et électronique                                                                                                          &           0.24 &          -1.18 \\
 681A      & Ouvriers non qualifiés du gros oeuvre du bâtiment                                                                                                                      &           0.24 &           0.13 \\
 526B      & Assistants dentaires, médicaux et vétérinaires, aides de techniciens médicaux                                                                                          &           0.24 &          -0.21 \\
 475B      & Techniciens de production et de contrôle-qualité des industries de transformation                                                                                      &           0.24 &           0.06 \\
 232A      & Chefs de moyenne entreprise, de 50 à 499 salariés                                                                                                                      &           0.24 &           0.32 \\
 475A      & Techniciens de recherche-développement et des méthodes de production des industries de transformation                                                                  &           0.23 &          -0.23 \\
 352A      & Journalistes (y c. rédacteurs en chef)                                                                                                                                 &           0.23 &          -0.78 \\
 472B      & Géomètres, topographes                                                                                                                                                 &           0.23 &          -1.23 \\
 481B      & Chefs de chantier (non cadres)                                                                                                                                         &           0.23 &          -0.01 \\
 423B      & Formateurs et animateurs de formation continue                                                                                                                         &           0.23 &           0.4  \\
 473B      & Techniciens de recherche-développement et des méthodes de fabrication en électricité, électromécanique et électronique                                                 &           0.23 &          -0.48 \\
 632F      & Plombiers et chauffagistes qualifiés                                                                                                                                   &           0.22 &          -0.29 \\
 486B      & Agents de maîtrise en maintenance, installation en électricité et électronique                                                                                         &           0.21 &          -1    \\
 524C      & Agents administratifs des collectivités locales                                                                                                                        &           0.2  &           0.74 \\
 674A      & Ouvriers de production non qualifiés : chimie, pharmacie, plasturgie                                                                                                   &           0.2  &          -0.33 \\
 451C      & Contrôleurs des Impôts, du Trésor, des Douanes et assimilés*                                                                                                           &           0.2  &          -1.22 \\
 434A      & Cadres de l'intervention socio-éducative                                                                                                                               &           0.19 &          -0.05 \\
 433D      & Préparateurs en pharmacie                                                                                                                                              &           0.19 &          -0.78 \\
 311D      & Psychologues, psychanalystes, psychothérapeutes (non médecins)                                                                                                         &           0.19 &          -0.45 \\
 637C      & Ouvriers et techniciens des spectacles vivants et audiovisuels                                                                                                         &           0.19 &          -0.62 \\
 621D      & Ouvriers des travaux publics en installations électriques et de télécommunications                                                                                     &           0.19 &          -0.85 \\
 477B      & Techniciens d'installation et de maintenance des équipements industriels (électriques, électromécaniques, mécaniques, hors informatique)                               &           0.18 &           0.38 \\
 479B      & Experts salariés de niveau technicien, techniciens divers                                                                                                              &           0.18 &           0.2  \\
 422C      & Maîtres auxiliaires et professeurs contractuels de l'enseignement secondaire                                                                                           &           0.18 &          -0.29 \\
 466B      & Responsables commerciaux et administratifs des transports de marchandises (non cadres)                                                                                 &           0.18 &          -0.92 \\
 562B      & Coiffeurs salariés                                                                                                                                                     &           0.17 &          -0.37 \\
 626A      & Pilotes d'installation lourde des industries de transformation : métallurgie, production verrière, matériaux de construction                                           &           0.17 &          -1.02 \\
 632H      & Soliers moquetteurs et ouvriers qualifiés de pose de revêtements souples sur supports horizontaux                                                                      &           0.16 &          -2.02 \\
 389A      & Ingénieurs et cadres techniques de l'exploitation des transports                                                                                                       &           0.16 &           0.02 \\
 625H      & Ouvriers qualifiés des autres industries (eau, gaz, énergie, chauffage)                                                                                                &           0.16 &          -0.86 \\
 484A      & Agents de maîtrise en fabrication : agroalimentaire, chimie, plasturgie, pharmacie.                                                                                    &           0.16 &          -0.41 \\
 473C      & Techniciens de fabrication et de contrôle-qualité en électricité, électromécanique et électronique                                                                     &           0.16 &          -0.33 \\
 685A      & Ouvriers non qualifiés divers de type artisanal                                                                                                                        &           0.16 &          -0.66 \\
 372F      & Cadres de la documentation, de l'archivage (hors fonction publique)                                                                                                    &           0.15 &          -1.21 \\
 621A      & Chefs d'équipe du gros oeuvre et des travaux publics                                                                                                                   &           0.15 &          -0.36 \\
 681B      & Ouvriers non qualifiés du second oeuvre du bâtiment                                                                                                                    &           0.15 &           0.03 \\
 434G      & Educateurs de jeunes enfants                                                                                                                                           &           0.13 &          -1.13 \\
 561E      & Employés de l'hôtellerie : réception et hall                                                                                                                           &           0.12 &          -0.12 \\
 342C      & Professeurs agrégés et certifiés en fonction dans l'enseignement supérieur                                                                                             &           0.12 &          -1.32 \\
 637D      & Ouvriers qualifiés divers de type artisanal                                                                                                                            &           0.11 &          -1.18 \\
 541B      & Agents d'accueil qualifiés, hôtesses d'accueil et d'information                                                                                                        &           0.11 &           0.24 \\
 622A      & Opérateurs qualifiés sur machines automatiques en production électrique ou électronique                                                                                &           0.11 &          -1.1  \\
 655A      & Autres agents et ouvriers qualifiés (sédentaires) des services d'exploitation des transports                                                                           &           0.11 &          -0.3  \\
 674D      & Ouvriers de production non qualifiés : métallurgie, production verrière, céramique, matériaux de construction                                                          &           0.11 &          -0.5  \\
 676D      & Agents non qualifiés des services d'exploitation des transports                                                                                                        &           0.11 &          -0.59 \\
 651A      & Conducteurs d'engin lourd de levage                                                                                                                                    &           0.1  &          -1.07 \\
 474B      & Techniciens de recherche-développement et des méthodes de fabrication en construction mécanique et travail des métaux                                                  &           0.1  &          -0.5  \\
 625G      & Autres ouvriers de production qualifiés ne travaillant pas sur machine : industrie agroalimentaire (hors transformation des viandes)                                   &           0.09 &          -1.2  \\
 676E      & Ouvriers non qualifiés divers de type industriel                                                                                                                       &           0.09 &           0.46 \\
 421B      & Professeurs des écoles                                                                                                                                                 &           0.08 &          -0.63 \\
 477C      & Techniciens d'installation et de maintenance des équipements non industriels (hors informatique et télécommunications)                                                 &           0.08 &          -0.07 \\
 562A      & Manucures, esthéticiens                                                                                                                                                &           0.08 &          -0.7  \\
 435B      & Animateurs socioculturels et de loisirs                                                                                                                                &           0.08 &           0.34 \\
 546C      & Employés administratifs d'exploitation des transports de marchandises                                                                                                  &           0.06 &          -0.48 \\
 472C      & Métreurs et techniciens divers du bâtiment et des travaux publics                                                                                                      &           0.06 &           0.13 \\
 384A      & Ingénieurs et cadres d'étude, recherche et développement en mécanique et travail des métaux                                                                            &           0.06 &           0.1  \\
 344B      & Médecins salariés non hospitaliers                                                                                                                                     &           0.06 &          -0.61 \\
 433C      & Autres spécialistes de l'appareillage médical salariés                                                                                                                 &           0.06 &          -1.31 \\
 524B      & Agents administratifs de l'Etat et assimilés (sauf Poste, France Télécom)                                                                                              &           0.05 &           0.36 \\
 468A      & Maîtrise de restauration : salle et service                                                                                                                            &           0.05 &          -0.08 \\
 451H      & Professions intermédiaires administratives des hôpitaux                                                                                                                &           0.05 &           0.03 \\
 478D      & Techniciens des télécommunications et de l'informatique des réseaux                                                                                                    &           0.04 &           0.15 \\
 382A      & Ingénieurs et cadres d'étude du bâtiment et des travaux publics                                                                                                        &           0.04 &           0.18 \\
 382B      & Architectes salariés                                                                                                                                                   &           0.04 &          -0.83 \\
 220X      & Commerçants et assimilés, salariés de leur entreprise                                                                                                                  &           0.03 &           0.31 \\
 621F      & Ouvriers qualifiés des travaux publics (salariés de l'Etat et des collectivités locales)                                                                               &           0.03 &           0.68 \\
 382C      & Ingénieurs, cadres de chantier et conducteurs de travaux (cadres) du bâtiment et des travaux publics                                                                   &           0.03 &           0.2  \\
 233D      & Chefs d'entreprise de services, de 10 à 49 salariés                                                                                                                    &           0.03 &           0.12 \\
 384B      & Ingénieurs et cadres de fabrication en mécanique et travail des métaux                                                                                                 &           0.02 &           0.26 \\
 671C      & Ouvriers non qualifiés des travaux publics et du travail du béton                                                                                                      &           0.02 &          -0.3  \\
 628G      & Ouvriers qualifiés divers de type industriel                                                                                                                           &           0.02 &           0.07 \\
 434F      & Educateurs techniques spécialisés, moniteurs d'atelier                                                                                                                 &           0.02 &          -1.47 \\
 487A      & Responsables d'entrepôt, de magasinage                                                                                                                                 &           0.02 &          -0.25 \\
 386B      & Ingénieurs et cadres d'étude, recherche et développement de la distribution d'énergie, eau                                                                             &           0.02 &          -0.19 \\
 486E      & Agents de maîtrise en entretien général, installation, travaux neufs (hors mécanique, électromécanique, électronique)                                                  &           0.01 &          -1.25 \\
 431D      & Infirmiers spécialisés (autres qu'infirmiers psychiatriques et puéricultrices)                                                                                         &           0.01 &          -1.25 \\
 374A      & Cadres de l'exploitation des magasins de vente du commerce de détail                                                                                                   &           0.01 &           0.46 \\
 674B      & Ouvriers de production non qualifiés de la transformation des viandes                                                                                                  &           0.01 &          -0.7  \\
 621B      & Ouvriers qualifiés du travail du béton                                                                                                                                 &          0.004 &          -0.5  \\
 488A      & Maîtrise de restauration  : cuisine/production                                                                                                                         &          0.002 &          -0.17 \\
 210X      & Artisans salariés de leur entreprise                                                                                                                                   &          0.002 &          -0.09 \\
 422B      & Professeurs de lycée professionnel                                                                                                                                     &         -0.002 &          -0.48 \\
 377A      & Cadres de l'hôtellerie et de la restauration                                                                                                                           &         -0.005 &          -0.004    \\
 383A      & Ingénieurs et cadres d'étude, recherche et développement en électricité, électronique                                                                                  &          -0.01 &           0.21 \\
 383C      & Ingénieurs et cadres technico-commerciaux en matériel électrique ou électronique professionnel                                                                         &          -0.01 &          -0.63 \\
 477A      & Techniciens de la logistique, du planning et de l'ordonnancement                                                                                                       &          -0.01 &           0.01 \\
 468B      & Maîtrise de l'hébergement : hall et étages                                                                                                                             &          -0.01 &          -0.92 \\
 488B      & Maîtrise de restauration  : gestion d'établissement                                                                                                                    &          -0.01 &          -0.61 \\
 479A      & Techniciens des laboratoires de recherche publique ou de l'enseignement                                                                                                &          -0.01 &          -0.73 \\
 625B      & Ouvriers qualifiés et agents qualifiés de laboratoire : agroalimentaire, chimie, biologie, pharmacie                                                                   &          -0.01 &          -0.53 \\
 481A      & Conducteurs de travaux (non cadres)                                                                                                                                    &          -0.02 &          -0.28 \\
 546E      & Autres agents et hôtesses d'accompagnement (transports, tourisme)                                                                                                      &          -0.02 &          -1.05 \\
 467D      & Professions intermédiaires techniques des organismes de sécurité sociale                                                                                               &          -0.02 &          -0.25 \\
 434C      & Conseillers en économie sociale familiale                                                                                                                              &          -0.03 &          -1.2  \\
 387F      & Ingénieurs et cadres techniques de l'environnement                                                                                                                     &          -0.03 &          -0.72 \\
 462B      & Maîtrise de l'exploitation des magasins de vente                                                                                                                       &          -0.03 &           0.26 \\
 385B      & Ingénieurs et cadres de fabrication des industries de transformation (agroalimentaire, chimie, métallurgie, matériaux lourds)                                          &          -0.03 &          -0.07 \\
 387E      & Ingénieurs et cadres de la maintenance, de l'entretien et des travaux neufs                                                                                            &          -0.03 &           0    \\
 541C      & Agents d'accueil non qualifiés                                                                                                                                         &          -0.04 &          -0.78 \\
 383B      & Ingénieurs et cadres de fabrication en matériel électrique, électronique                                                                                               &          -0.04 &          -0.9  \\
 233A      & Chefs d'entreprise du bâtiment et des travaux publics, de 10 à 49 salariés                                                                                             &          -0.04 &          -0.49 \\
 545D      & Employés des services techniques des organismes de sécurité sociale et assimilés                                                                                       &          -0.05 &           0.02 \\
 332B      & Ingénieurs des collectivités locales et des hôpitaux                                                                                                                   &          -0.06 &          -0.07 \\
 386D      & Ingénieurs et cadres de la production et de la distribution d'énergie, eau                                                                                             &          -0.06 &          -0.57 \\
 385A      & Ingénieurs et cadres d'étude, recherche et développement des industries de transformation (agroalimentaire, chimie, métallurgie, matériaux lourds)                     &          -0.07 &           0.16 \\
 553C      & Autres vendeurs non spécialisés                                                                                                                                        &          -0.07 &           0.24 \\
 342F      & Directeurs et chargés de recherche de la recherche publique                                                                                                            &          -0.07 &          -1.05 \\
 554B      & Vendeurs en ameublement, décor, équipement du foyer                                                                                                                    &          -0.07 &          -0.04 \\
 463E      & Techniciens commerciaux et technico-commerciaux, représentants auprès de particuliers (hors banque, assurance, informatique)                                           &          -0.07 &           0.22 \\
 375A      & Cadres de la publicité                                                                                                                                                 &          -0.08 &          -0.29 \\
 472D      & Techniciens des travaux publics de l'Etat et des collectivités locales                                                                                                 &          -0.08 &          -0.42 \\
 376B      & Cadres des opérations bancaires                                                                                                                                        &          -0.08 &           0.04 \\
 487B      & Responsables du tri, de l'emballage, de l'expédition et autres responsables de la manutention                                                                          &          -0.08 &          -0.99 \\
 525D      & Agents de service hospitaliers                                                                                                                                         &          -0.08 &           0.5  \\
 466A      & Responsables commerciaux et administratifs des transports de voyageurs et du tourisme (non cadres)                                                                     &          -0.08 &          -0.39 \\
 387D      & Ingénieurs et cadres du contrôle-qualité                                                                                                                               &          -0.09 &          -0    \\
 385C      & Ingénieurs et cadres technico-commerciaux des industries de transformations (biens intermédiaires)                                                                     &          -0.09 &          -0.13 \\
 386C      & Ingénieurs et cadres d'étude, recherche et développement des autres industries (imprimerie, matériaux souples, ameublement et bois)                                    &          -0.09 &          -0.87 \\
 526A      & Aides-soignants                                                                                                                                                        &          -0.09 &           0.11 \\
 433B      & Opticiens lunetiers et audioprothésistes salariés                                                                                                                      &          -0.09 &          -1.29 \\
 233C      & Chefs d'entreprise commerciale, de 10 à 49 salariés                                                                                                                    &          -0.09 &          -0.36 \\
 464A      & Assistants de la publicité, des relations publiques                                                                                                                    &          -0.09 &          -0.64 \\
 467A      & Chargés de clientèle bancaire                                                                                                                                          &          -0.1  &           0.43 \\
 382D      & Ingénieurs et cadres technico-commerciaux en bâtiment, travaux publics                                                                                                 &          -0.11 &          -0.27 \\
 387A      & Ingénieurs et cadres des achats et approvisionnements industriels                                                                                                      &          -0.11 &          -0.19 \\
 422D      & Conseillers principaux d'éducation                                                                                                                                     &          -0.12 &          -1.84 \\
 376A      & Cadres des marchés financiers                                                                                                                                          &          -0.12 &          -0.53 \\
 525C      & Agents de service de la fonction publique (sauf écoles, hôpitaux)                                                                                                      &          -0.13 &          -1.68 \\
 545C      & Employés des services techniques des assurances                                                                                                                        &          -0.13 &           0.29 \\
 684A      & Nettoyeurs                                                                                                                                                             &          -0.14 &           0.67 \\
 233B      & Chefs d'entreprise de l'industrie ou des transports, de 10 à 49 salariés                                                                                               &          -0.14 &          -0.62 \\
 435A      & Directeurs de centres socioculturels et de loisirs                                                                                                                     &          -0.14 &          -1.13 \\
 384C      & Ingénieurs et cadres technico-commerciaux en matériel mécanique professionnel                                                                                          &          -0.14 &          -0.66 \\
 521A      & Employés de la Poste                                                                                                                                                   &          -0.14 &          -1.42 \\
 478C      & Techniciens d'installation, de maintenance, support et services aux utilisateurs en informatique                                                                       &          -0.14 &           0.07 \\
 376C      & Cadres commerciaux de la banque                                                                                                                                        &          -0.14 &           0.24 \\
 333F      & Personnels administratifs de catégorie A des collectivités locales et hôpitaux publics (hors Enseignement, Patrimoine)                                                 &          -0.14 &           0.31 \\
 333C      & Cadres de la Poste*                                                                                                                                                    &          -0.14 &          -0.88 \\
 541D      & Standardistes, téléphonistes                                                                                                                                           &          -0.15 &          -0.35 \\
 332A      & Ingénieurs de l'Etat (y.c. ingénieurs militaires) et assimilés                                                                                                         &          -0.15 &          -0.64 \\
 372E      & Juristes                                                                                                                                                               &          -0.15 &           0.03 \\
 463C      & Techniciens commerciaux et technico-commerciaux, représentants en biens de consommation auprès d'entreprises                                                           &          -0.16 &          -0.11 \\
 561F      & Employés d'étage et employés polyvalents de l'hôtellerie                                                                                                               &          -0.16 &          -0.14 \\
 636D      & Cuisiniers et commis de cuisine                                                                                                                                        &          -0.16 &           0.33 \\
 387C      & Ingénieurs et cadres des méthodes de production                                                                                                                        &          -0.17 &          -0.09 \\
 556A      & Vendeurs en gros de biens d'équipement, biens intermédiaires                                                                                                           &          -0.17 &           0.04 \\
 632K      & Ouvriers qualifiés d'entretien général des bâtiments                                                                                                                   &          -0.17 &          -0.17 \\
 333A      & Magistrats*                                                                                                                                                            &          -0.18 &          -1.34 \\
 451G      & Professions intermédiaires administratives des collectivités locales                                                                                                   &          -0.18 &          -0.04 \\
 467B      & Techniciens des opérations bancaires                                                                                                                                   &          -0.18 &          -0.09 \\
 563B      & Aides à domicile, aides ménagères, travailleuses familiales                                                                                                            &          -0.19 &           0.55 \\
 555A      & Vendeurs par correspondance, télévendeurs                                                                                                                              &          -0.19 &          -0.49 \\
 523D      & Adjoints administratifs des hôpitaux publics                                                                                                                           &          -0.19 &          -0.68 \\
 526D      & Aides médico-psychologiques                                                                                                                                            &          -0.2  &          -0.25 \\
 231A      & Chefs de grande entreprise de 500 salariés et plus                                                                                                                     &          -0.21 &          -1.25 \\
 545A      & Employés administratifs des services techniques de la banque                                                                                                           &          -0.21 &          -0.18 \\
 333B      & Inspecteurs et autres personnels de catégorie A des Impôts, du Trésor et des Douanes                                                                                   &          -0.21 &          -0.24 \\
 463D      & Techniciens commerciaux et technico-commerciaux, représentants en services auprès d'entreprises ou de professionnels (hors banque, assurance, informatique)            &          -0.22 &          -0.03 \\
 387B      & Ingénieurs et cadres de la logistique, du planning et de l'ordonnancement                                                                                              &          -0.23 &          -0    \\
 374D      & Cadres commerciaux des petites et moyennes entreprises (hors commerce de détail)                                                                                       &          -0.24 &           0.94 \\
 462A      & Chefs de petites surfaces de vente                                                                                                                                     &          -0.24 &          -0.4  \\
 463B      & Techniciens commerciaux et technico-commerciaux, représentants en biens d'équipement, en biens intermédiaires, commerce interindustriel (hors informatique)            &          -0.24 &          -0.03 \\
 424A      & Moniteurs et éducateurs sportifs, sportifs professionnels                                                                                                              &          -0.25 &          -0.5  \\
 375B      & Cadres des relations publiques et de la communication                                                                                                                  &          -0.25 &          -0.07 \\
 554A      & Vendeurs en alimentation                                                                                                                                               &          -0.25 &           0.31 \\
 478A      & Techniciens d'étude et de développement en informatique                                                                                                                &          -0.25 &           0.26 \\
 554E      & Vendeurs en habillement et articles de sport                                                                                                                           &          -0.25 &           0.29 \\
 563A      & Assistantes maternelles, gardiennes d'enfants, familles d'accueil                                                                                                      &          -0.26 &          -0.15 \\
 462C      & Acheteurs non classés cadres, aides-acheteurs                                                                                                                          &          -0.26 &          -0.4  \\
 462D      & Animateurs commerciaux des magasins de vente, marchandiseurs (non cadres)                                                                                              &          -0.26 &          -0.7  \\
 525A      & Agents de service des établissements primaires                                                                                                                         &          -0.26 &          -0.48 \\
 478B      & Techniciens de production, d'exploitation en informatique                                                                                                              &          -0.26 &          -0.23 \\
 461E      & Maîtrise et techniciens administratifs des services juridiques ou du personnel                                                                                         &          -0.29 &           0.34 \\
 467C      & Professions intermédiaires techniques et commerciales des assurances                                                                                                   &          -0.29 &          -0.01 \\
 545B      & Employés des services commerciaux de la banque                                                                                                                         &          -0.29 &           0.01 \\
 374B      & Chefs de produits, acheteurs du commerce et autres cadres de la mercatique                                                                                             &          -0.29 &           0.33 \\
 376D      & Chefs d'établissements et responsables de l'exploitation bancaire                                                                                                      &          -0.29 &          -0.02 \\
 542A      & Secrétaires                                                                                                                                                            &          -0.29 &           1.03 \\
 652A      & Ouvriers qualifiés de la manutention, conducteurs de chariots élévateurs, caristes                                                                                     &          -0.29 &           0.43 \\
 676C      & Ouvriers du tri, de l'emballage, de l'expédition, non qualifiés                                                                                                        &          -0.3  &           0.73 \\
 554C      & Vendeurs en droguerie, bazar, quincaillerie, bricolage                                                                                                                 &          -0.3  &          -0.41 \\
 376E      & Cadres des services techniques des assurances                                                                                                                          &          -0.3  &           0.12 \\
 543B      & Employés qualifiés des services comptables ou financiers                                                                                                               &          -0.3  &           0.87 \\
 551A      & Employés de libre service du commerce et magasiniers                                                                                                                   &          -0.31 &           0.75 \\
 376F      & Cadres des services techniques des organismes de sécurité sociale et assimilés                                                                                         &          -0.32 &          -0.48 \\
 561D      & Aides de cuisine, apprentis de cuisine et employés polyvalents de la restauration                                                                                      &          -0.32 &           0.57 \\
 373B      & Cadres des autres services administratifs des grandes entreprises                                                                                                      &          -0.32 &           0.73 \\
 561B      & Serveurs, commis de restaurant, garçons qualifiés                                                                                                                      &          -0.32 &           0.31 \\
 561C      & Serveurs, commis de restaurant, garçons non qualifiés                                                                                                                  &          -0.33 &           0.49 \\
 461F      & Maîtrise et techniciens administratifs des autres services administratifs                                                                                              &          -0.35 &           0.65 \\
 374C      & Cadres commerciaux des grandes entreprises (hors commerce de détail)                                                                                                   &          -0.35 &           0.62 \\
 544A      & Employés et opérateurs d'exploitation en informatique                                                                                                                  &          -0.35 &          -0.23 \\
 543F      & Employés qualifiés des services commerciaux des entreprises (hors vente)                                                                                               &          -0.36 &           0.74 \\
 373D      & Cadres des autres services administratifs des petites et moyennes entreprises                                                                                          &          -0.38 &           0.91 \\
 331A      & Personnels de direction de la fonction publique (Etat, collectivités locales, hôpitaux)                                                                                &          -0.38 &           0.04 \\
 676A      & Manutentionnaires non qualifiés                                                                                                                                        &          -0.39 &           0.5  \\
 563C      & Employés de maison et personnels de ménage chez des particuliers                                                                                                       &          -0.4  &           0.03 \\
 653A      & Magasiniers qualifiés                                                                                                                                                  &          -0.4  &           0.43 \\
 462E      & Autres professions intermédiaires commerciales (sauf techniciens des forces de vente)                                                                                  &          -0.4  &           0.58 \\
 636C      & Boulangers, pâtissiers (sauf activité industrielle)                                                                                                                    &          -0.41 &          -0.55 \\
 552A      & Caissiers de magasin                                                                                                                                                   &          -0.42 &           0.15 \\
 372A      & Cadres chargés d'études économiques, financières, commerciales                                                                                                         &          -0.42 &           0.69 \\
 372C      & Cadres spécialistes des ressources humaines et du recrutement                                                                                                          &          -0.43 &           0.39 \\
 525B      & Agents de service des autres établissements d'enseignement                                                                                                             &          -0.44 &          -0.22 \\
 543H      & Employés administratifs non qualifiés                                                                                                                                  &          -0.45 &           0.73 \\
 543E      & Employés qualifiés des services du personnel et des services juridiques                                                                                                &          -0.46 &           0.03 \\
 372B      & Cadres de l'organisation ou du contrôle des services administratifs et financiers                                                                                      &          -0.46 &           0.83 \\
 372D      & Cadres spécialistes de la formation                                                                                                                                    &          -0.47 &          -0.26 \\
 388A      & Ingénieurs et cadres d'étude, recherche et développement en informatique                                                                                               &          -0.47 &           1.06 \\
 376G      & Cadres de l'immobilier                                                                                                                                                 &          -0.47 &          -0.22 \\
 534A      & Agents civils de sécurité et de surveillance                                                                                                                           &          -0.48 &          -0.36 \\
 388E      & Ingénieurs et cadres spécialistes des télécommunications                                                                                                               &          -0.49 &          -0.11 \\
 461D      & Maîtrise et techniciens des services financiers ou comptables                                                                                                          &          -0.51 &           0.44 \\
 523C      & Adjoints administratifs des collectivités locales                                                                                                                      &          -0.52 &           0.01 \\
 373C      & Cadres des services financiers ou comptables des petites et moyennes entreprises                                                                                       &          -0.53 &           0.64 \\
 388B      & Ingénieurs et cadres d'administration, maintenance, support et services aux utilisateurs en informatique                                                               &          -0.53 &           0.33 \\
 388C      & Chefs de projets informatiques, responsables informatiques                                                                                                             &          -0.54 &           0.84 \\
 543G      & Employés administratifs qualifiés des autres services des entreprises                                                                                                  &          -0.54 &           1.04 \\
 564B      & Employés des services divers                                                                                                                                           &          -0.59 &           0.48 \\
 461B      & Secrétaires de direction, assistants de direction (non cadres)                                                                                                         &          -0.62 &           0.15 \\
 543C      & Employés non qualifiés des services comptables ou financiers                                                                                                           &          -0.63 &          -0.46 \\
 373A      & Cadres des services financiers ou comptables des grandes entreprises                                                                                                   &          -0.64 &           0.08 \\
 461C      & Secrétaires de niveau supérieur (non cadres, hors secrétaires de direction)                                                                                            &          -0.64 &          -0.26 \\
 388D      & Ingénieurs et cadres technico-commerciaux en informatique et télécommunications											    &    -0.64            &        -0.09       
    
\end{longtblr}

\section{Inter- vs. Intra-Community Transitions}\label{appx:intravsinter}

\begin{figure}[H]
\centering
\includegraphics[width=0.9\textwidth]{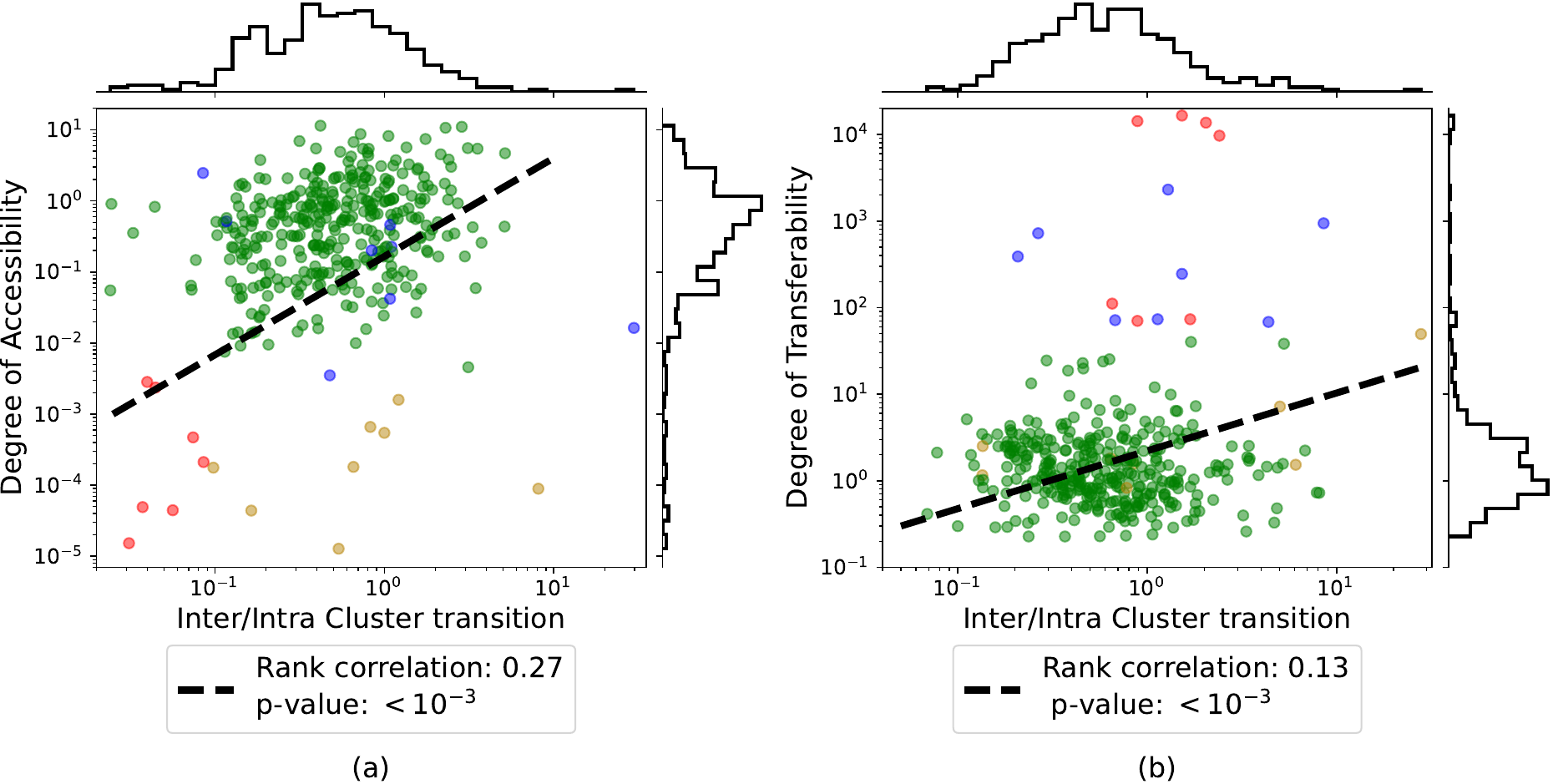}
\caption{\textbf{Correlation of Accessibility and Transferability Metrics with Inter- vs. Intra-Community Transitions}. (a) Scatter plot depicting the relationship between the accessibility of an occupation and weighted inter- vs. intra-community transitions, showing a significant positive rank correlation of 0.27. (b) Scatter plot depicting the relationship between the transferability of an occupation and weighted inter- vs. intra-community transitions, revealing a significant positive rank correlation of 0.13. The color of the points corresponds to the respective accessibility and transferability taxonomy described in Section \ref{sec:fitnessandcomplex}.  A dashed line serves as a visual guide and represents the linear fit of the points.}
\label{fig:intervsintra}
\end{figure}

\section{Influences of Age and Gender on Accessibility and Transferability}

\subsection{Gender Effect on Accessibility and Transferability}\label{sec:gendereff}

\begin{figure}[H]
    \centering
    \includegraphics[width=1\textwidth]{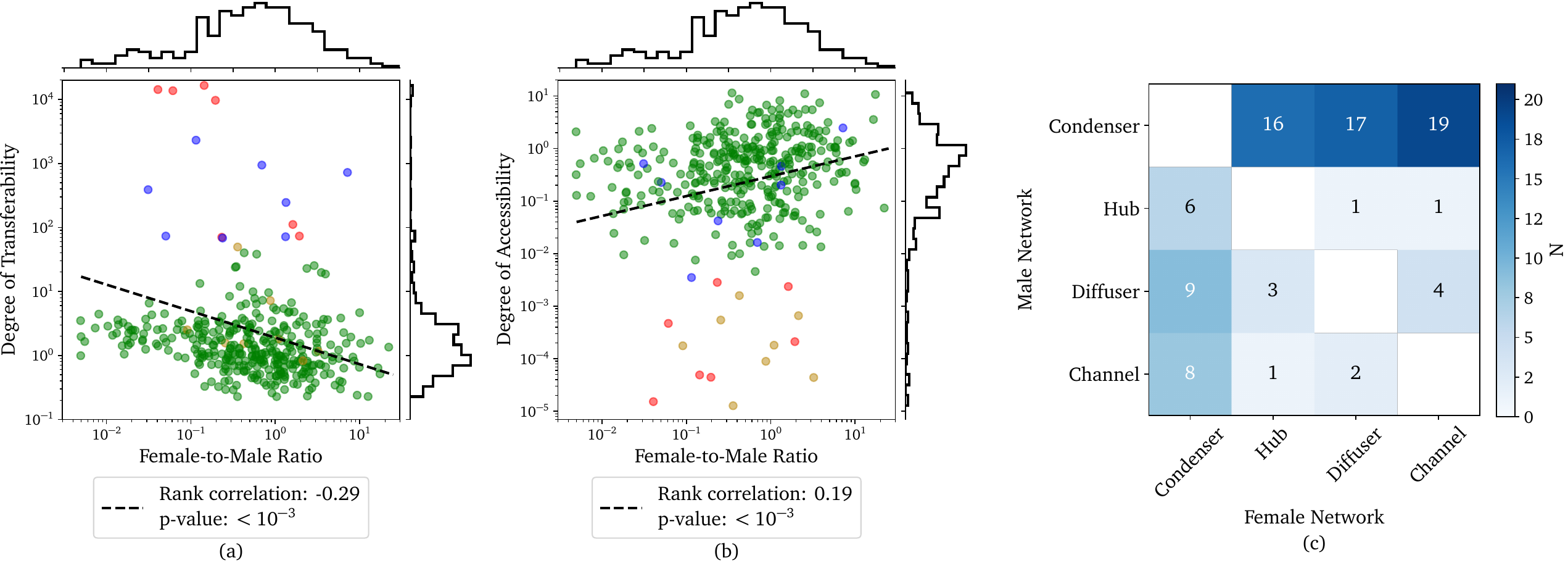}
    \caption{\textbf{Gender Effect on Accessibility and Transferability.} Scatter plot depicts the relationship between Female-to-Male ratio (a) the degree of transferability and (b) the degree of accessibility for each occupation. Color coding corresponds to the introduced occupation taxonomy. A dashed line serves as a visual guide and represents the linear fit of the points. There is a significant and negative rank correlation of -0.29 observed between Female-to-Male ratio and transferability and a significant and positive rank correlation of 0.19 between Female-to-Male ratio and accessibility (legend of (a,b)). Panel (c) illustrates the confusion between taxonomy clustering for the labor market within female and male network. Occupations assigned to the same cluster in both networks are excluded (diagonal elements). Notably, there is a substantial number of occupations in the Condenser cluster for male network that are distributed across the Hubs, Diffuser, and Channels clusters for the female network. This indicates a broader dispersion in the accessibility and transferability space for the female network.}
    \label{fig:gender}
\end{figure}

The female-to-male ratio of employees in an occupation has a significant influence on the accessibility and transferability metrics. The rank correlation between transferability and the gender ratio is at -0.29, while for accessibility the correlation is 0.19, both with significant p-values of $<10^{-3}$(see Figure \ref{fig:gender} (a,b)). Occupations with lower transferability tend to exhibit higher female-to-male ratios, whereas those with higher accessibility tend to have higher female-to-male ratios. This discrepancy between the transferability and accessibility and gender ratios in different occupations is captured by our metrics.

To assess gender differences, we construct separate male and female networks from the data and independently calculate their accessibility and transferability metrics. Figure \ref{fig:gender} (c) shows that within the female network, 52 occupations that are classified as Condensers in the male network are classified as Hub, Diffuser, and Channel in the female network. This means that the dispersion of transferability and accessibility of occupations is much wider, and the female labor flow network reveals more occupations with fewer transitions between them and more occupations that are challenging to access. Since only the relative difference between the accessibility and transferability matters, when it comes to reallocating workers in response to shocks, female workers face higher constraints than male workers. 

Using the accessibility transferability metric, we can identify exactly the bottleneck occupations in the female labor flow network. In the male network, occupations like IT project managers (388C), Computer operating employees (544A), and Hospitality Staff in Educational Institutions have very low transferability (525B). Occupations in the Channel category with the least accessibility include salaried hairdressers (562B), visual artists (354A), and babysitters (563A), which, on average, have a higher proportion of female workers than male.

Conversely, in the female network, Secretaries (461C), Qualified personnel from legal departments (543E), and Unskilled employees of accounting and financial services (543C) have the lowest transferability (see Figure \ref{fig:gender}). Occupations with the lowest accessibility in the Channel category are male dominated, including sea fishers and aquaculture workers (692A), Transport Controllers (546A), and Qualified drivers of guided transport vehicles (654B). 

\subsection{Age Effect on Accessibility and Transferability}\label{sec:ageeff} 

\begin{figure}[H]
    \centering
    \includegraphics[width=1\textwidth]{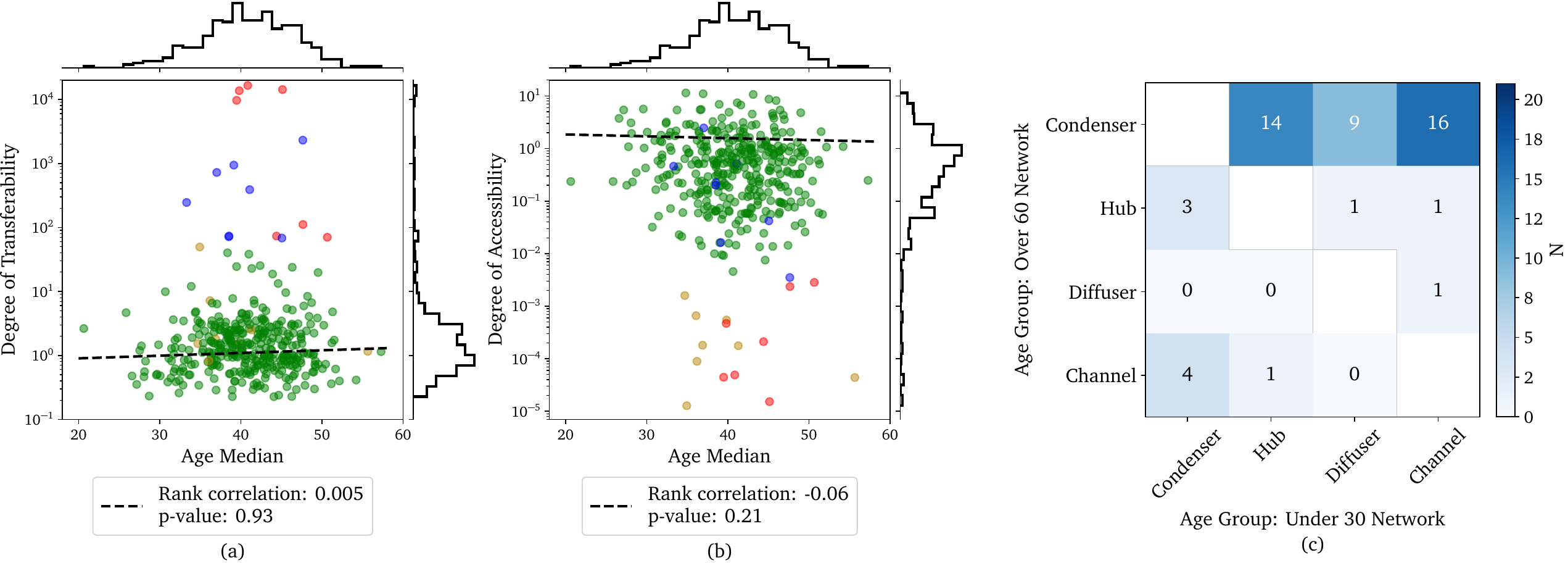}
    \caption{\textbf{Age Effect on Accessibility and Transferability.} Scatter plot depicts the relationship between age median and (a) the degree of transferability and (b) the degree of accessibility for each occupation. Color coding corresponds to the introduced occupation taxonomy. A dashed line serves as a visual guide and represents the linear fit of the points. There is no significant rank correlation observed between age and transferability nor between age and accessibility (legend of (a,b)). Panel (c) illustrates the confusion between taxonomy clustering for the labor market within the age groups below 30 and above 60. Occupations assigned to the same cluster in both networks are excluded (diagonal elements). Notably, there is a substantial number of occupations in the Condenser cluster for the age group above 60 that are distributed across the Hubs, Diffuser, and Channels clusters for the age group below 30. This indicates a broader dispersion in the accessibility and transferability space for the age group below 30.}
    \label{fig:age}
\end{figure}

Figures \ref{fig:age} (a) and (b), show the relationship between the median age of an occupation and its (a) degree of transferability and (b) accessibility. There is no significant rank correlation between the median age and either transferability or accessibility. This indicates that our metrics for accessibility and transferability are note correlated to age-related effects.

To investigate the impact of age on our findings, we construct a labor flow network separately for workers below the age of 30 who are typically still in the learning or training phase for specific occupations and for workers above 60 who might experience seniority effects \citep{eurostat2018being}. 

When analyzing transitions separately for workers under 30 and those over 60, we can discover on a single occupational level the presence of entry positions and those offering more opportunities for older workers. Figure \ref{fig:age} (c), shows the confusion matrix illustrating the classification of the same occupations across different age groups\footnote{We exclude diagonal elements, representing occupations classified the same in both age groups.}. Fifty occupations are classified differently between the age groups. The majority of shifts in classification occur from the "Condenser" category in the over-60 network to other categories in the under-30 network. Suggesting a more dispersed spectrum of accessibility and transferability for the under-30 age group, which indicates a higher prevalence of occupations that are less accessible but possess a greater degree of transferability. 

If an occupation in the network of workers below 30 is in the Hub (Condenser) category and switches to a Diffuser position (Channel) in the network above 60, it suggests higher accessibility to younger workers, indicating potential entry-level roles. Examples include Unskilled employees of accounting or financial services (543C) or Employed eyewear opticians and audioprosthetists (433B).

Conversely, if transferability of an occupation increases with age, leading to a shift from Condenser (Channel) to Hub (Diffuser), it suggests that opportunities for specific occupations tend to expand with age. Examples of such jobs include Qualified employees of accounting or financial services (543B) and Sellers of beauty, luxury and optical products (554F) (see Figure \ref{fig:age}). For the first occupation, one might argue that specific skills necessitate time for acquisition, whereas in the latter occupation, acquiring skills like selling products to individuals and cultivating a network of trusted customers could create new opportunities for occupational transitions.

\section{Policy Evaluation with Matrix Perturbation}\label{appx:policyperturbation}
The occupational transition matrix $\transmat[][]$ is a right stochastic matrix where each entry 
\begin{equation}
    \transmat = \mathbf{Pr}(X_i = i | X_j = j)
\end{equation}
corresponds to the probability of transitioning to occupation j, given an occupation i, with $\sum_j \transmat = 1$. Then there exists one eigenvalue $\lambda_1=1$ corresponding to an eigenvector $\Vec{v}_1 = \Vec{\pi}$ that is the stationary distribution $\Vec{\pi}$ for a markov process of $N \to \infty$ steps.

Let $\transmat[][]$ be a symmetric matrix with eigenvectors $\vec v_i$ and $\lambda_i$, then
\begin{equation}
    \transmat[][] \vec v_i = \lambda_i \vec v_i
\end{equation}
Now let us consider a perturbation $\epsilon V$, then 
\begin{equation}
 (\transmat[][] + \epsilon V) \vec v^\star_2 = \lambda^\star_2 \vec v^\star_2    
\end{equation}
We can write $\vec v^\star_2$ and $ \lambda^\star_2$ as 
\begin{align}
    \vec v^\star_2 &= \vec v_2 + \epsilon \vec v + ...\\
    \lambda^\star_2 &= \lambda_2 + \epsilon \lambda + ...
\end{align}
with this 
\begin{equation}
    (\transmat[][] + \epsilon V)(\vec v_2 + \epsilon \vec v) = (\lambda_2 + \epsilon \lambda)( \vec v_2 + \epsilon \vec v)
\end{equation}
If we separate the equation for each order of $\epsilon$, neglect higher orders of $\epsilon$ and use the symmetry of $\transmat[][]$, such that $\vec v_2^T\transmat[][]\vec v = \vec v_2^T\lambda_2\vec v$, then
\begin{align}
    \lambda &= \frac{\vec v_2^TV\vec v_2}{\vec v_2^T\vec v_2} \\
    \lambda_2^\star &= \lambda_2 + \epsilon\frac{\vec v_2^TV\vec v_2}{\vec v_2^T\vec v_2} \\
    \Delta \lambda &= \lambda_2^\star - \lambda_2 =  \epsilon\frac{\vec v_2^TV\vec v_2}{\vec v_2^T\vec v_2} \label{eq:product}
\end{align}

with $\vec v_i^T \vec v_i = 1$. Knowing the eigenvectors $\vec v_i$ of the unperturbed occupational transition matrix $\transmat[][]$, we can calculate the change of eigenvalue $\Delta \lambda$ w.r.t. a small perturbation of a rate of a single link. We are looking for a policy $V$ where $\Delta \lambda_2 < 0$ as with this, the convergence of the perturbed matrix is faster since the spectral gap of the perturbed matrix gets larger.  Let's say we denote the target occupation with $k$ and increase the flow between $k$ and a destination occupation $l^+$, simultaneously we decrease the flow from target occupation $k$ to source occupation $l^-$. Then the perturbation $V$ is

\begin{align}
    V &= 
    \begin{pmatrix}
    0 & ... & ... & ... & 0 \\
    ... & ... & ... & ... & ... \\
    -1 & ... & 1 &  ... &  0\\
    ... & ... & ... & ... & ... \\
    0 & ... & ... & ... & 0 \\
  \end{pmatrix} \\
   \mathrm{or} \\
  (V)_{i,j} &= 
  \begin{cases}
      -1, \quad i=l^-, j=k\\
      1, \quad i=l^+, j=k\\
      0, \quad \mathrm{else}
  \end{cases}
\end{align}
with this the product in eq. \ref{eq:product} gets
\begin{align}
    \vec v_2^TV\vec v_2 &= (\vec v_2^T)_k \cdot \sum_i V_{i,k} (\vec v_2)_i \\
    &= (\vec v_2^T)_k \cdot ((\vec v_2)_{l^+} - (\vec v_2)_{l^-}) \label{eq:final_perturbation}
\end{align}
Here we can derive four conditions for the sign of $\Delta \lambda$ 
\begin{align}
    \Delta \lambda  
    \begin{cases}
      > 0, \quad \mathrm{if} \; (\vec v_2)_{l^+}>(\vec v_2)_{l^-} \wedge (\vec v_2)_{k} > 0\\
      > 0, \quad \mathrm{if} \; (\vec v_2)_{l^+}<(\vec v_2)_{l^-} \wedge (\vec v_2)_{k} < 0\\
      < 0, \quad \mathrm{if} \; (\vec v_2)_{l^+}>(\vec v_2)_{l^-} \wedge (\vec v_2)_{k} < 0\\
      < 0, \quad \mathrm{if} \; (\vec v_2)_{l^+}<(\vec v_2)_{l^-} \wedge (\vec v_2)_{k} > 0 \label{eq:signcases}
    \end{cases}
\end{align}

\begin{figure}[H]
    \centering
    \includegraphics[width=1\textwidth]{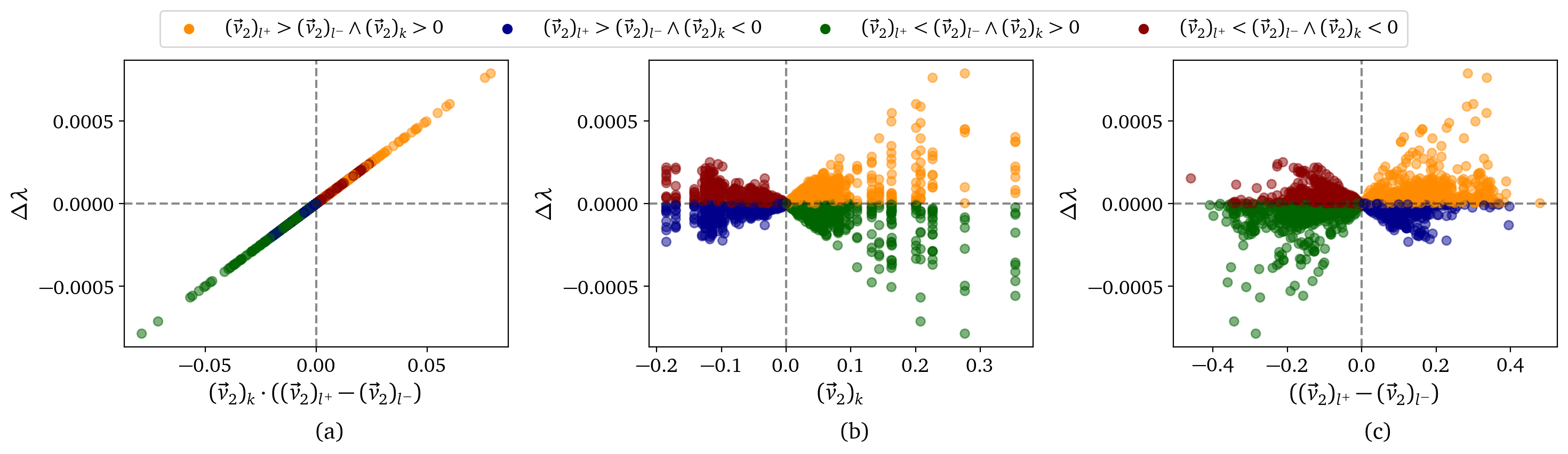}
    \caption{\textbf{Matrix Perturbation of the French Labor Market.} (a) Scatter plot showing eq. \ref{eq:final_perturbation}, demonstrating a strong positive correlation that supports the validity of the assumptions made for first-order matrix perturbation. The color coding of points corresponds to the sign cases of $\Delta \lambda$ (see eq. \ref{eq:signcases}). (b) Scatter plot showing the relationship between the eigenvalue difference and the value of the eigenvector for the target occupation, $(\vec{v}_2)_k$. (c) Scatter plot showing the eigenvalue difference against the difference in eigenvector values between the destination and source occupations, $((\vec{v}_2)_{l^+} - (\vec{v}_2)_{l^-})$.}
    \label{fig:confirmation_theory}
\end{figure}

\begin{figure}[H]
    \centering
    \includegraphics[width=1\textwidth]{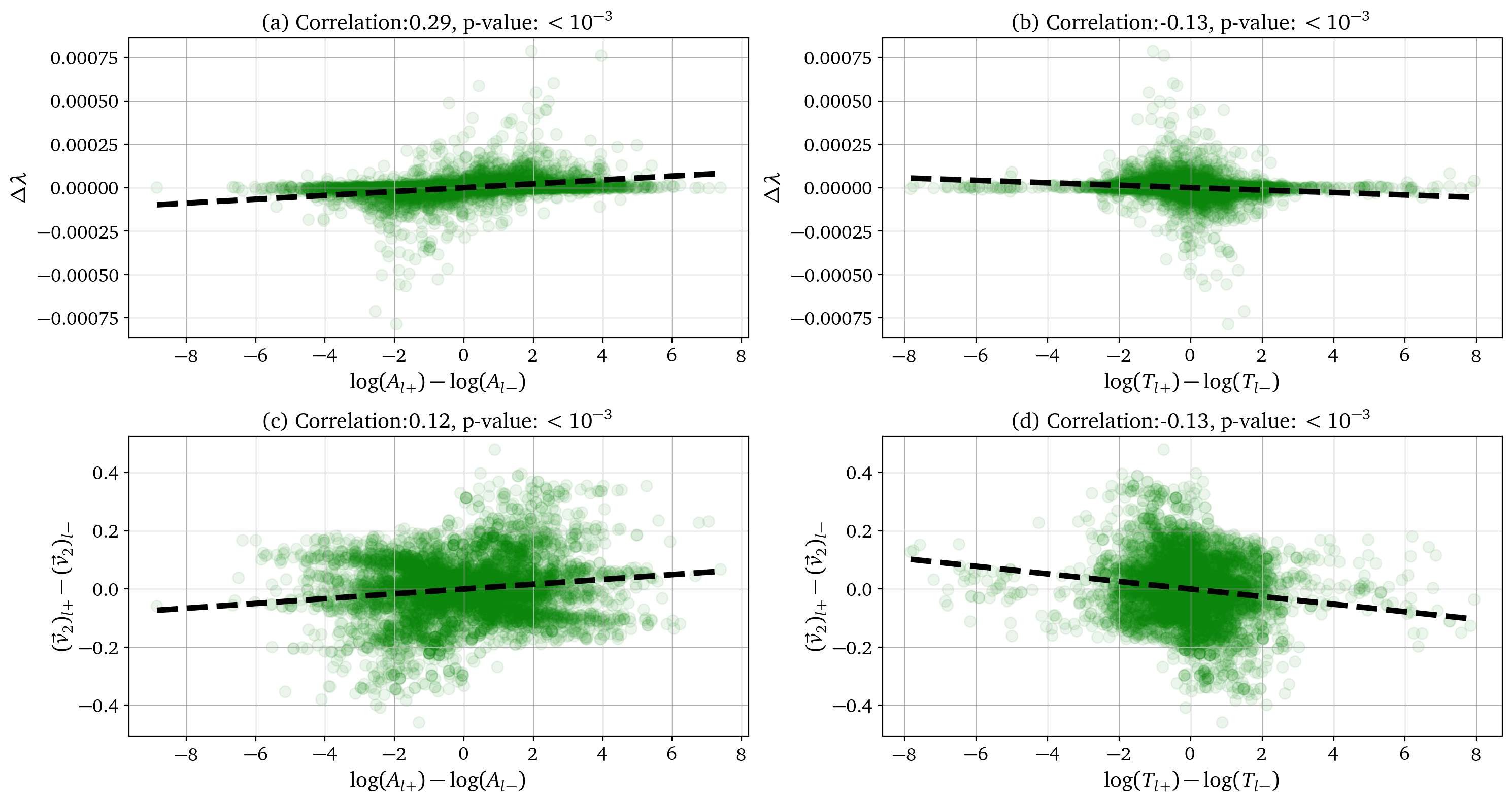}
    \caption{\textbf{Correlation of Accessibility and Transferability Metrics with Changes in Eigenvalue} (a) Scatter plot showing the log difference in accessibility metrics between destination and source occupations against the change in the second eigenvalue. (b) Scatter plot showing the log difference in transferability metrics between destination and source occupations against the change in the second eigenvalue. (c) Scatter plot showing the log difference in accessibility metrics between destination and source occupations against the difference in the corresponding values of the second eigenvector. (d) Scatter plot showing the log difference in transferability metrics between destination and source occupations against the difference in the corresponding values of the second eigenvector.}
    \label{fig:correlation_at}
\end{figure}

\end{document}